\documentclass[a4paper,11pt]{article}
\pdfoutput=1
\usepackage{jheppub}

\usepackage{amssymb,amsmath}
\usepackage[normalem]{ulem}
\usepackage[utf8x]{inputenc}
\usepackage{slashed}
\usepackage{graphicx}
\usepackage{here}

\usepackage{csquotes} 
\usepackage{comment}
\usepackage{mathrsfs}
\usepackage{float}
\usepackage{ascmac}

\usepackage{multirow}
\usepackage{longtable}

\usepackage[hang,small,bf]{caption}
\usepackage[subrefformat=parens]{subcaption}
\def\lsim{\mathrel{\rlap{\lower4pt\hbox{\hskip1pt$\sim$}}
    \raise1pt\hbox{$<$}}}         
\def\gsim{\mathrel{\rlap{\lower4pt\hbox{\hskip1pt$\sim$}}
    \raise1pt\hbox{$>$}}}         

\numberwithin{equation}{section}

\preprint{
\begin{minipage}{5cm}
\small
\flushright
KYUSHU-HET-257
\end{minipage}}

\title{Exploring the flavor structure of quarks and leptons with reinforcement learning}

\author{Satsuki Nishimura,} 
\author{Coh Miyao,} 
\author{Hajime Otsuka} 
\affiliation{
Department of Physics, Kyushu University, 744 Motooka, Nishi-ku, Fukuoka 819-0395, Japan\\
}
\emailAdd{nishimura.satsuki@phys.kyushu-u.ac.jp}
\emailAdd{miyao.coh@phys.kyushu-u.ac.jp}
\emailAdd{otsuka.hajime@phys.kyushu-u.ac.jp}

\abstract{
We propose a method to explore the flavor structure of quarks and leptons with reinforcement learning. 
As a concrete model, we utilize a basic value-based algorithm for models with $U(1)$ flavor symmetry. 
By training neural networks on the $U(1)$ charges of quarks and leptons, 
the agent finds 21 models to be consistent with experimentally measured masses and mixing angles of quarks and leptons. 
In particular, an intrinsic value of normal ordering tends to be larger than that of inverted ordering, and the normal ordering is well fitted with the current experimental data in contrast to 
the inverted ordering. 
A specific value of effective mass for the neutrinoless double beta decay and a sizable leptonic CP violation induced by an angular component of flavon field are predicted by autonomous behavior of the agent. 
Our finding results indicate that the reinforcement learning can be a new method for understanding the flavor structure.
}
\makeatletter
\gdef\@fpheader{}
\makeatother

\begin{document}

\maketitle

\section{Introduction}

The origin of flavor structure of quarks and leptons is one of the unsolved problems in particle physics. 
To understand the peculiar pattern of fermion masses and mixings, the flavor symmetries have been utilized to explain these flavor puzzle.\footnote{See, e.g., Refs. \cite{Altarelli:2010gt,Ishimori:2010au,Hernandez:2012ra,King:2013eh,King:2014nza,Tanimoto:2015nfa,King:2017guk,Petcov:2017ggy,Feruglio:2019ybq,Kobayashi:2022moq} for a review.} An attractive feature of the flavor symmetry is that it may connect the bottom-up approach of flavor model building with the top-down approach based on an ultra-violet completion of the Standard Model such as the string theory. 
In this paper, we adopt a bottom-up approach to explore the flavor structure of quarks and leptons. 

\medskip

In most traditional approaches to address the flavor structure of quarks and leptons, 
one assumes a certain representation of quarks and leptons under the flavor symmetry among all possible configurations. 
Indeed, it will be difficult to exhaust all possible realistic flavor patterns from a 
broad theoretical landscape. 
For instance, in a global $U(1)$ flavor symmetric model using the Froggatt-Nielsen (FN) mechanism \cite{Froggatt:1978nt}, 
we have a degree of freedom of $U(1)$ charge assignment of each quark and lepton. When we consider the flavor dependent $U(1)$ charges of the quarks $q_i$ within the range $-9 \leq q_i \leq 9$, it results in ${\cal O}(10^{14})$ patterns of $U(1)$ charges even for the quark sector. 
When we combine with the lepton sector, we are faced with the problem of doing
a brute-force search over a higher-dimensional parameter space. 
This is a simple flavor model using the continuous flavor symmetry, but in general, it is difficult to find a realistic flavor pattern from a huge amount of possibilities in flavor models with discrete symmetries. 
Thus, it motivates us to apply recent machine learning techniques for an exhaustive search of flavor models. 
 
\medskip

In order to explore such a huge landscape of flavor models, in this paper, we will deal with a reinforcement learning (RL), which is known as one of machine learning techniques
\footnote{In addition to RL, supervised learning and unsupervised learning are known as machine learning methods. 
The supervised learning can estimate the correspondence between reference data and teaching signals, while the unsupervised learning can find the similarity among reference data. 
While these methods require a large amount of data, RL can find best solutions from a small amount of data by repeatedly trying to solve the problem. 
This feature makes RL useful not only in the search for flavor models but also in the field of particle theory.}. 
In the framework of RL, an agent autonomously discovers desirable behavior to solve given problems, where a systematic search is impossible. 
So far, such a technique was utilized to find the parameter space of FN models with an 
emphasis on the quark sector \cite{Harvey:2021oue}, where only the experimental values of quark masses and mixing angles are efficiently reproduced. 
However, it is quite important to see whether one can reproduce the flavor structure of all the fermion masses and mixings. Throughout this paper, we assume the Type-I see-saw mechanism to realize active neutrino masses and large mixing angles in the lepton sector. 
We will utilize a basic value-based algorithm, where the neural network is trained by data given by an environment. 
To find the flavor structure of quarks and leptons efficiently, we set the environment where the inputs consist of $U(1)$ charges of quarks and leptons under the $U(1)$ flavor symmetry and the coefficients appearing in Yukawa couplings are randomly fixed as ${\cal O}(1)$ real values. 
The outputs of the neural network are probabilities for the action determined by a policy. 
Here, the action of agent is given by increasing or decreasing one of the $U(1)$ charges by one, and 
the agent receives the reward (punishment) for this action when the fermion masses and mixings determined by the $U(1)$ charges approach (deviate from) the experimental values. 
Specifically, the reward function is defined by the intrinsic value consisting of fermion masses and elements of CKM and PMNS matrices whose values are minimized under a vacuum expectation value (VEV) of complex flavon field.

\medskip

In addition to reproducing the experimental values, parameter search with RL will provide new insights on the neutrino mass ordering and CP phase in the lepton sector. 
Note that a source of CP violation is assumed to be originating from the phase of complex flavon field. 
By training neural networks without specifying the neutrino mass orderings, 
RL can help to find whether the neutrinos are in the normal ordering or in the inverted ordering. 
From the results of trained network, we find that the normal ordering is statistically favored by the 
agent. Furthermore, the sizable Majorana CP phases and effective mass for the neutrinoless double beta decay 
are predicted around specific values.

\medskip

This paper is organized as follows. 
After briefly reviewing RL with an emphasis on Deep Q-network in Sec. \ref{sec:DQN}, 
we establish the FN model with RL in Sec. \ref{sec:model}. 
We begin with the model building with RL by focusing on the quark sector in Sec. \ref{sec:quark}, and 
the training of the lepton sector is performed in Sec. \ref{sec:lepton}. 
In particular, we analyze two scenarios for the neutrino sector. 
In Sec. \ref{sec:neutrino1}, we implement the FN model with fixed neutrino mass ordering to the neural network, but 
the neutrino mass ordering is not specified in the analysis of Sec. \ref{sec:neutrino2}. 
Sec. \ref{sec:con} is devoted to the conclusion and discussion. 
In Appendix \ref{app}, we list our finding $U(1)$ charge assignment of quarks and leptons.

\section{Reinforcement learning with deep Q-network}
\label{sec:DQN}

In this section, we briefly review RL with the Deep Q-Network (DQN) used in the analysis
of this paper. For more details, see, e.g., Ref. \cite{RL}. The RL is constructed by an agent and an environment.
At a certain time, the agent observes the environment and takes some action.
Depending on the change of the environment caused by the action, the agent will receive rewards or penalties.
By repeating those processes and searching for actions that maximize the total rewards,
the agent is designed to exhibit autonomous behavior in the environment.

\medskip

To determine the action, we utilize the neural network model. 
In the multi-layer perceptrons, a $n$-th layer with $N_{n-1}$-dimensional vector $\Vec{x}_{n-1} = (x_{n-1,1},x_{n-1,2},\cdots, x_{n-1,N_{n-1}})$ in multi-layer perceptrons transforms into a $N_n$-dimensional vector $\Vec{x}_n = (x_{n,1},x_{n,2},\cdots, x_{n,N_{n}})$: 
\begin{align}
x_{n,i} = h_n (w_{ij}^n x_{n-1,j} + b_i^n), \label{eq:activation}
\end{align}
with $h, w, b$ being the activation function, the weight and the bias, respectively. 
In the analysis of this paper, we employ a fully-connected layer. 
Then, the DQN known as one of the RL methods is characterized by Q network, target network, and experience replay.
In this paper, we consider the neural networks whose output can be constructed by a softmax layer. Note that Q network and target network have same structures, but weights and biases in those are generically different from each other.

\medskip

RL using the DQN proceeds through the following 5 steps:
\begin{enumerate}
    \item An agent observes the environment state $s$ which is given as an input in the target neural network, as shown in Fig. \ref{fig:input}. 
    The target network (TN) gives the probabilities $p$ as an output. 
    Since we adopt the softmax layer defined by
    \begin{align}
        f : \mathbb{R}^n \rightarrow [0,1]^n,
        \label{eq:softmax}
    \end{align}
    with $f({\bf x})_i = e^{x_i} / \sum_{i=1}^n e^{x_i}$, the output will also be regarded as probabilities. 

\begin{figure}[H]
    \centering
    \includegraphics[width=75mm]{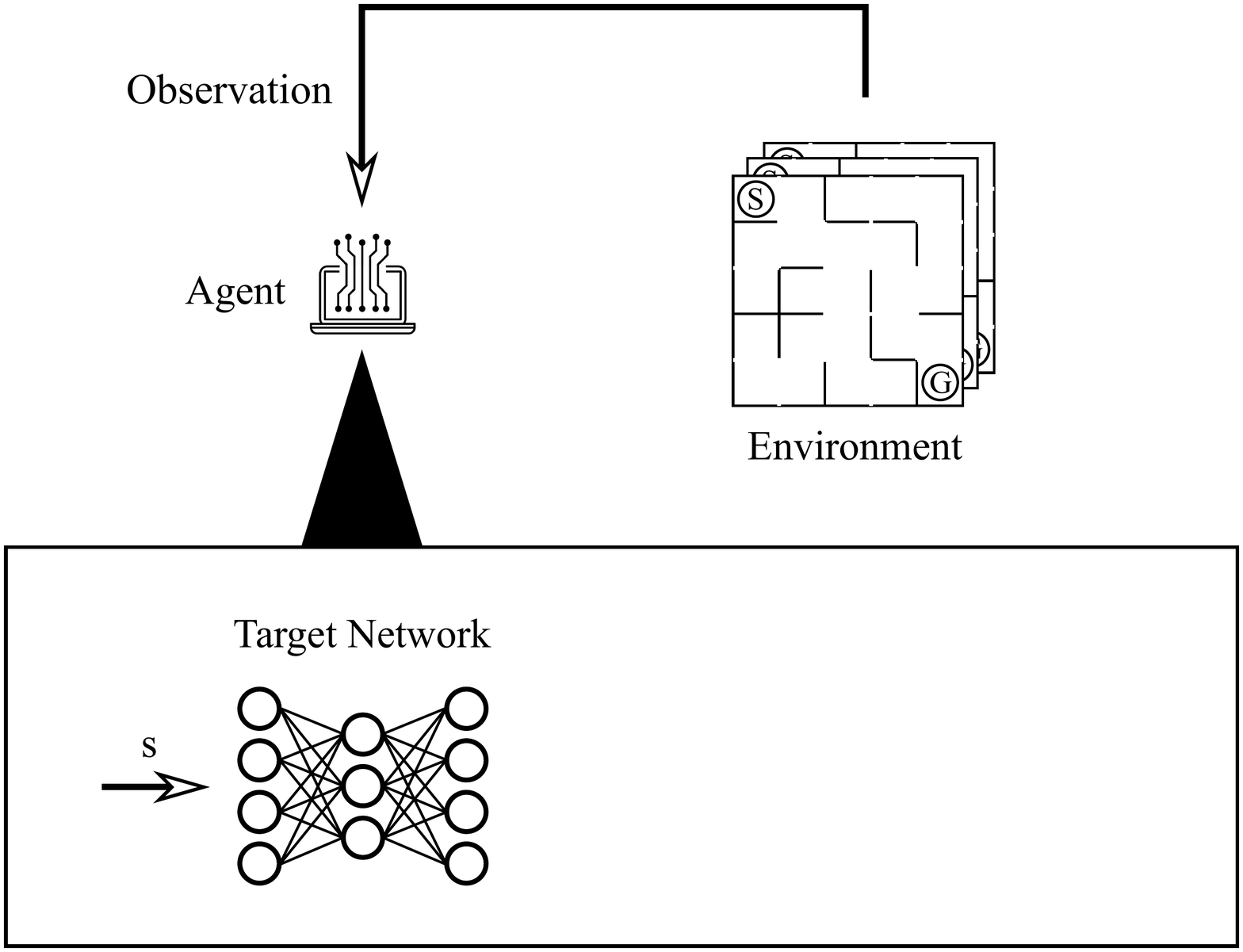}
    \caption{As a first step, the state $s$ observed by the agent is given as input in the target network.}
    \label{fig:input}
\end{figure}

    \item In a second step, the agent determine the {\it action} $\mathfrak{a}$, taking into account the probabilities $p$ given by the first step.
    At the initial stage, the neural network cannot judge whether the action is an appropriate one. Let us denote the action with the highest probability $\mathfrak{b}$. To acquire the ability of autonomous behavior for the agent, we adopt the $\epsilon$-greedy method, where the greedy action $\mathfrak{b}$ is selected with probability $1-\epsilon$ and a random action $\mathfrak{c}$ is selected with probability $\epsilon$ (see Fig. \ref{fig:output}), that is,
    \begin{align}
        \mathfrak{a} = 
        \left\{
        \begin{array}{l}
             \mathfrak{b}\quad (\text{with}\,1-\epsilon)  \\
             \mathfrak{c}\quad (\text{with}\,\epsilon) 
        \end{array}
        \right.
        .
    \end{align}

    \begin{figure}[H]
        \centering
        \includegraphics[width=75mm]{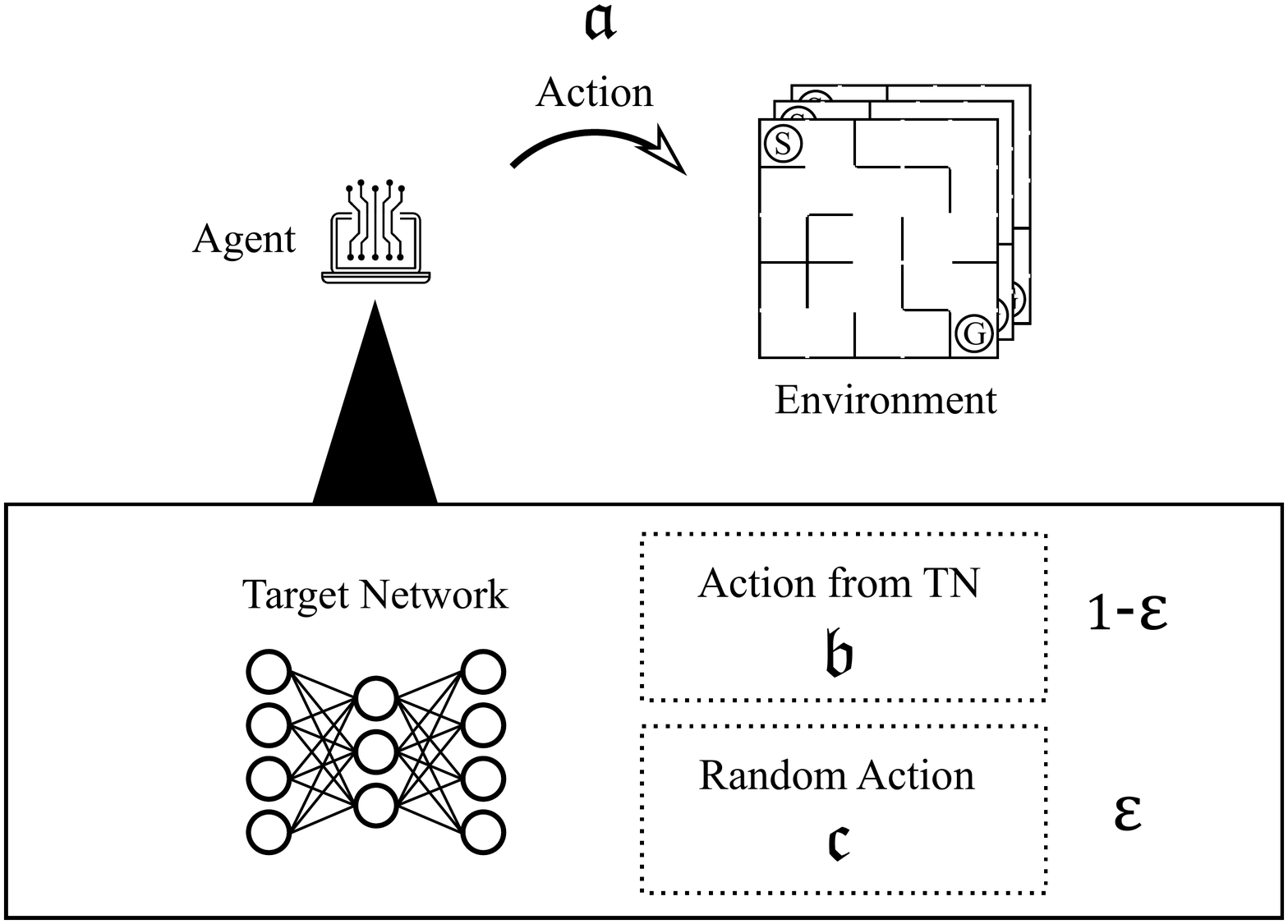}
        \caption{In the second step, the agent selects the action $\mathfrak{a}$ through the $\epsilon$-greedy policy.}
        \label{fig:output}
    \end{figure}

    By repeating this process, a sequence of the states is defined as follows: 
    \begin{align}
        s_{1} \xrightarrow{\mathfrak{a}_{1}}
        s_{2} \xrightarrow{\mathfrak{a}_{2}}
        s_{3} \xrightarrow{\mathfrak{a}_{3}} \cdots,
    \end{align}
    and this chain is called an {\it episode}. The initial environment state $s_{1}$ is chosen randomly.
    The number of actions is specified by $N_{\rm step}$ for one episode and the agent repeats this step $N_{\rm ep}$ times as shown in Table \ref{tab:episode}. 
    Note that the greedy action is determined by taking into account the probabilities $p$ in the first step. 
    The value of $\epsilon$ is chosen to ensure that the agent gradually takes the greedy action, 
    whose explicit form will be given by
    \begin{align}
        \epsilon = \max \left( \epsilon_0 r^{k-1},\  \epsilon_{\rm min}\right),
    \end{align}
    with $k=1,2,...,N_{\rm ep}$. 
    In the following analysis, we adopt $\epsilon_0 = 1,\ r=0.99999$ and $\epsilon_{\rm min}=0.01$. 
    This definition means that the agent gains various experiences for the large $\epsilon$, using which the agent gradually takes a plausible action.

    \begin{table}[H]
        \centering
        \begin{tabular}{|c|c|c|c|c|}\hline
             & Step 1 & Step 2 & $\cdots$ & Step $N_{\rm step}$ \\ \hline
         Episode 1    & $s_1^1$ & $s_2^1$ & $\cdots$ & $s_{N_{\rm step}}^1$\\ \hline
         Episode 2    & $s_1^2$ & $s_2^2$ & $\cdots$ & $s_{N_{\rm step}}^2$\\ \hline
         $\vdots$    & $\vdots$ & $\vdots$ & $\cdots$ & $\vdots$ \\ \hline
         Episode $N_{\rm ep}$    & $s_1^{N_{\rm ep}}$ & $s_2^{N_{\rm ep}}$ & $\cdots$ & $s_{N_{\rm step}}^{N_{\rm ep}}$ \\ \hline
        \end{tabular}
        \caption{The environment states $s$ are changed by the actions. The agent performs at most $N_{\rm step}$ step for one episode.}
        \label{tab:episode}
    \end{table}

    \item The state $s$ is updated to $s'$ through the action $\mathfrak{a}$. Depending on the states $s'$, the agent receives a reward ${\cal R}$. In a third step, the transition $e=\langle s, \mathfrak{a}, s', {\cal R}\rangle$ corresponding to trajectories of experience is stored in the replay buffer as seen in Fig. \ref{fig:buffer}.

\begin{figure}[H]
    \centering
    \includegraphics[width=75mm]{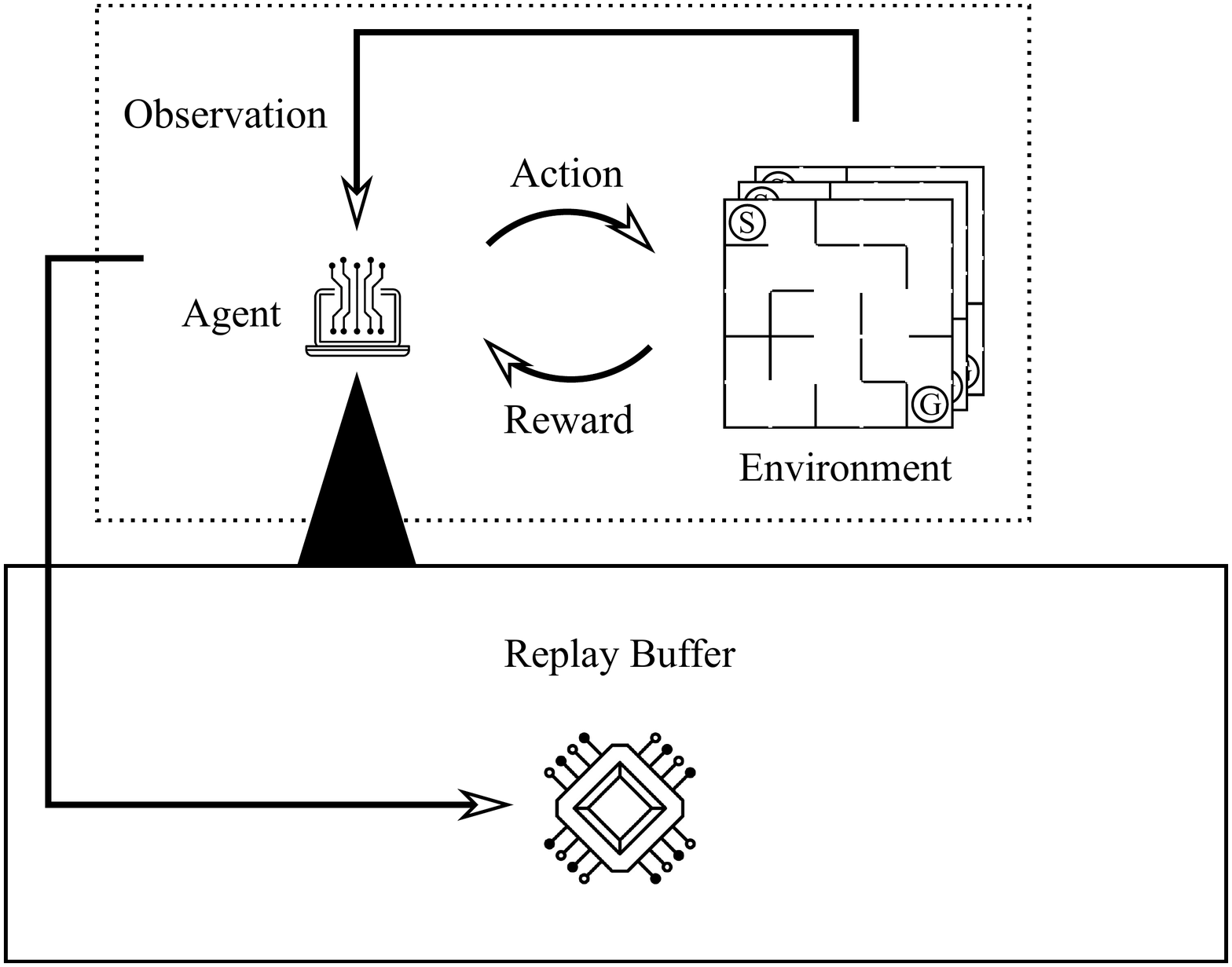}
    \caption{In the third step, the transition $e=\langle s, \mathfrak{a}, s', {\cal R}\rangle$ is stored in the replay buffer.}
    \label{fig:buffer}
\end{figure}

    \item A fourth step consists of ``experience replay'' and ``stochastic gradient method''. 
    The experience replay is to extract a mini-batch of transitions randomly sampled from the replay buffer, 
    where the Q network is optimized by using at most a batch of transitions times epoch number. 
    The advantage of this experience replay is twofold. First, the transitions in a batch are uncorrelated due to the random selection of past experiences. Second, one can reuse each transition in the training because all the experience is stored in the replay buffer.   

    \newpage
    \medskip

    In the framework of DQN, there are two neural networks: Q network and target network. 
    The Q network is updated by the stochastic gradient method where the mini-batch of transitions is used in the training data (see Fig. \ref{fig:replay}). 
    When we denote outputs of the Q network and the target network by $y(s)$ and $y'(s')$, respectively, the weights and the biases are updated by minimizing a loss function $L(y,y')$. 
    In this paper, we adopt the Huber function:
    \begin{align}
    \begin{split}
        L_{\rm Huber}(y, y') = 
        \left\{
        \begin{array}{l}
             \frac{1}{2} (y'_{\left( {\cal R} \right)i} - y_i)^2\qquad \qquad\,\,\,\, \text{if}\ |y'_{\left( {\cal R} \right)i} - y_i| \leq \delta   \\
             \delta \cdot |y'_{\left( {\cal R} \right)i} - y_i| - \frac{1}{2}\delta^2\qquad \text{if}\ |y'_{\left( {\cal R} \right)i} - y_i| > \delta
        \end{array}
        \right.
        ,
    \end{split}
    \label{eq:Huber}
    \end{align}
    with $y'_{\left( {\cal R} \right)} = {\cal R} + \gamma y',\ \gamma = 0.99$ and $\delta = 1$, which combines a mean squared error and a mean absolute error
    \footnote{The inputs of $y$ and $y'$ are the state $s$ and $s'$ from the transition $e$, respectively. This construction of the loss function is grounded in the Bellman equation, which is the formulation of RL. The relation between the formulation of RL and DQN is described in Appendix \ref{app_RL}.}. 
    Note that the training of Q network is carried out at the end of one episode, including at most the $N_{\rm step}$ step as shown in Table \ref{tab:episode}. 

\begin{figure}[h]
    \centering
    \includegraphics[width=75mm]{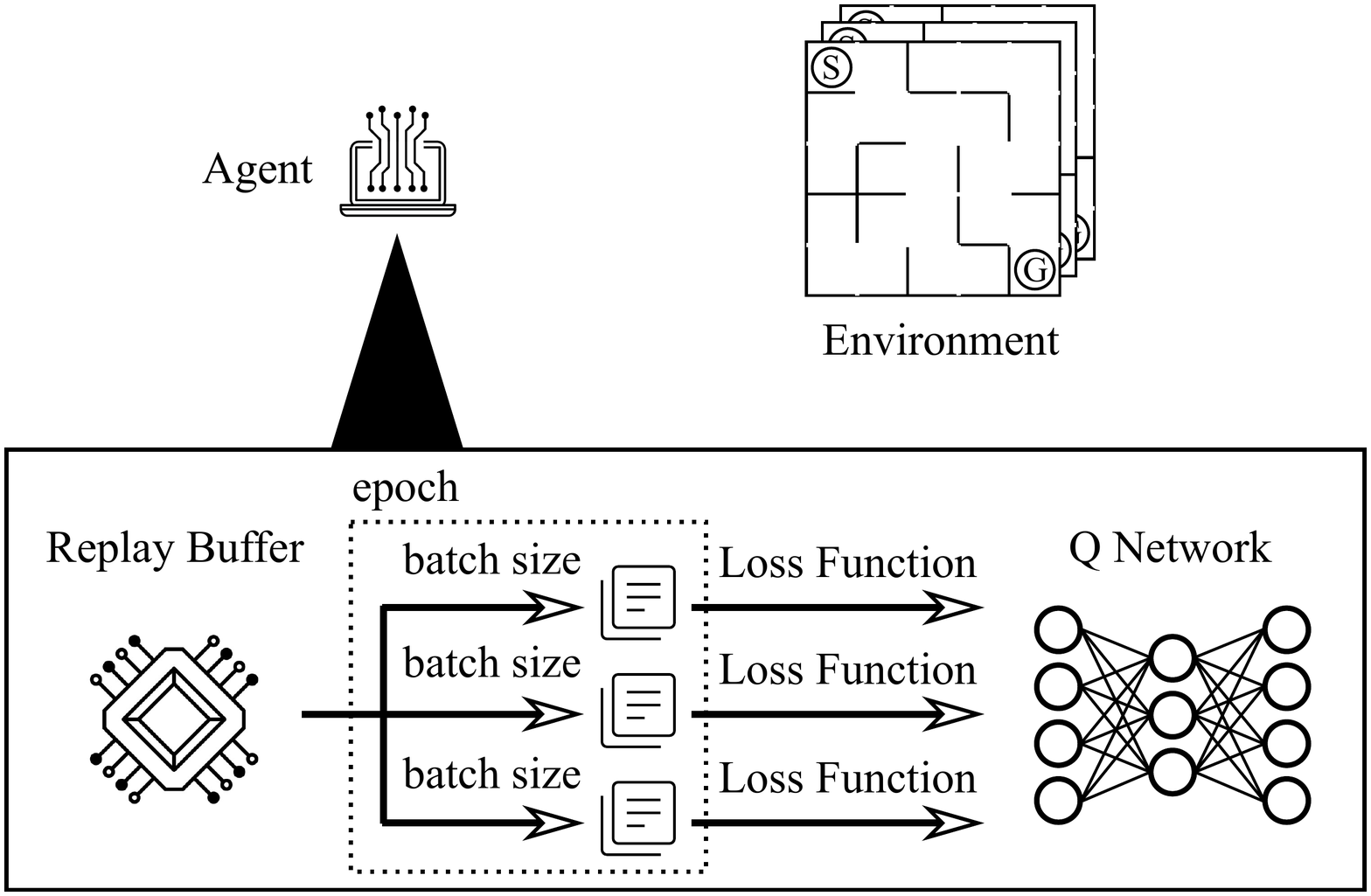}
    \caption{In the fourth step, we randomly pick up the transitions with batch size from the replay buffer, and the weights and the biases of the Q network are updated by the stochastic gradient method in terms of these transitions.}
    \label{fig:replay}
\end{figure}

    \item The Q network and the target network have different parameters $\Theta=\{w,b\}$ and $\Theta'=\{w',b'\}$, respectively. 
    Lastly, the parameters $\Theta$ in the Q network are slightly reflected in $\Theta'$ in the target network (see Fig. \ref{fig:update}). 
    Specifically, in the case of a soft update, this reflection proceeds as follows:
    \begin{align}
        \Theta' \leftarrow (1-\alpha)\Theta' + \alpha \Theta,
    \end{align}
    where $\alpha$ is called the ``learning rate''. This procedure can suppress rapid parameter changes and update the target network while maintaining learning stability. 
    When $\alpha$ is large, the stability of the learning will be lost, but the small $\alpha$ will lead to a slow learning. In this paper, we adopt $\alpha = 2.5\times 10^{-4}$. 
    Since the stochastic gradient method is not used in updating the parameters of the target network as described above, no loss function is defined for this network.

\begin{figure}[H]
    \centering
    \includegraphics[width=75mm]{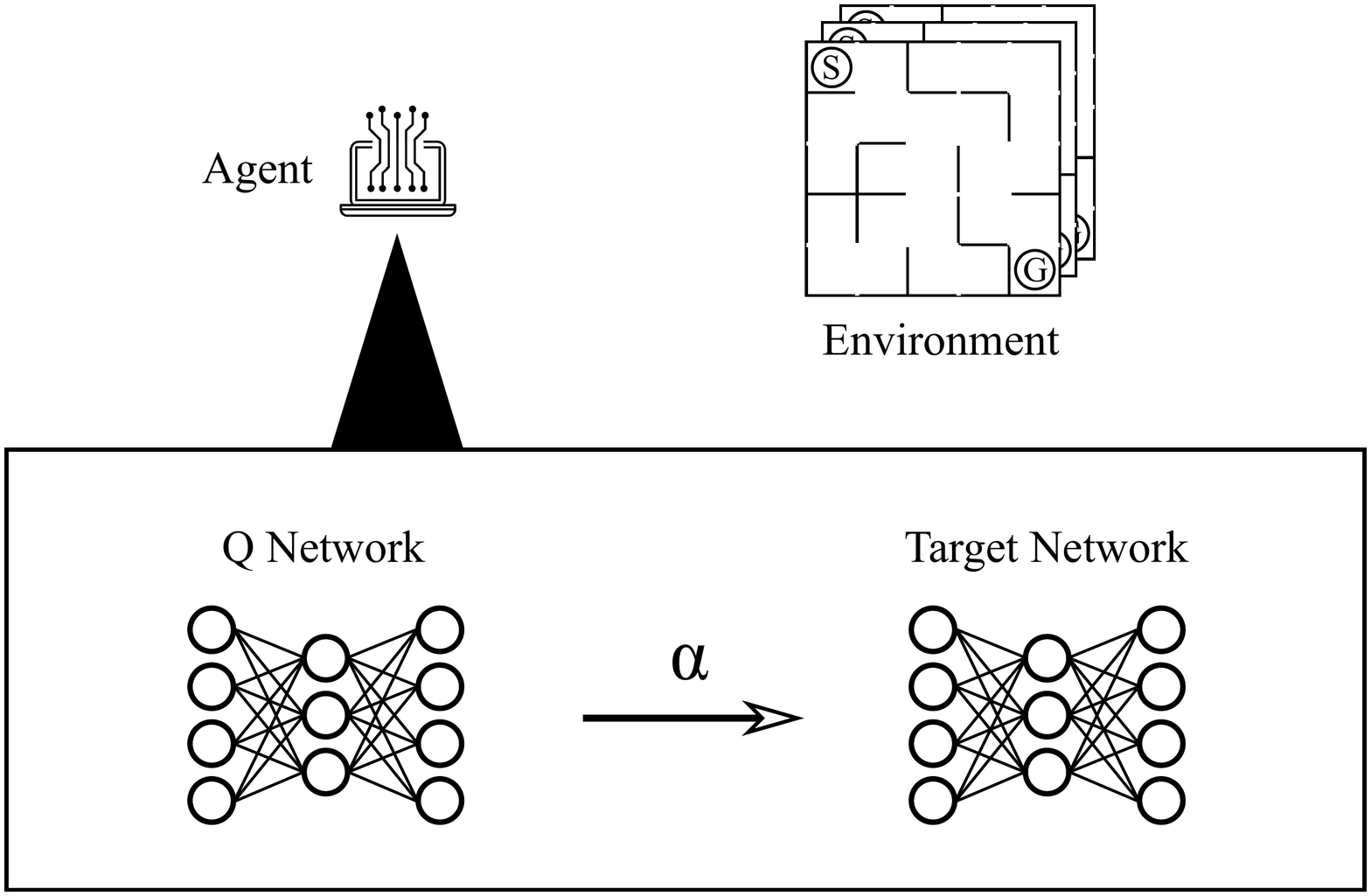}
    \caption{In the last step, the weights and the biases of the target network are updated, following the soft update with the learning rate $\alpha$.}
    \label{fig:update}
\end{figure}
    
\end{enumerate}

\section{Froggatt-Nielsen model with reinforcement learning}
\label{sec:model}

\subsection{The environment}
\label{sec:env}

In FN models, the hierarchical structure of fermion masses and the flavor structure 
are addressed by the global $U(1)$ symmetry. 
For simplicity, we introduce only one complex scalar field (so-called flavon field), charged 
under $U(1)$. 
The relevant Yukawa terms of quarks and leptons are given by
\begin{align}
    {\cal L} &= y^u_{ij}\left(\frac{\phi}{M}\right)^{n^u_{ij}}\Bar{Q}_i H^c u_j
              + y^d_{ij}\left(\frac{\phi}{M}\right)^{n^d_{ij}}\Bar{Q}_i H d_j
              + y^l_{ij}\left(\frac{\phi}{M}\right)^{n^l_{ij}}\Bar{L}_i H l_j
              \nonumber\\
              &+ y^\nu_{ij}\left(\frac{\phi}{M}\right)^{n^\nu_{ij}}\Bar{L}_i H^c N_j
              + \frac{y^N_{ij}}{2}\left(\frac{\phi}{M}\right)^{n^N_{ij}}M\Bar{N}^{c}_i N_j
              + \text{h.c.},
\label{eq:Lagrangian}
\end{align}
where $\{Q_i, u_i, d_i, L_i, l_i, N_i, H\}$ denote the left-handed quarks, 
the right-handed up-type quarks, the right-handed down-type quarks, the left-handed leptons, 
the right-handed charged leptons, the right-handed neutrinos, and the SM Higgs doublet with $H^c = i\sigma_2 H^\ast$, respectively. 
Here, we assume three right-handed neutrinos and tiny neutrino masses are generated by Type-I seesaw mechanism where 
the parameter $M$ is chosen as $M=10^{15}$\,GeV throughout the analysis of this paper, and the Yukawa couplings $\{y^u_{ij},y^d_{ij},y^l_{ij},y^\nu_{ij},y^N_{ij}\}$ are ${\cal O}(1)$ real coefficients. 
Since the SM fields and the flavon field are also charged under $U(1)$, 
let us denote their $U(1)$ charges by
\begin{align}
    \{q(Q_i),\,q(u_i),\,q(d_i),\,q(L_i),\,q(l_i),\,q(N_i),\,q(H),\,q(\phi)\}.
\end{align}

\medskip

To be invariant under the $U(1)$ symmetry, the integers $n_{ij}$ satisfy the following relations:
\begin{align}
\begin{split}
    n^u_{ij} &= -\frac{q(\Bar{Q}_iH^c u_j)}{q(\phi)} = -\frac{-q(Q_i)-q(H)+q(u_j)}{q(\phi)},
    \\
    n^d_{ij} &= -\frac{q(\Bar{Q}_i H d_j)}{q(\phi)} = -\frac{-q(Q_i) + q(H) + q(d_j)}{q(\phi)},
\end{split}
\label{eq:nij_quark}
\end{align}
\begin{align}
\begin{split}
    n^l_{ij} &= -\frac{q(\Bar{L}_i H l_j)}{q(\phi)} = -\frac{-q(L_i) + q(H) + q(l_j)}{q(\phi)},
    \\
    n^\nu_{ij} &= -\frac{q(\Bar{L}_i H^c N_j)}{q(\phi)} = -\frac{-q(L_i) - q(H) + q(N_j)}{q(\phi)},
    \\
    n^N_{ij} &= -\frac{q(\Bar{N}_i^{c} N_j)}{q(\phi)} = -\frac{q(N_i)  + q(N_j)}{q(\phi)},
\end{split}
\label{eq:nij_lepton}
\end{align}
where $n_{ij}$ are considered positive integers throughout this paper.\footnote{See, e.g., Ref.\cite{Alonso:2018bcg}, for the possibility of negative integers by introducing vector-like fermions.} 
Furthermore, we require the presence of Yukawa term $\Bar{Q}_3H^c u_3$, irrelevant to $q(\phi)$:
\begin{align}
    q(\Bar{Q}_3H^c u_3) = 0 
    \leftrightarrow
    q(H) = q(u_3) - q(Q_3);
    \label{eq:Higgscharge}
\end{align}
otherwise one cannot realize the value of top quark mass. Once $\phi$ and $H$ develop VEVs, $ \langle \phi \rangle = v_\phi$ and $\langle H\rangle = v_{\rm EW}=174\,$GeV, 
the Dirac mass matrices of quarks and leptons as well as the Majorana mass matrix are given by
\begin{align}
\begin{split}
    m_{ij}^u &= y^u_{ij} \epsilon^{n^u_{ij}} v_{\rm EW},\qquad
    m_{ij}^d = y^d_{ij} \epsilon^{n^d_{ij}} v_{\rm EW},\qquad    
    \\
    m_{ij}^l &= y^l_{ij} \epsilon^{n^l_{ij}} v_{\rm EW},\qquad
    m_{Dij}^\nu = y^\nu_{ij} \epsilon^{n^\nu_{ij}} v_{\rm EW},\qquad 
    m_{ij}^N =  M y^N_{ij} \epsilon^{n^N_{ij}}.
\end{split}
\end{align}
The light neutrino mass matrix is obtained by integrating out heavy right-handed neutrinos:
\begin{align}
    m_{ij}^\nu = -\left(m^\nu \cdot (m^N)^{-1}\cdot (m^\nu)^T\right)_{ij}.
\end{align}
The quark and lepton mass matrices are diagonalized as
\begin{align}
\begin{split}
    m^u &= U^u {\rm diag}(m^u) (V^u)^\dagger,\qquad
    m^d = U^d {\rm diag}(m^d) (V^d)^\dagger,\qquad
    \\
    m^l &= U^l {\rm diag}(m^l) (V^l)^\dagger,\qquad
    m^\nu = U^\nu {\rm diag}(m^\nu) (U^\nu)^T,
\end{split}
\end{align}
and the flavor mixings are given by the difference between mass eigenstates and flavor eigenstates:
\small
\begin{align}
\begin{split}
    V_{\rm PMNS} &= (U^l)^\dagger V^\nu
    \\
    &= 
    \begin{pmatrix} c_{12} c_{13} & s_{12} c_{13} & s_{13} e^{-i \delta_{\rm CP}} \\ 
-s_{12} c_{23} - c_{12} s_{23} s_{13} e^{i \delta_{\text{CP}}} & c_{12} c_{23} - s_{12} s_{23} s_{13} e^{i \delta_{\text{CP}}} & s_{23} c_{13} \\
s_{12} s_{23} - c_{12} c_{23} s_{13} e^{i \delta_{\text{CP}}} & -c_{12} s_{23} - s_{12} c_{23} s_{13} e^{i \delta_{\text{CP}}} & c_{23} c_{13} 
\end{pmatrix}
\begin{pmatrix} 1 & 0 & 0 \\ 0 & e^{i \frac{\alpha_{21}}{2}} & 0 \\ 0 & 0 & e^{i \frac{\alpha_{31}}{2}} 
\end{pmatrix},
\end{split}
\end{align}
\normalsize
with $c_{ij}= \cos \theta_{ij}$ and $s_{ij} = \sin \theta_{ij}$, which also holds for the CKM matrix 
$V_{\rm CKM} = (U^u)^\dagger U^d$ in the quark sector except the Majorana phases $\{\alpha_{21}, \alpha_{31}\}$. 
Since the quarks and the leptons are charged under $U(1)$, 
the flavon VEV $ \langle \phi \rangle = v_\phi$ 
will lead to the flavor structure due to the smallness of $|\epsilon|$:
\begin{align}
    \epsilon := \frac{v_\phi}{M}. 
\end{align}

\medskip

Recalling that the $U(1)$ charge of Higgs doublet is determined by Eq. (\ref{eq:Higgscharge}), 
the flavor structure of quarks and leptons is specified by the following charge 
vector:
\begin{align}
    {\cal Q}_a := \{q(Q_i),\,q(u_i),\,q(d_i),\,q(L_i),\,q(l_i),\,q(N_i),\,q(\phi)\},
\end{align}
consisting of 19 elements. This is the input for target network and Q network as the state $s$. 
In the following analysis using RL, we restrict ourselves to the following range of $U(1)$ charge:
\begin{align}
    -9 \leq {\cal Q}_a \leq 9,
\end{align}
corresponding to total $19^{19}\sim 10^{24}$ possibilities for the charge assignment.\footnote{In this counting, a permutation symmetry among the charge assignment is not taken into account, and we will not incorporate this effect for RL analysis.} It will be a challenging issue to find a realistic flavor pattern by the brute force approach. 
Furthermore, it is generally difficult to use the gradient descent method, because the $U(1)$ charges are discrete and even a small difference in the charges will result in exponential differences in calculated values such as masses. 
Against those backgrounds, the necessity of applying reinforcement learning arises. 
Note that a generic $U(1)$ charge of flavon will lead to the non-integer $n_{ij}$; thereby 
we focus on $q(\phi)=1$ or $-1$ with 50\% probability in the following analysis.

\subsection{Neural Network}

A state $s$ given by the charge assignment ${\cal Q}$ will be updated to $s^\prime$ through the action $\mathfrak{a}$. 
To determine the action, we utilize the neural network as shown in Table \ref{tab:network}.
The activation function $h$ (in Eq.\eqref{eq:activation}) is chosen as a SELU function for hidden layers 1,2,3 and 
 the softmax function (\ref{eq:softmax}) for the output layer. 
We employ the ADAM optimizer in TensorFlow \cite{DBLP:journals/corr/AbadiABBCCCDDDG16}\footnote{We use the ``gym'' proposed by the OpenAI.}, where the weights and biases are chosen 
to minimize the loss function given by the 
Huber function \eqref{eq:Huber}. 

\medskip

In the FN model, the flavor structure of quarks and leptons is determined by 
the charge vector ${\cal Q}_a$, including total $10^{24}$ possibilities for the 
charge assignment as pointed out before. 
When we focus on only the quark sector, the parameter spaces of $U(1)$ charges reduce 
to $19^{10}\sim 10^{12}$ possibilities. 
To achieve a highly efficient learning in a short time, it is better to perform a separate training for the $U(1)$ charge assignment of quarks and leptons. 
Note that only the flavon $U(1)$ charge connects the quark sector with the lepton sector since 
the $U(1)$ charge of the Higgs is determined by the charge of third generation quarks (\ref{eq:Higgscharge}). 
As mentioned before, we focus on $q(\phi)=1$ or $-1$ with 50\% probability in the following analysis. 
Thus, we first analyze the parameter space of quark $U(1)$ charges by RL as will be discussed in detail in Sec. \ref{sec:quark}, and move to the 
lepton sector with fixed $U(1)$ charge of Higgs fields as will be discussed in detail in Sec. \ref{sec:lepton}. 

\medskip

The hyperparameters are set as $N_{\rm ep}=10^5$ and $N_{\rm ep}=6\times 10^4$ for the episode number in the quark and lepton sector respectively, $N_{\rm step}=32$ for the step number, 
batch sizes of 32, epoch number of 32, and the learning rate $\alpha= 2.5\times 10^{-4}$, respectively. 
The hyperparameters in $\epsilon$-greedy method are described in the previous section. About the step number $N_{\rm step}=32$, the same value was used in the previous research that focuses on only the quark sector \cite{Harvey:2021oue}. 
In Ref. \cite{Harvey:2021oue}, it was shown that terminal states can be reached after a sufficient amount of learning. 
Therefore, it is expected that $N_{\rm step}=32$ is enough to achieve terminal states in the current situation where the quark sector and the lepton sector are searched separately.

\begin{table}[H]
    \centering
    \begin{tabular}{|c||c|c|c|c|c|}\hline
       layer  &  Input & Hidden 1 & Hidden 2 & Hidden 3 & Output\\
       \hline
       Dimension  & $\mathbb{Z}^A$ & $\mathbb{R}^{64}$ & $\mathbb{R}^{64}$ & $\mathbb{R}^{64}$ & $\mathbb{R}^{2A}$
       \\
       \hline
    \end{tabular}
    \caption{In the neural network, the input is the charge assignment ${\cal Q}_a$ with dimension $A$, and 
    the activation functions are the SELU function for hidden layers 1,2,3. Since we use the softmax function (\ref{eq:softmax}) for the output layer, the output with dimension $2A$ is interpreted as probabilities. 
    The dimension of the output layer is twice that of the input layer due to the action of the agent \eqref{eq:action_agent}.}
    \label{tab:network}
\end{table}

\subsection{Agent}
\label{sec:agent}

\medskip

\begin{figure}[H]
    \centering
    \includegraphics[width=70mm]{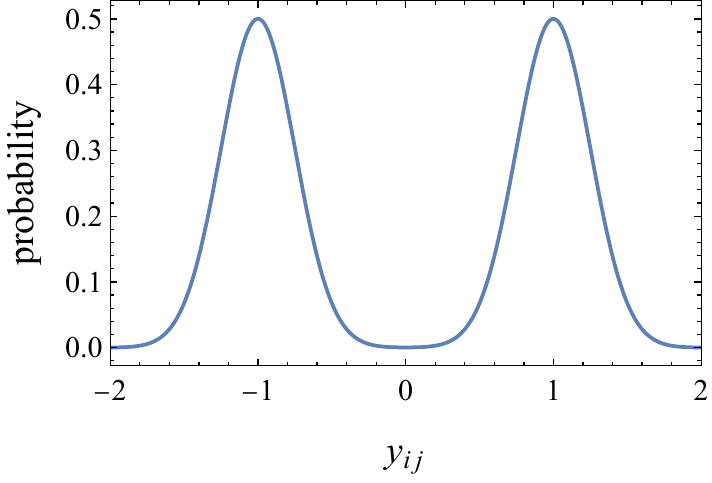}
    \caption{Distribution of ${\cal O}(1)$ coefficients in Yukawa terms (\ref{eq:Lagrangian}).}
    \label{fig:random}
\end{figure}

To implement the FN model in the context of RL with DQN, we specify the following action $\mathfrak{a}$ of the agent at each step:
\begin{align}
    \mathfrak{a}\,:\, {\cal Q}_a \rightarrow {\cal Q}_a \pm 1\,\,(a \in A),
    \label{eq:action_agent}
\end{align}
where $A$ corresponds to $\{Q_i,u_i,d_i,\phi\}$ in the analysis of Sec.~\ref{sec:quark} and $\{L_i,l_i,N_i,\phi \}$ in the analysis of Sec.~\ref{sec:lepton}. 
These two candidates of the action make the dimension of the output layer $2A$ in Table \ref{tab:network}.
At the initial stage, the ${\cal O}(1)$ coefficients in Yukawa terms (\ref{eq:Lagrangian}) are 
picked up from the two Gaussian distribution with an average $\pm 1$ and standard deviation 0.25 (see Fig. \ref{fig:random}) and after the training by neural network introduced in the previous section, they are optimized to proper values by the Monte-Carlo simulation.

Thus, once the charges are fixed, one can compare the masses and mixings of quarks or leptons given by the action $\mathfrak{a}$ with the experimental values. 
Specifically, we define the intrinsic value:
\begin{align}
\begin{split}
    {\cal V}({\cal Q}) = 
        \left\{
        \begin{array}{ll}    
    -{\rm min}_{v_\phi}\bigl[{\cal M}_{\rm quark} + {\cal C} \bigl]&\quad ({\rm used\,in\,Sec.\,\ref{sec:quark}})
    \\
    -\bigl[{\cal M}_{\rm lepton} + {\cal M}_{\rm neutrino} + {\cal P} \bigl]&\quad 
    ({\rm used\,in\,Sec.\,\ref{sec:lepton}})
        \end{array}
        \right.
\end{split}
,
\label{eq:intrinsic_value}
    \end{align}
whose components will be defined below. 
Note that the flavon VEV is chosen to maximize the intrinsic value relevant for the quark sector in Sec. \ref{sec:quark}; thereby there is no flavon dependence in the intrinsic value of the lepton sector.

\begin{enumerate}
    \item Quark and lepton masses:

${\cal M}_{\rm quark}$ (${\cal M}_{\rm lepton}$) consists of the ratio of the predicted quark (lepton) masses by the agent to the experimental values:
\begin{align}
    {\cal M}_{\rm quark} = \sum_{\alpha = u,d}E_\alpha,\qquad
    {\cal M}_{\rm lepton} = \sum_{\alpha = l}E_\alpha, 
\end{align}
with
\begin{align}
    E_\alpha = \biggl| \log_{10}\left(\frac{|m_\alpha|}{|m_{\alpha,{\rm exp}}|}\right) \biggl|.
\end{align}
The experimental values are listed in Tables \ref{tab:data_quark} and \ref{tab:data_lepton} for 
quarks and leptons, respectively.

\item Neutrino masses:

Since the ordering of neutrino masses has not been confirmed yet, we search the neutrino structure in two cases: RL with fixed neutrino mass ordering in Sec. \ref{sec:neutrino1} and RL without specifying the neutrino mass ordering in Sec. \ref{sec:neutrino2}. In each case, the intrinsic value relevant to the neutrino masses is defined as:
\begin{align}
    {\cal M}_{\rm neutrino} = 
    \left\{
    \begin{array}{ll}
         \sum_{\alpha = \nu_{21},\nu_{31}}E_\alpha^{\nu}
         &\quad ({\rm normal\,ordering\,in\,Sec.\,\ref{sec:neutrino1}})  \\
         \sum_{\alpha = \nu_{21},\nu_{32}}E_\alpha^{\nu}
         &\quad({\rm inverted\,ordering\,in\,Sec.\,\ref{sec:neutrino1}})\\
         0 &\quad ({\rm unspecified\,mass\,ordering\,in\,Sec.\,\ref{sec:neutrino2}})
    \end{array}
    \right.
    ,
\end{align}
with
\begin{align}
    E_\alpha^{\nu} = \biggl| \log_{10}\left(\frac{|\Delta m_\alpha^2|}{|\Delta m^2_{\alpha,{\rm exp}}|}\right) \biggl|,
\end{align}
where the experimental values are listed in Table \ref{tab:data_lepton}.

\item Mixing angles:

In addition, the intrinsic value includes the information of quark mixings and lepton mixings in ${\cal C}$ and ${\cal P}$:
\begin{align}
    {\cal C} = \sum_{i,j} E_{\cal C}^{ij},\qquad
    {\cal P} = \sum_{i,j} E_{\cal P}^{ij},    
\end{align}
with
\begin{align}
    E_{\cal C}^{ij} = \biggl| \log_{10}\left(\frac{|V_{\rm CKM}^{ij}|}{|V_{\rm CKM,\,exp}^{ij}|}\right) \biggl|,
    \quad
    E_{\cal P}^{ij} = \biggl| \log_{10}\left(\frac{|V_{\rm PMNS}^{ij}|}{|V_{\rm PMNS,\,exp}^{ij}|}\right) \biggl|,
\end{align}
where $E_{\cal C}^{ij}$ and $E_{\cal P}^{ij}$ represent the ratio of the predicted quark and lepton mixings by the agent to the experimental values, respectively. 
From Tables \ref{tab:data_quark} and \ref{tab:data_lepton}, the CKM and PMNS matrices are of 
the form:
\begin{align}
\begin{split}
    |V_{\rm CKM,\,exp}| &=  
    \left(\begin{array}{lll}
    0.97435\pm 0.00016 \,\,&\,\, 0.22500 \pm 0.00067 \,\,&\,\, 0.00369 \pm 0.00011 \\
    0.22486 \pm 0.00067 \,\,&\,\, 0.97349 \pm 0.00016 \,\,&\,\, 0.04182_{-0.00074}^{+0.00085} \\
    0.00857_{-0.00018}^{+0.00020} \,\,&\,\, 0.0410_{-0.00072}^{+0.00083} \,\,&\,\, 0.999118_{-0.000036}^{+0.000031}
    \end{array}\right)
    ,
\\
    |V_{\rm PMNS,\,exp}|_{3\sigma} &=  
    \left(\begin{array}{lll}
    0.803 \rightarrow 0.845 \,\,&\,\, 0.514 \rightarrow 0.578 \,\,&\,\, 0.143 \rightarrow 0.155 \\
    0.244 \rightarrow 0.498 \,\,&\,\, 0.502 \rightarrow 0.693 \,\,&\,\, 0.632 \rightarrow 0.768 \\
    0.272 \rightarrow 0.517 \,\,&\,\, 0.473 \rightarrow 0.672 \,\,&\,\, 0.623 \rightarrow 0.761 
    \end{array}\right).
\end{split}
\end{align}

\end{enumerate}

\begin{table}[H]
\renewcommand{\arraystretch}{1.25}
\centering
\scalebox{0.88}{
   \begin{tabular}{|c|c|c|c|c|c|}\hline
         $m_u$/\text{MeV} & $m_d$/\text{MeV} & $m_s$/\text{MeV} & $m_c$/\text{GeV} & $m_b$/\text{GeV} & $m_t$/\text{GeV}\\\hline
         $2.16^{+0.49}_{-0.26}$ & ${4.67}^{+0.48}_{-0.17}$ & ${93.4}^{+8.6}_{-3.4}$ & ${1.27}\pm{0.02}$ & $4.18^{+0.03}_{-0.02}$ & $172.69 \pm 0.30$ \\\hline\hline
          $s_{12}$ & $s_{13}$ &  $s_{23}$ & $\delta_{\rm CP}$ &  &\\\hline
         $0.22500 \pm 0.00067$ & $0.00369 \pm 0.00011$ & $0.04182^{+0.00085}_{-0.00074}$ & $1.144 \pm 0.027$ &  &\\
         \hline
      \end{tabular}
      }
  \caption{Masses, mixing angles, and CP phase in the quark sector \cite{ParticleDataGroup:2022pth}, where we show the top-quark mass from direct measurements.}
    \label{tab:data_quark}  
\end{table}
     \renewcommand{\arraystretch}{1}

     \renewcommand{\arraystretch}{1.25}
\begin{table}[H]
 \centering
\small
   \begin{tabular}{|c||c|c||c|c|} \hline
     \multirow{2}{*}{Observables} & \multicolumn{2}{c||}{Normal Ordering (NO)} & \multicolumn{2}{c|}{Inverted Ordering (IO)}  \\
     \cline{2-5}
       & $1\sigma$ range & $3\sigma$ range & $1\sigma$ range & $3\sigma$ range  \\
     \hline
     $\sin^{2}\theta_{12}$ & $0.303_{-0.011}^{+0.012}$ & $0.270\rightarrow 0.341$ & $0.303_{-0.011}^{+0.012}$ & $0.270\rightarrow 0.341$ \\
     \hline
     $\sin^{2}\theta_{13}$ & $0.02225_{-0.00059}^{+0.00056}$ & $0.02052\rightarrow 0.02398$ & $0.02223_{-0.00058}^{+0.00058}$ & $0.02048\rightarrow 0.02416$ \\
     \hline
      $\sin^{2}\theta_{23}$ & $0.451_{-0.016}^{+0.019}$ & $0.408\rightarrow 0.603$ & $0.569_{-0.021}^{+0.016}$ & $0.412\rightarrow 0.613$ \\
     \hline
      $\delta_{\text{CP}}/\pi$ & $1.29_{-0.14}^{+0.20}$ & $0.80\rightarrow 1.94$ & $1.53_{-0.16}^{+0.12}$ & $1.08\rightarrow 1.91$ \\
     \hline
     $\cfrac{\Delta m_{21}^{2}}{10^{-5} \mathrm{eV}^{2}}$ & $7.41_{-0.20}^{+0.21}$ & $6.82\rightarrow 8.03$ & $7.42_{-0.20}^{+0.21}$ & $6.82\rightarrow 8.04$ \\
     \hline
     $\dfrac{\Delta m_{3l}^{2}}{10^{-3} \mathrm{eV}^{2}}$ & $2.507_{-0.027}^{+0.026}$ & $2.427\rightarrow 2.590$ & $-2.486_{-0.028}^{+0.025}$ & $-2.570\rightarrow -2.406$ \\ \hline
     $m_{e}$/\text{MeV} & \multicolumn{4}{c|} {$0.510999$} \\ 
     \hline
     $m_{\mu}$/\text{MeV} & \multicolumn{4}{c|}{$105.658$} \\
     \hline
     $m_{\tau}/\text{MeV}$ & \multicolumn{4}{c|}{$1776.86$} \\
     \hline
 \end{tabular}
     \renewcommand{\arraystretch}{1}
\caption{Experimental values for the lepton sector obtained from global analysis of the data, 
where $\Delta m_{3l}^{2}\equiv \Delta m_{31}^{2}=m^2_3 -m^2_1>0$ for NO and $\Delta m_{3l}^{2}\equiv \Delta m_{32}^{2}=m^2_3 -m^2_2<0$ for IO. 
Here, we use the data from Ref. \cite{ParticleDataGroup:2022pth} for charged lepton masses and NuFIT v5.2 results with Super-Kamiokande atmospheric data for the lepton mixing angles and CP phase \cite{Esteban:2020cvm}.}
\label{tab:data_lepton}
\end{table}

\medskip

The flavon VEV is defined to maximize the intrinsic value, and we search for the VEV within
\begin{align}
    0.01 \leq |v_\phi| \leq 0.3,\qquad -\pi \leq {\rm arg}(v_\phi) \leq \pi,
\end{align}
where the angular component of the flavon VEV determines the CP phase. 
The large intrinsic value indicates that the obtained charge assignment well reproduces the experimental values. 
Such a charge assignment is called {\it terminal state}. 
Specifically, the terminal state is defined to satisfy the following requirement:
\begin{align}
    |{\cal V}({\cal Q})| < V_0,\qquad
    E_\alpha, E_\alpha^\nu <V_1\quad ({\rm for}\,\forall\alpha),\qquad
    E_{{\cal C},{\cal P}}^{ij}<V_2\quad ({\rm for}\,\forall i,j).
\end{align}
In this paper, we adopt $V_0 = 10.0$, $V_1 = 1.0$ and $V_2 = 0.2$. 
Here, $V_1=1.0$ ($V_2=0.2)$ means that the ratio of the predicted fermion masses (mixings) 
to the observed masses (mixings) is considered within $0.1 \leq r_{\rm mass} \leq 10$ ($0.63 \leq r_{\rm mixings} \leq 1.58$).

\medskip

Let us denote the charge assignment ${\cal Q}$ observed by the agent and ${\cal Q}'$ after 
the action $\mathfrak{a}$. 
For the action of the agent $({\cal Q}, \mathfrak{a})$, 
we will give the reward $\mathcal{R}$ in the following prescription:
\begin{screen}
\begin{enumerate}

    \item Give the basic point $\mathcal{R}_{\rm base}$, depending on the value of intrinsic value:
    \begin{align}
        \mathcal{R}_{\rm base}=
        \left\{
        \begin{array}{ll}
             {\cal V}({\cal Q}') -{\cal V}({\cal Q})&\qquad {\rm if}\,{\cal V}({\cal Q}') -{\cal V}({\cal Q})>0 \\
             \mathcal{R}_{\rm offset}&\qquad {\rm if}\,{\cal V}({\cal Q}') -{\cal V}({\cal Q})\leq 0         
        \end{array}
        \right.
        ,
    \end{align}
    where $\mathcal{R}_{\rm offset}$ corresponds to a penalty, chosen as $\mathcal{R}_{\rm offset}=-10$. 

    \item When the ${\cal Q}'$ lies outside $-9\leq {\cal Q}' \leq 9$ or the flavon charge satisfies $q(\phi)=0$, we give the penalty ${\cal R}_{\rm offset}$ and the environment comes back to the original charge assignment ${\cal Q}$. 

    \item When the ${\cal Q}'$ is turned out to be a terminal state, we give the bonus point ${\cal R}_{\rm term}$, chosen as $\mathcal{R}_{\rm term}=100$. 

    \item Summing up the above points, we define the reward ${\cal R}({\cal Q},\mathfrak{a})$.
\end{enumerate}
\end{screen}

The structure of the neural network and the design of the reward largely determine the behavior of RL. 
Therefore, to facilitate comparison with Ref. \cite{Harvey:2021oue}, we use the same architecture
\footnote{Indeed, the method of giving a large positive reward when the desired behavior (terminal states) is achieved is an empirically technique for successful RL. 
On the other hand, another technique called as reward clipping is known to improve learning efficiency by clipping rewards in the range of $-1$ to $+1$. 
In order to apply the large rewards for reaching the terminal states, this technique was not used in this study.}.

\newpage

\section{Learning the quark sector}
\label{sec:quark}

\begin{figure}[H]
\begin{minipage}{0.32\hsize}
  \begin{center}
  \includegraphics[height=52mm]{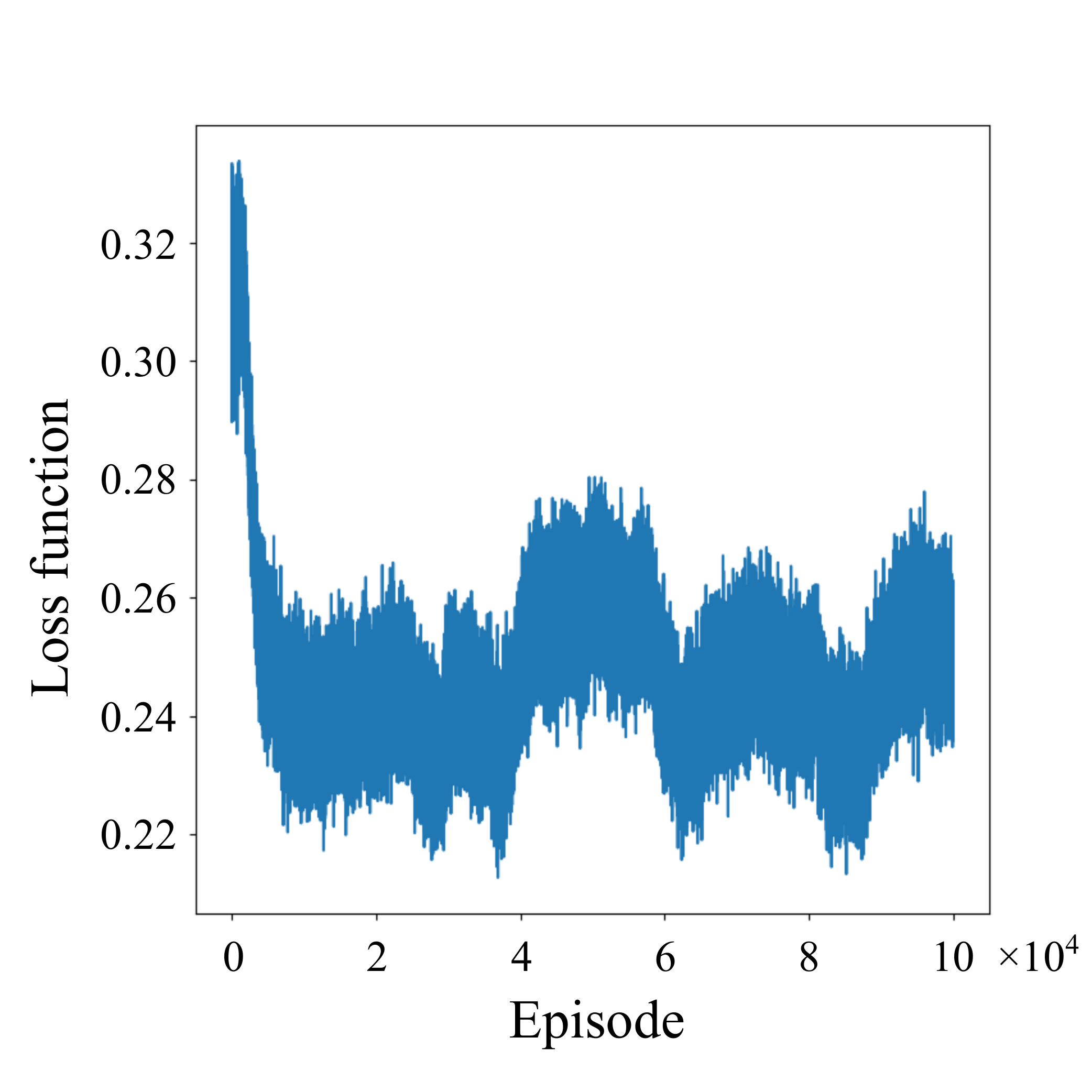}
  \end{center}
 \end{minipage}
 \begin{minipage}{0.32\hsize}
  \begin{center}
   \includegraphics[height=52mm]{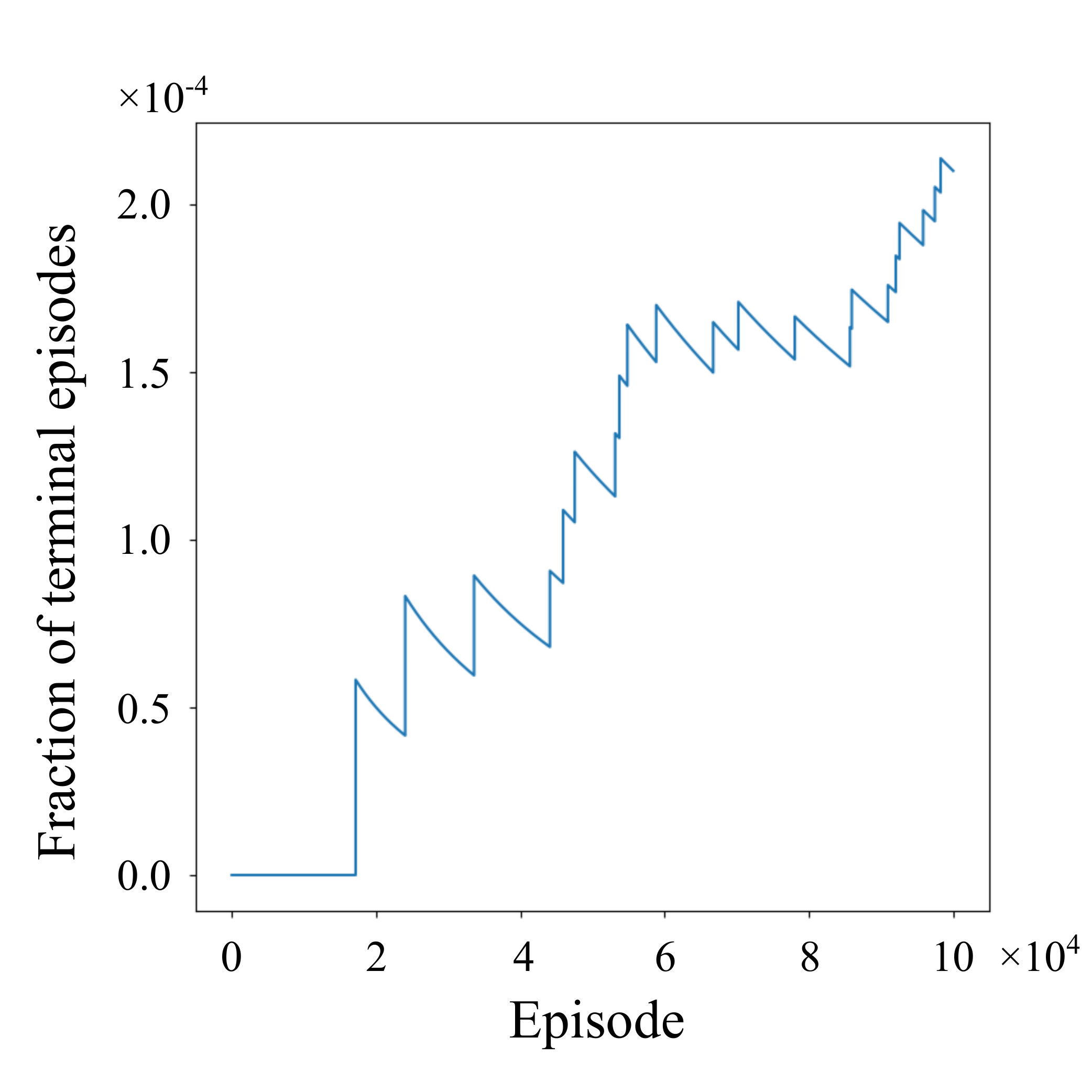}
  \end{center}
 \end{minipage}
 \begin{minipage}{0.32\hsize}
  \begin{center}
   \includegraphics[height=52mm]{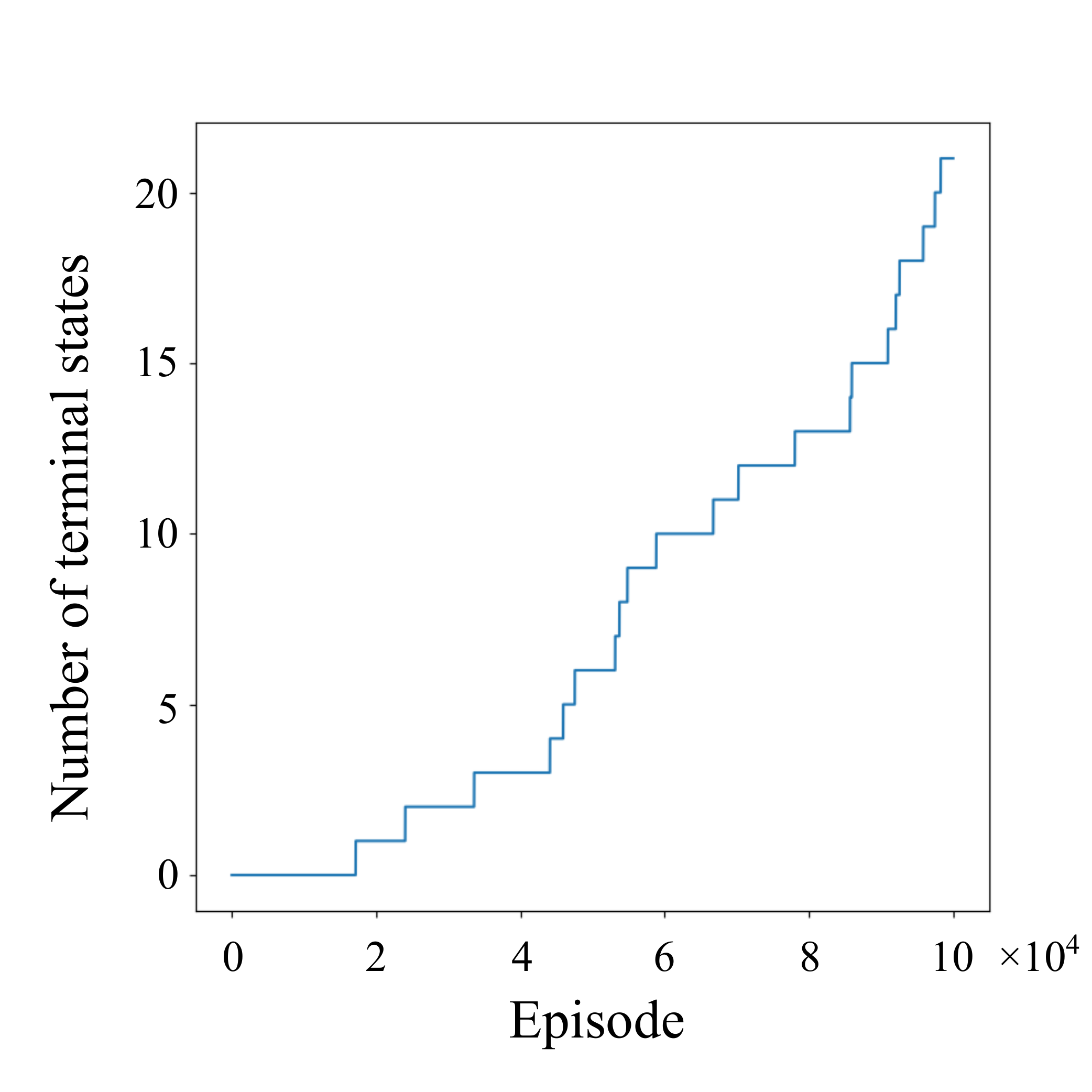}
  \end{center}
 \end{minipage}
  \caption{Learning results for the quark sector. 
  The results are the output of neural network leading to the best-fit model shown in Table \ref{tab:benchmark_quark}. 
  From left to right, three panels show (a) the loss function vs episode number (b) the fraction of terminal episodes vs episode number (c) the number of terminal states vs episode number, respectively.}
\label{fig:result_quark}
\end{figure}

In this section, we analyze the charge assignment of the quark sector, following 
the RL with DQN introduced in the previous section. 
Even for the quark sector within the range of $U(1)$ charges $-9\leq q \leq 9$ , 
there exists $19^{10}\sim 6.1\times 10^{12}$ possible states in the environment. 
By training the neural network about 15 hours on a single CPU
\footnote{Computation time can be reduced by using GPUs. On the other hand, as described in Sec. \ref{sec:neutrino1}, an excessive number of episodes may cause overtraining. Thus, it cannot be said that GPUs will generally lead to improved results.}, it turned out that 
terminal states are found after ${\cal O}(20,000)$ episodes as shown 
in Fig. \ref{fig:result_quark}. 
The loss function tends to be minimized as in Fig. \ref{fig:result_quark}, where 
the small positive loss corresponds to the existence of various paths to terminal states 
as commented in Ref. \cite{Harvey:2021oue}. 
We also check that the reward increases when the loss function decreases. 
The network leads to terminal states in >6\% of all cases for total episode $N_{\rm ep}=10^5$. 
Then, after removing the negative integers of $n_{ij}$ in Eq. \eqref{eq:nij_quark}, it results in 21 independent terminal states. 
When we focus on only the quark masses in the training of neural network, we will obtain terminal states in 90\%. 
It implies that the implementation of masses and mixings will be a more difficult task for 
the agent to find a realistic flavor pattern. 

\medskip

By performing the Monte-Carlo search with the  Gaussian distribution shown in Fig. \ref{fig:random}, the ${\cal O}(1)$ coefficients $y_{ij}$ are optimized to more realistic ones, according to which the intrinsic value is also optimized. 
We show the benchmark point of charge assignment with the highest intrinsic value in Table \ref{tab:benchmark_quark}, where the masses 
and mixing angles are well fitted with observed values up to ${\cal O}(0.1)$\%. This will be improved by a further brute-force search over the parameter space of ${\cal O}(1)$ coefficients. 
Furthermore, from Eq. \eqref{eq:intrinsic_value}, the averaged intrinsic value of the terminal states in \cite{Harvey:2021oue} is calculated as $\mathcal{V} \simeq -0.646$. 
Thus, we argue that the reinforcement learning constructed in this work is able to search for $U(1)$ charges with the equivalent accuracy as previous research, even when $v_{\phi}$ is extended to be a complex number. 
Note that there is no CP phase in the quark sector. 
Even when the angular component of flavon is non-zero, the CP phase is chosen to 0 due to the phase rotation of quark fields. Nonvanishing CP phase in the quark sector will be realized by introducing multiple flavon fields \cite{Leurer:1993gy}, but it will be left for future work. 

\medskip

The above fact that the realistic model is a very rarefied distribution means that learning results can change with changes in the random number seed. 
The RL algorithm developed in this work involves a large random numbers (such as the initial Yukawa couplings, the initial $U(1)$ charges, the choice of greedy actions, and the behavior in random actions). 
While the agent aims to maximize the reward, it does not always reach the terminal states due to the weak distribution of such states. 
Therefore, if the random seed is changed, there is no guarantee that the same terminal states described in this paper will be discovered.
Nevertheless, even in that case, the discovery of different terminal states would be expected.\footnote{Indeed, other sets of charges that reproduce the experimental results can be obtained in retrainings.} 

\newpage
\vspace*{\stretch{1}}

\vspace{\stretch{1}}

\begin{table}[H]
\centering
\scalebox{0.9}{
\begin{tabular}{l|c}
    \hline
    \begin{tabular}{l} Charges \end{tabular} &
    ${\cal Q}=\left(\begin{array}{ccc|ccc|ccc|cc}
        Q_{1} & Q_{2} & Q_{3} & u_{1} & u_{2} & u_{3} & d_{1} & d_{2} & d_{3} & H & \phi \\ 
        \hline
        9 & 8 & 6 & 1 & 3 & 4 & 6 & 5 & 5 & -2 & 1 \\ 
    \end{array}\right)$ \\ 
    \hline
    \begin{tabular}{l} $\mathcal{O}\left(1\right)$ coeff. \end{tabular} & $y^{u} \simeq
    \left(\begin{array}{rrr}
        0.430 & 0.806 & -1.220 \\ 
        -0.962 & -0.598 & -0.747 \\ 
        -1.328 & -1.172 & 1.018
    \end{array}\right) \ ,\ 
    y^{d} \simeq
    \left(\begin{array}{rrr}
        -0.996 & -0.747 & -1.068 \\ 
        0.958 & 1.441 & 1.033 \\ 
        0.765 & -0.500 & -1.029
    \end{array}\right)$ \\ 
    \hline
    \begin{tabular}{l} VEV\end{tabular} & $v_{\phi}\simeq 0.181\cdot e^{-0.863i}$ \\ 
    \hline
    \begin{tabular}{l} Intrinsic value \end{tabular} & $\mathcal{V}_{\mathrm{opt}}\simeq-0.701$ \\ 
    \hline
    \begin{tabular}{l} Masses\\(output) \end{tabular} & $\left(\begin{array}{lll}
    m_{u} & m_{c} & m_{t}\\
    m_{d} & m_{s} & m_{b}
    \end{array}\right)
    \simeq \left(\begin{array}{lll}
    0.002 & 1.468 & 180.945 \\
    0.003 & 0.102 & 4.501
    \end{array}\right)
    \ \mathrm{GeV}$ \\
    \hline
    \begin{tabular}{l} Ratios\\(masses) \end{tabular}
    & $\left(\begin{array}{lll}
    E_{u} & E_{c} & E_{t}\\
    E_{d} & E_{s} & E_{b}
    \end{array}\right)
    \simeq \left(\begin{array}{lll}
    0.008 & 0.063 & 0.020 \\
    0.149 & 0.037 & 0.032
    \end{array}\right)$ \\
    \hline
    \begin{tabular}{l} CKM matrix\\(output) \end{tabular} & $\left|V_{\mathrm{CKM}}\right| \simeq
    \left(\begin{array}{lll}
    0.973 & 0.229 & 0.004 \\
    0.229 & 0.972 & 0.057 \\
    0.009 & 0.057 & 0.998
    \end{array}\right)$
    \\
    \hline
    \begin{tabular}{l} Ratios\\(mixings) \end{tabular}
    &
    $E_{\cal C} \simeq
    \left(\begin{array}{lll}
    0.000 & 0.005 & 0.047 \\
    0.004 & 0.001 & 0.149 \\
    0.033 & 0.152 & 0.000
    \end{array}\right)$ \\
    \hline
\end{tabular}
}
\caption{Benchmark point for the quark sector.}
\label{tab:benchmark_quark}
\end{table}

\vspace{\stretch{1}}

\newpage

\section{Learning the neutrino structure}
\label{sec:lepton}

In this section, we move to the numerical analysis of the lepton sector, following 
the RL with DQN introduced in the Secs. \ref{sec:DQN} and \ref{sec:model}. 
Based on the analysis in Sec. \ref{sec:quark}, 
we fix the Higgs $U(1)$ charges and the VEV $v_{\phi}$ to realize the 21 realistic FN models in the quark sector. 
However, there still exists $19^{9}\sim 3.2\times 10^{11}$ possible states within the range of $U(1)$ charges $-9\leq q \leq 9$ in the environment. 
We first analyze the lepton sector with fixed neutrino mass ordering; normal ordering or 
inverted ordering in Sec. \ref{sec:neutrino1}. In the next analysis of Sec. \ref{sec:neutrino2}, 
the neutrino mass ordering has not been fixed yet. Thus, one can find plausible FN models 
whether the neutrino masses are in the normal ordering or in the inverted ordering.

\subsection{Fixed ordering of neutrino masses}
\label{sec:neutrino1}

\begin{figure}[H]
\begin{minipage}{0.32\hsize}
  \begin{center}
  \includegraphics[height=52mm]{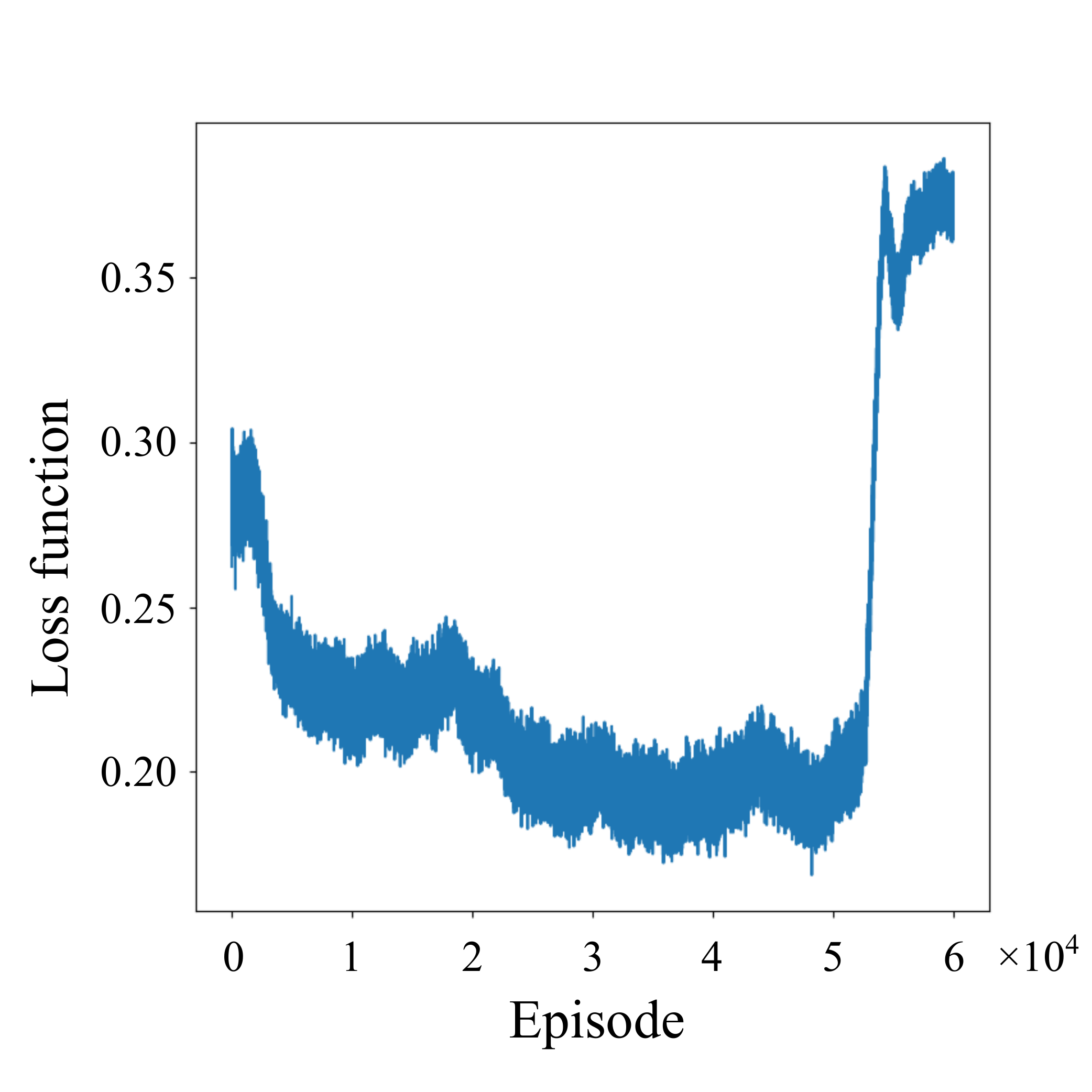}
  \end{center}
 \end{minipage}
 \begin{minipage}{0.32\hsize}
  \begin{center}
   \includegraphics[height=52mm]{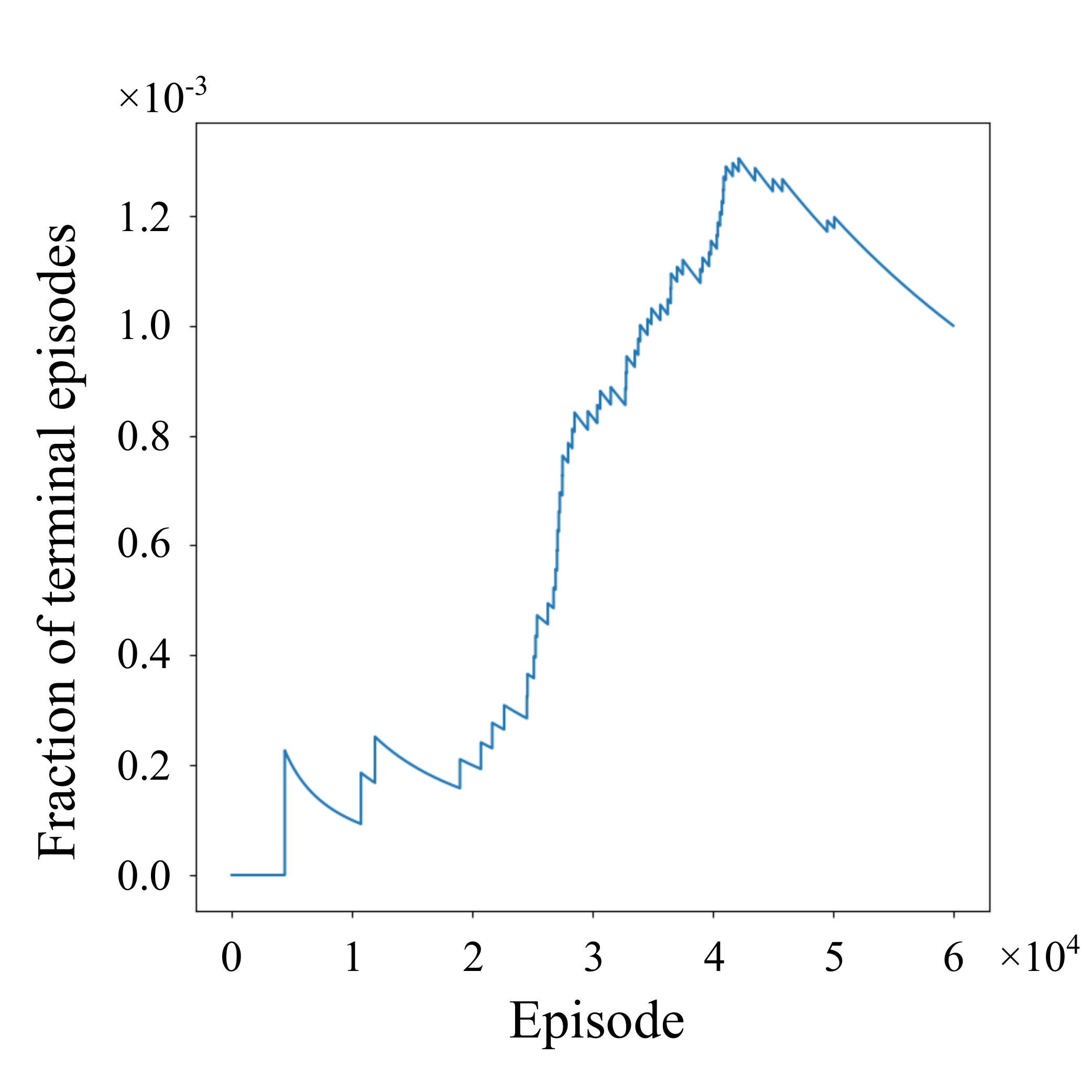}
  \end{center}
 \end{minipage}
 \begin{minipage}{0.32\hsize}
  \begin{center}
   \includegraphics[height=52mm]{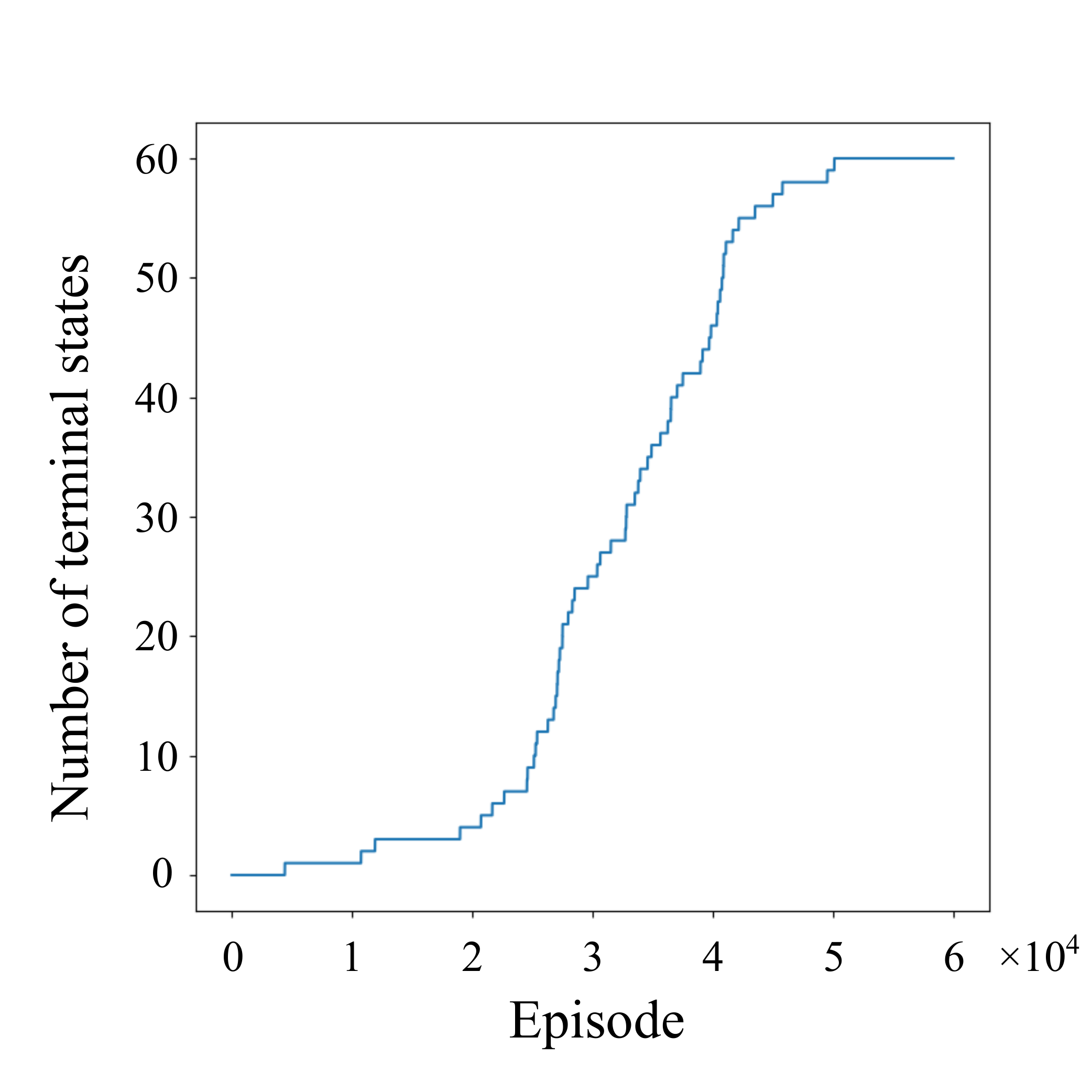}
  \end{center}
 \end{minipage}
  \caption{Learning results for the lepton sector with fixed NO of neutrino masses. 
  The results are the output of neural network leading to the best-fit model (the square in Figs. \ref{fig:data_lepton_fixed_NO1} and \ref{fig:data_lepton_fixed_NO2}). 
  From left to right, three panels show (a) the loss function vs episode number (b) the fraction of terminal episodes vs episode number (c) the number of terminal states vs episode number, respectively.}
\label{fig:result_lepton_NO}
\end{figure}

By training the neural network about 8 hours on a single CPU, it turned out that 
terminal states are found after ${\cal O}(5,000)$ episodes as shown 
in Fig. \ref{fig:result_lepton_NO} with the normal ordering. 
The loss function tends to be minimized as in Fig. \ref{fig:result_lepton_NO} until ${\cal O}(50,000)$ episodes.\footnote{We obtain similar results in the case of inverted ordering.} 
It is notable that the reward increases when the loss function decreases. 
After these critical numbers of episodes, the loss function increases, indicating a sign of overtraining. 
This is because the lepton sector rapidly leads to the terminal states compared with the quark sector. 
Indeed, the network leads to terminal states in >0.06\% of all cases for total episode $N_{\rm ep}=6\times 10^4$. 
Therefore, for the lepton sector, we provided an upper bound for the computational cost, which is that ${\cal O}(50,000)$ episodes are sufficient for the agent to acquire the optimal behavior. 
After removing the negative integers of $n_{ij}$ in Eq. \eqref{eq:nij_lepton} and picking up flavon $U(1)$ charge to be consistent with quark sector, 
we arrive at 63 and 121 terminal states with normal ordering and inverted ordering, respectively. 
By performing the Monte-Carlo search over the ${\cal O}(1)$ coefficients $y_{ij}$ with the  Gaussian distribution shown in Fig. \ref{fig:random}, 
the lepton masses and mixings are further optimized to more realistic ones, according to which the intrinsic value is also optimized. 
Specifically, we performed the Monte-Carlo search 10 times to search the realistic values within $3\sigma$. In the first 10,000 trials, the ${\cal O}(1)$ coefficients $y_{ij}$ are optimized by using the Gaussian distribution shown in Fig. \ref{fig:random}. 
Then, for the ${\cal O}(1)$ coefficients with highest intrinsic value among them, we performed the second 10,000 trials with the Gaussian distribution where an average is the coefficients obtained by the first Monte-Carlo search and the standard deviation is 0.25. 

\begin{figure}[H]
\begin{minipage}{0.49\hsize}
  \begin{center}
  \includegraphics[height=60mm]{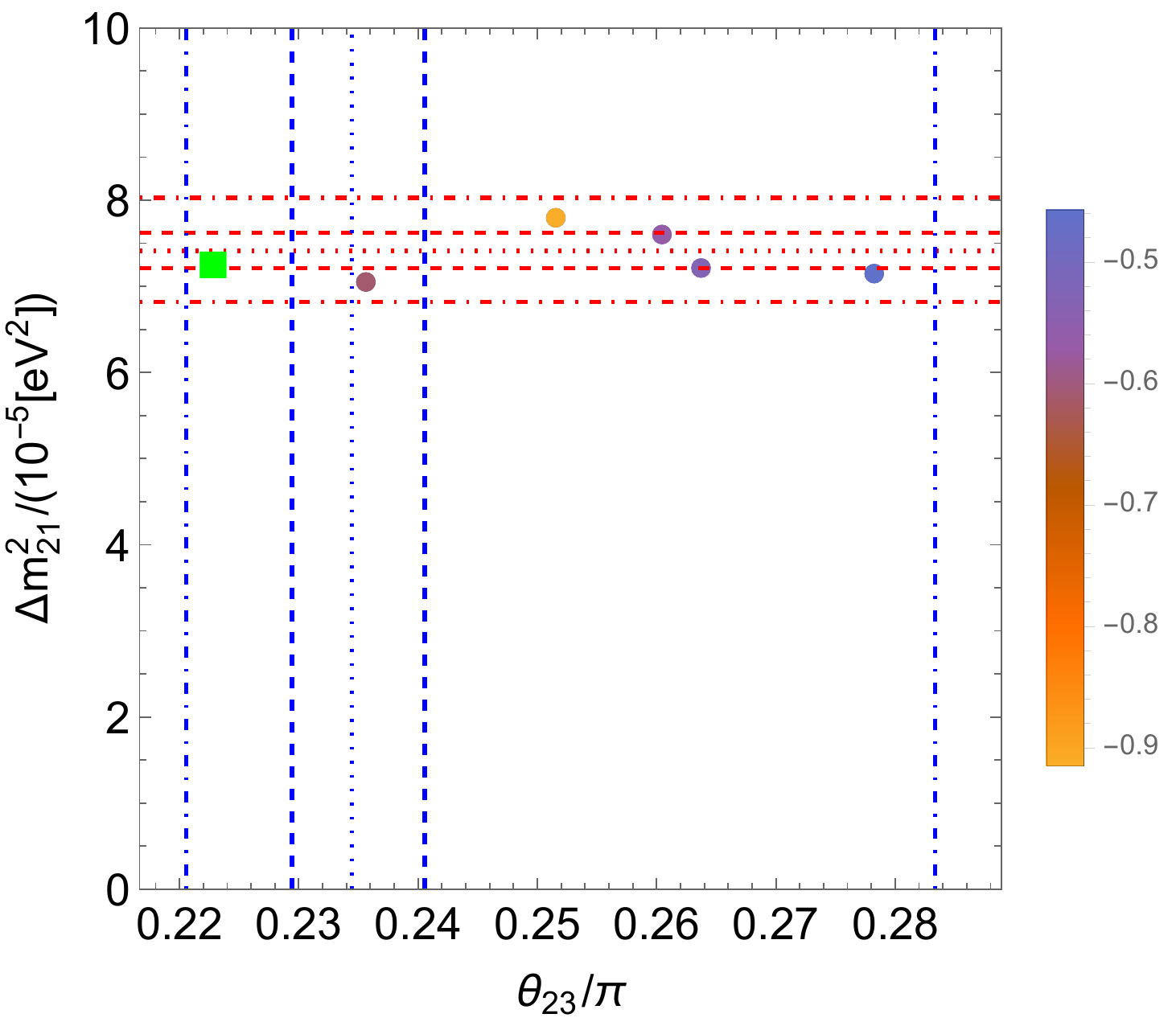}
  \end{center}
 \end{minipage}
\begin{minipage}{0.49\hsize}
  \begin{center}
  \includegraphics[height=60mm]{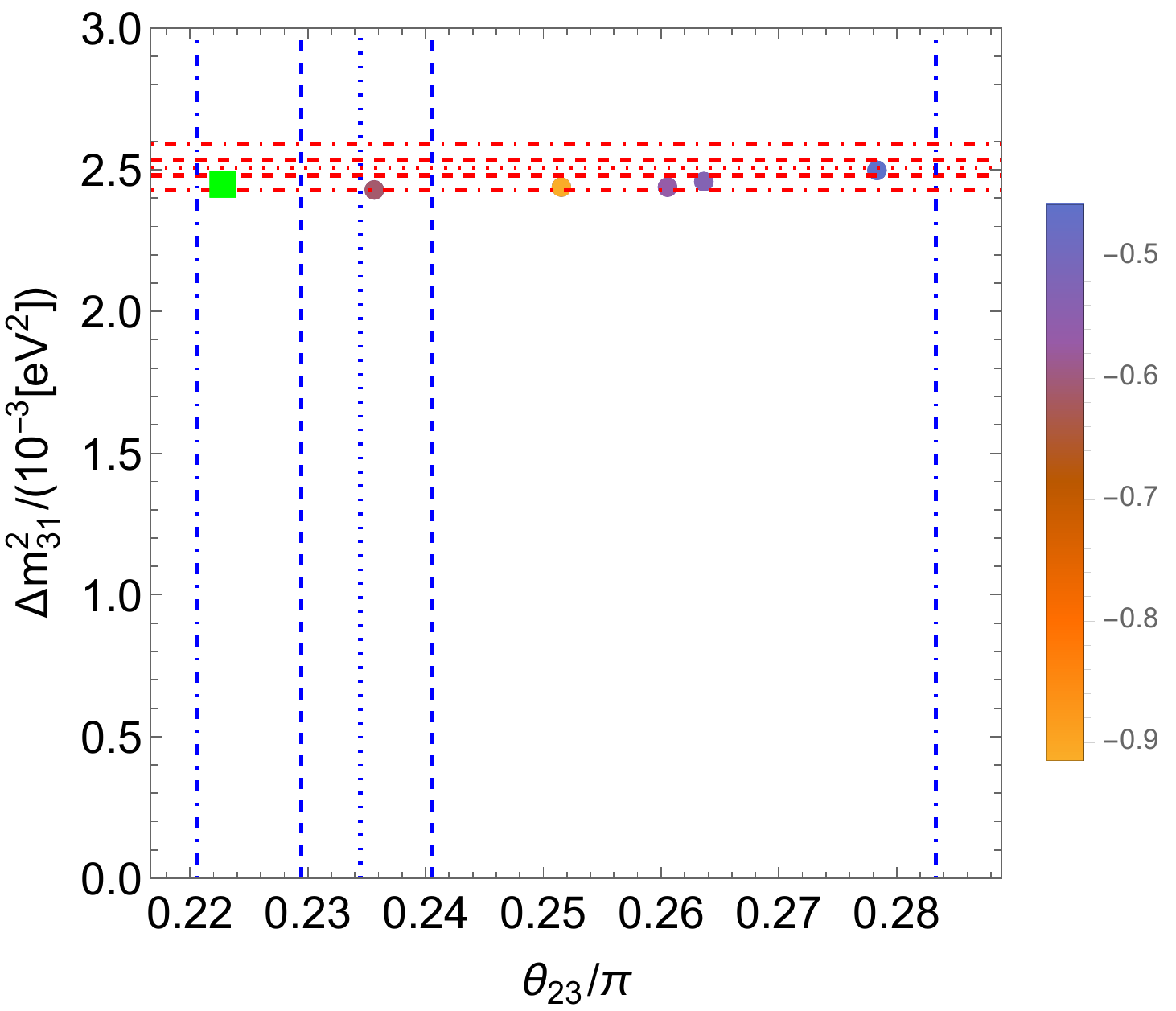}
  \end{center}
 \end{minipage}
\begin{minipage}{0.49\hsize}
  \begin{center}
  \includegraphics[height=60mm]{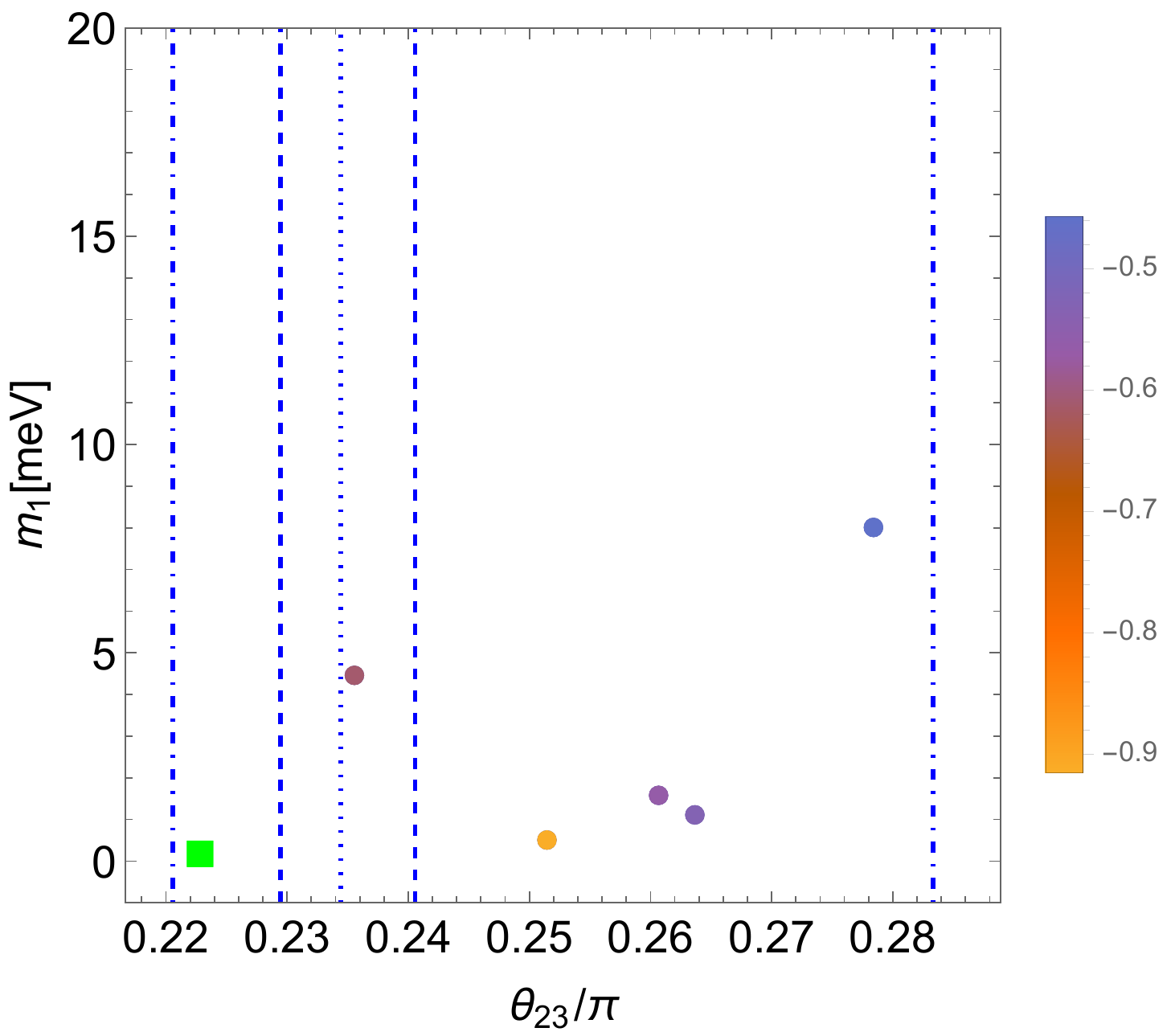}
  \end{center}
 \end{minipage}
\begin{minipage}{0.49\hsize}
  \begin{center}
  \includegraphics[height=60mm]{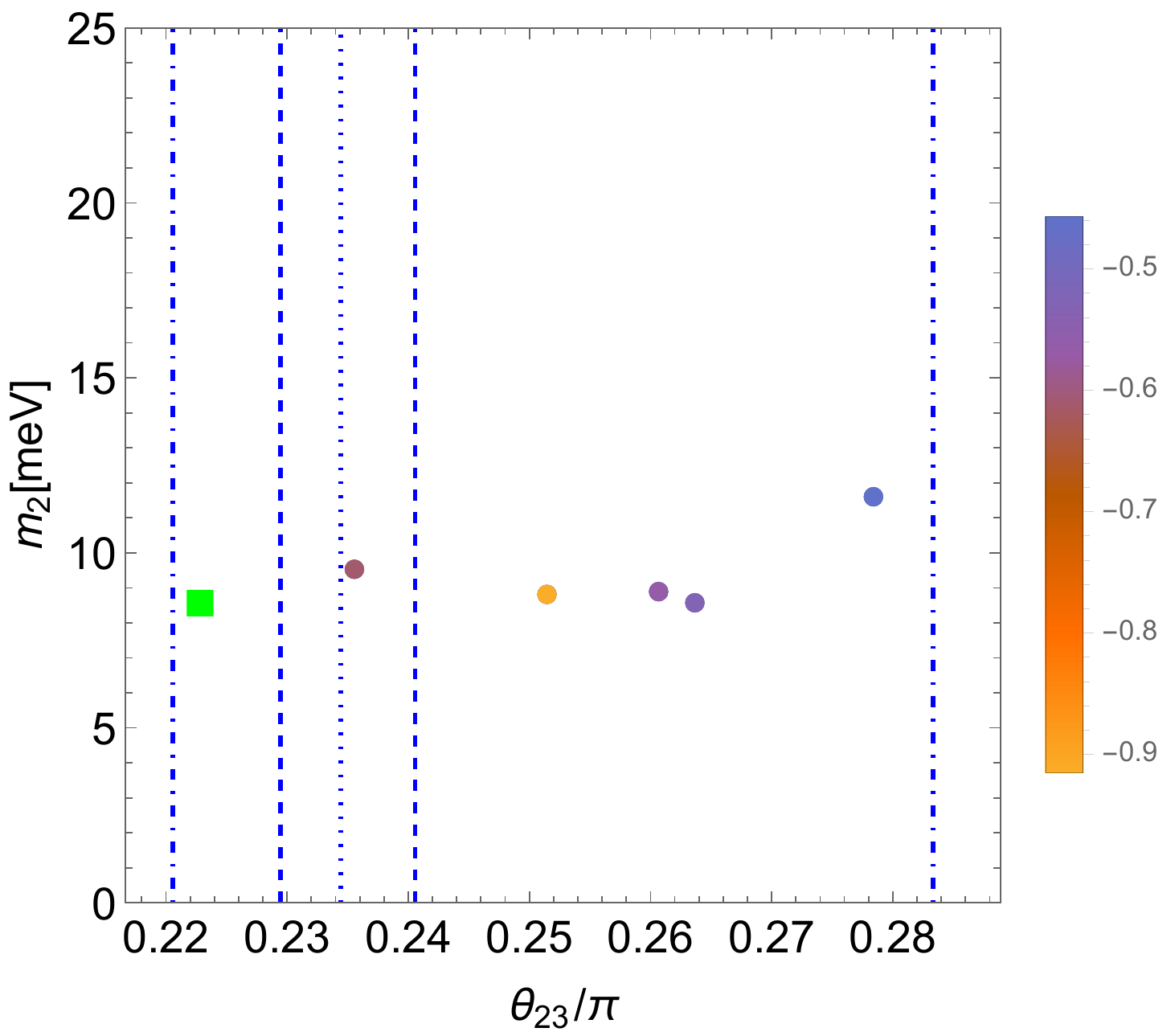}
  \end{center}
 \end{minipage}
\begin{minipage}{0.49\hsize}
  \begin{center}
  \includegraphics[height=60mm]{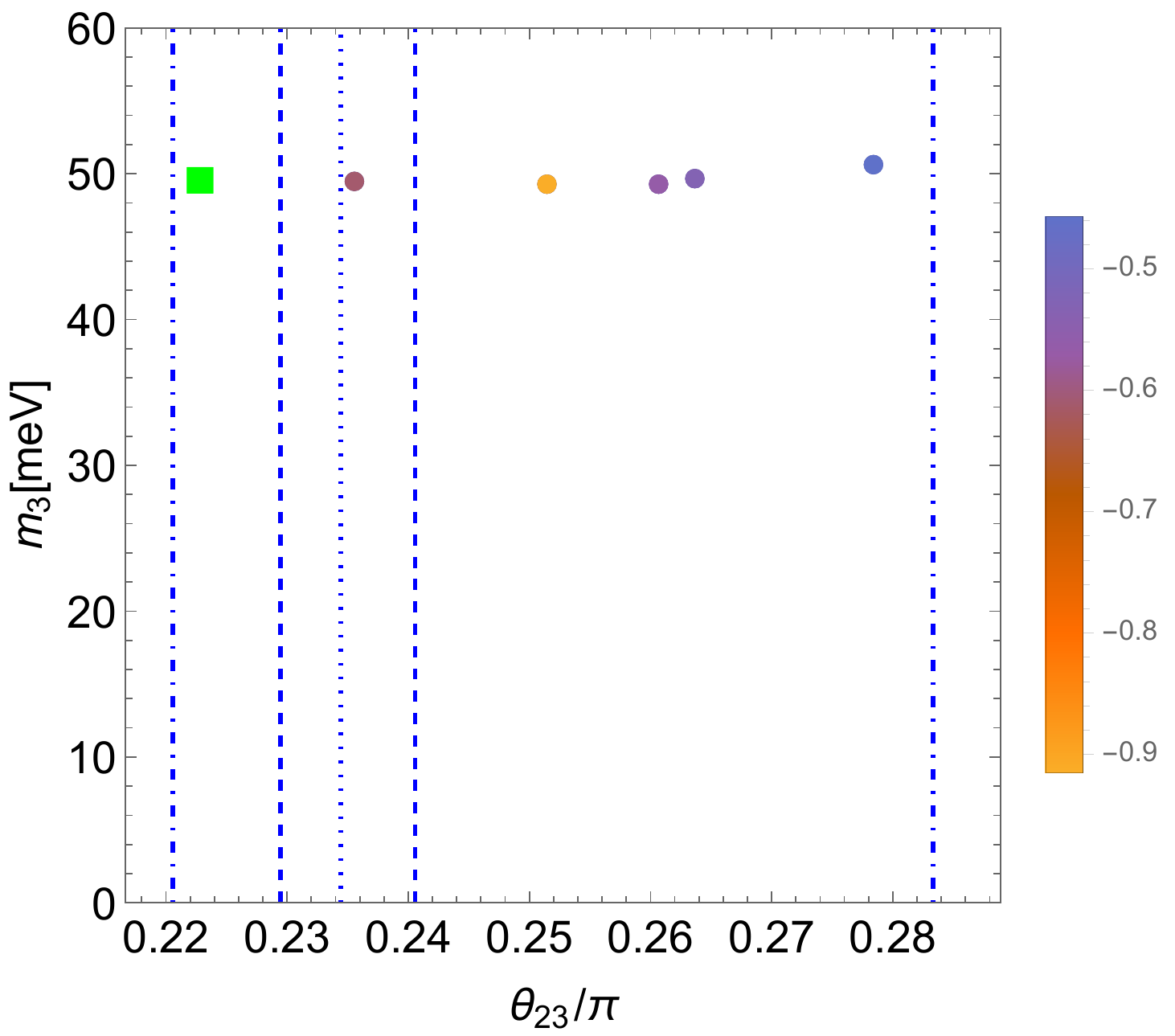}
  \end{center}
 \end{minipage}
 \begin{minipage}{0.49\hsize}
  \begin{center}
  \includegraphics[height=60mm]{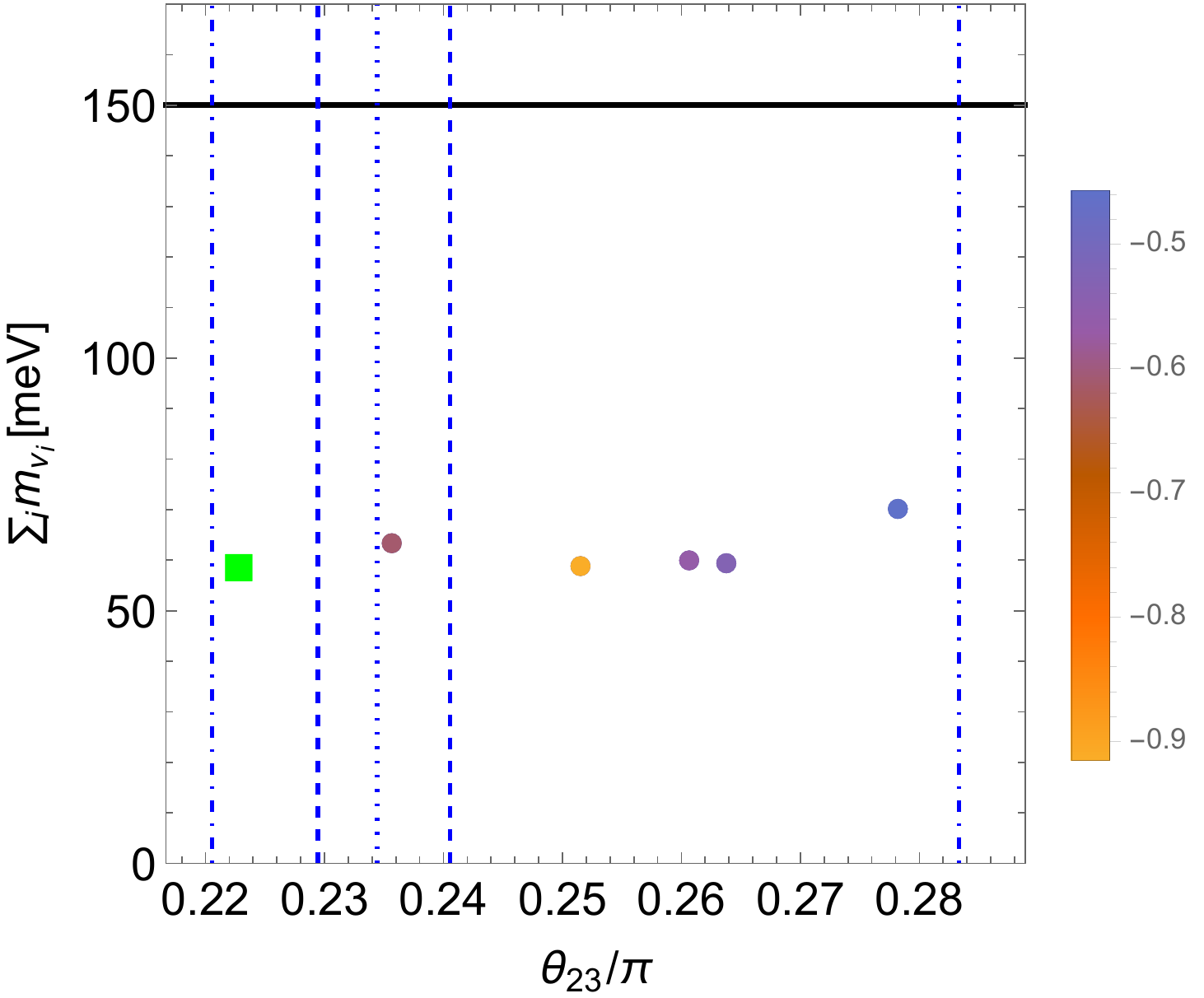}
  \end{center}
 \end{minipage}
 \caption{Neutrino masses vs mixing angle $\theta_{23}$, where the dotted line represents the global best fit value in NuFIT v5.2 results with Super-Kamiokande atmospheric data \cite{Esteban:2020cvm}, and the inside region of each line represents dashed line $\leq 1\sigma$, dotdashed line $\leq 3\sigma$ CL, respectively. 
 The sum of neutrino masses is constrained by $0.15$ eV (95\% CL) corresponding to the black solid line in the case of $\Lambda$CDM model \cite{RoyChoudhury:2019hls}. 
 We denote a best-fit point within $3\sigma$ by a square, and the intrinsic value \eqref{eq:intrinsic_value} is written in the legend. 
 Note that the neutrino mass ordering is fixed as NO in the training of the neural network.}
\label{fig:data_lepton_fixed_NO1}
\end{figure}

\begin{figure}[H]
 \begin{minipage}{0.49\hsize}
  \begin{center}
  \includegraphics[height=60mm]{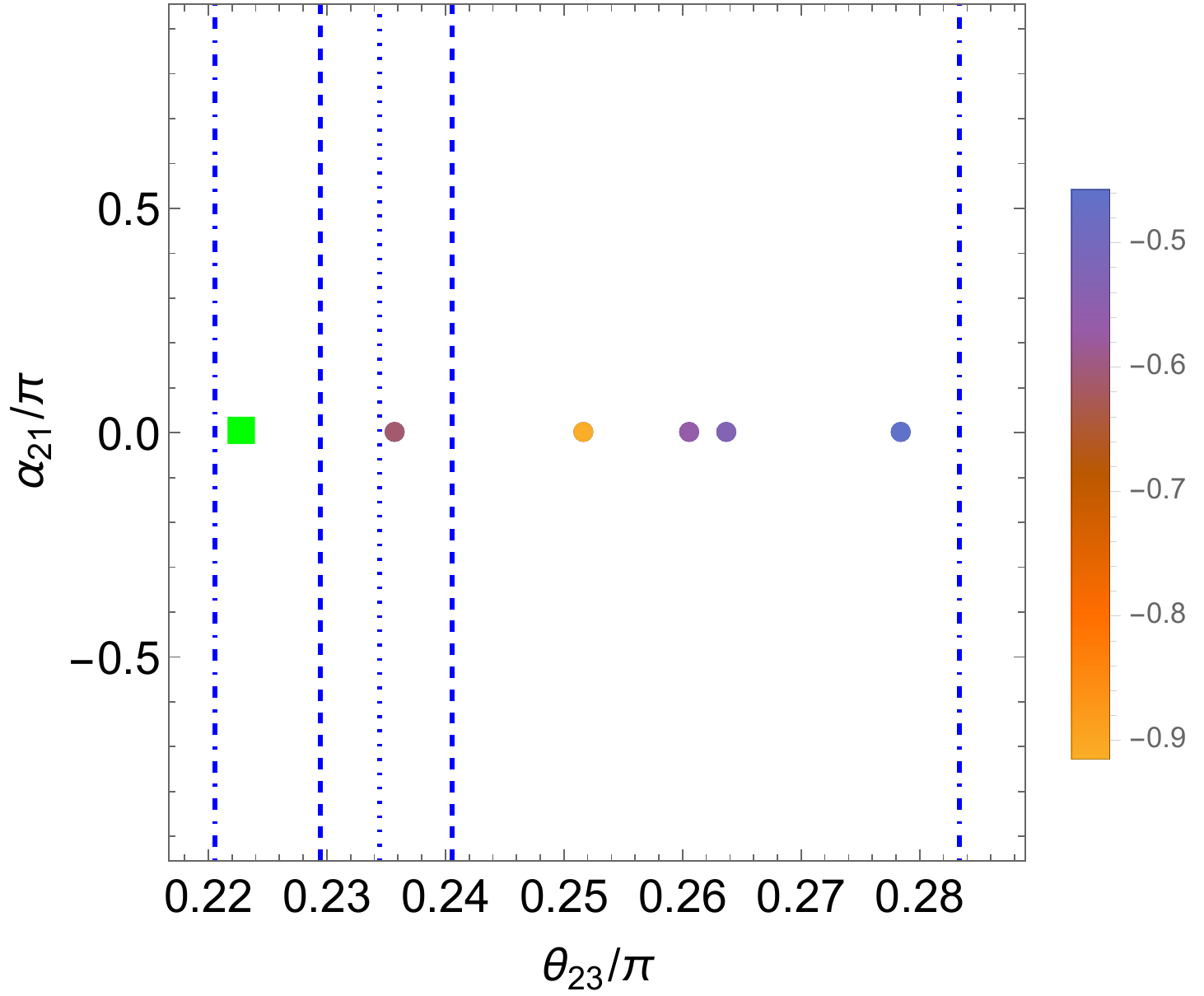}
  \end{center}
 \end{minipage} 
  \begin{minipage}{0.49\hsize}
  \begin{center}
  \includegraphics[height=60mm]{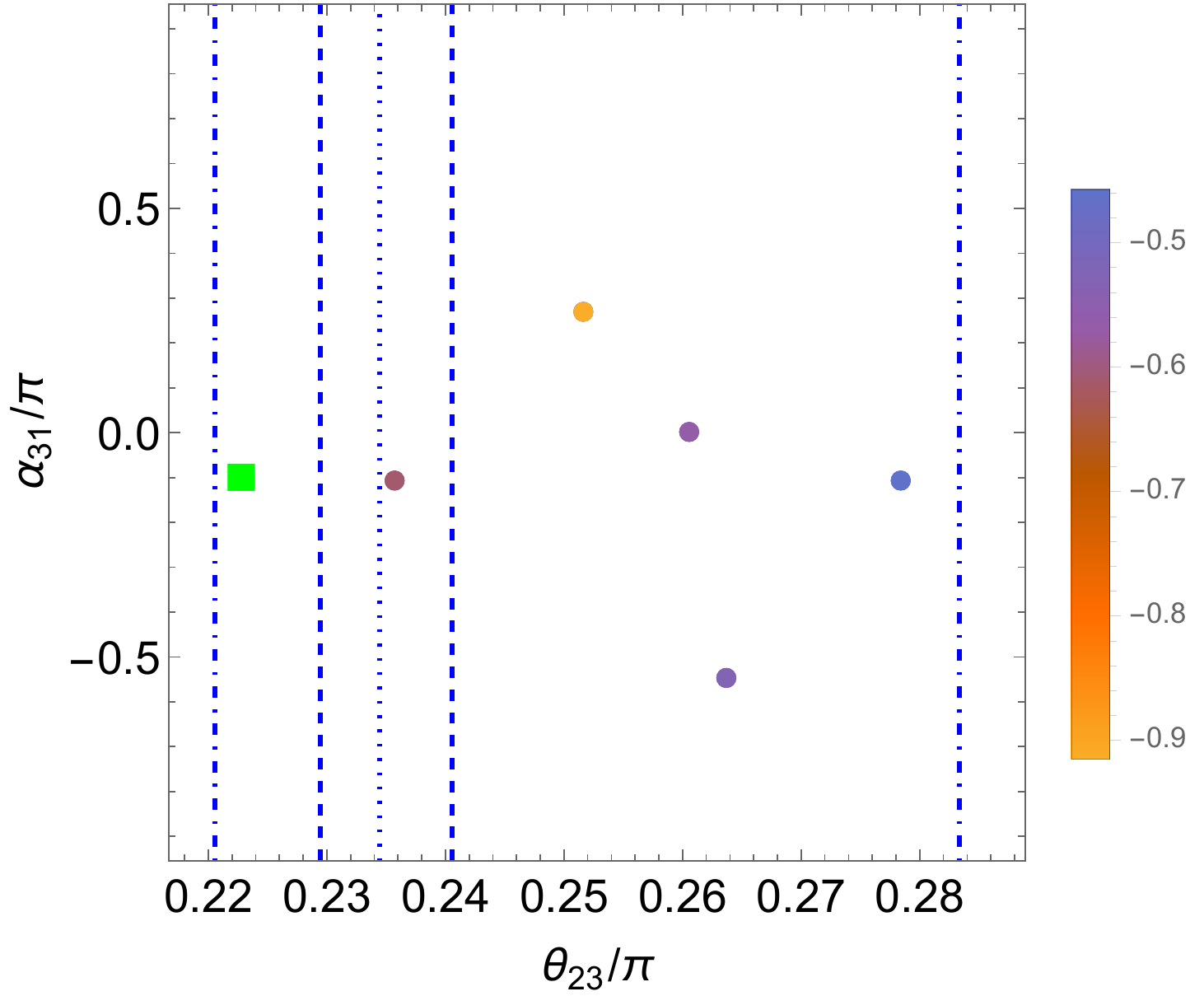}
  \end{center}
 \end{minipage}
  \begin{minipage}{0.49\hsize}
  \begin{center}
  \includegraphics[height=60mm]{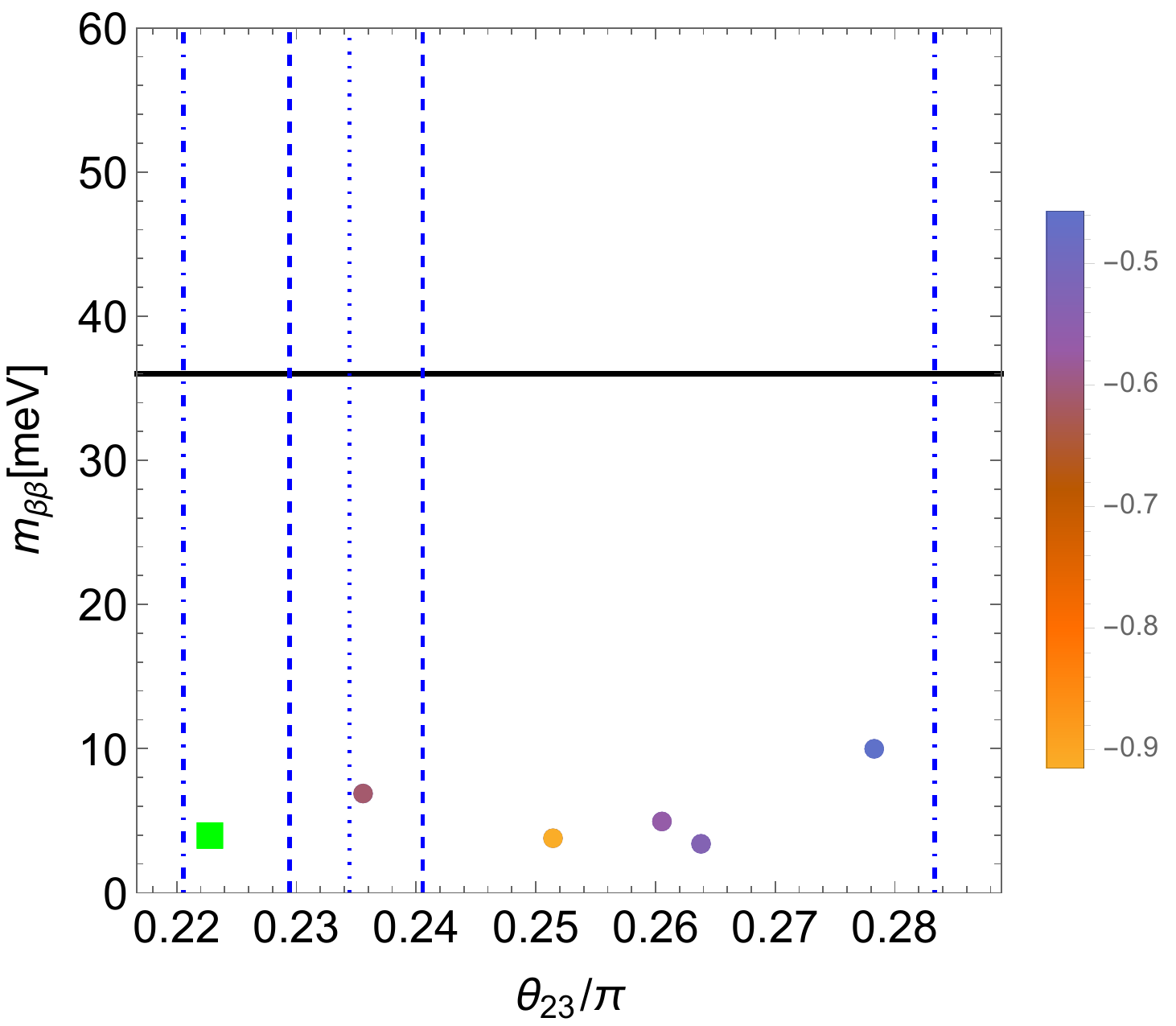}
  \end{center}
 \end{minipage}
 \caption{Majorana phases $\alpha_{21}, \alpha_{31}$ and effective Majorana neutrino mass $m_{\beta\beta}$ vs mixing angle $\theta_{23}$, where the dotted line represents the global best fit value in NuFIT v5.2 results with Super-Kamiokande atmospheric data \cite{Esteban:2020cvm}, and the inside region of each line represents dashed line $\leq 1\sigma$, dotdashed line $\leq 3\sigma$ CL, respectively. 
 The effective Majorana neutrino mass is upper bounded by $0.036$ eV (90\% CL) corresponding to the black solid line \cite{KamLAND-Zen:2022tow}. 
 We denote a best-fit point within $3\sigma$ by a square, and the intrinsic value \eqref{eq:intrinsic_value} is written in the legend. 
 Note that the neutrino mass ordering is fixed as NO in the training of the neural network.}
\label{fig:data_lepton_fixed_NO2}
\end{figure}

After carrying out the same procedure 10 times in total, we find that the results of 6 models with normal ordering are in agreement with experimental values within $3\sigma$. 
We show the benchmark point with the highest intrinsic value in Table \ref{tab:lepton_fixed_NO} for the normal ordering. Here, we list the effective Majorana neutrino mass $m_{\beta \beta}$ for the neutrinoless double beta decay: 
\begin{align}
    m_{\beta\beta}=\left|m_{1}\cos^{2}{\theta_{12}}\cos^{2}{\theta_{13}}+m_{2}\sin^{2}{\theta_{12}}\cos^{2}{\theta_{13}}e^{i\alpha_{21}}+m_{3}\sin^{2}{\theta_{13}}e^{i(\alpha_{31}-2\delta_{\mathrm{CP}})}\right|,
    \label{eq:m_beta}
\end{align}
which would be measured by the KamLAND-Zen experiment \cite{KamLAND-Zen:2022tow}. 
In this analysis, we assume the parameter $M=10^{15}$ GeV to realize the tiny neutrino masses with ${\cal O}(1)$ coefficients of Yukawa couplings, but we leave the detailed study with different values of $M$ for future work. 
Note that the angular component of flavon leads to the nonvanishing Majorana CP phases in contrast to the quark sector.\footnote{The Dirac CP phase is chosen to 0 due to the same reason as in the quark sector} Thus, one can analyze the correlation between mixing angles and CP phase as shown in Figs. \ref{fig:data_lepton_fixed_NO1} and \ref{fig:data_lepton_fixed_NO2} for the normal ordering, in which all the terminal states within $3\sigma$ are shown. 
The CP phase $\alpha_{21}$ is predicted at 0. 
Note that the information of the CP phase has not been implemented in the learning of neural network. 

\begin{figure}[H]
 \begin{minipage}{0.49\hsize}
  \begin{center}
   \includegraphics[height=60mm]{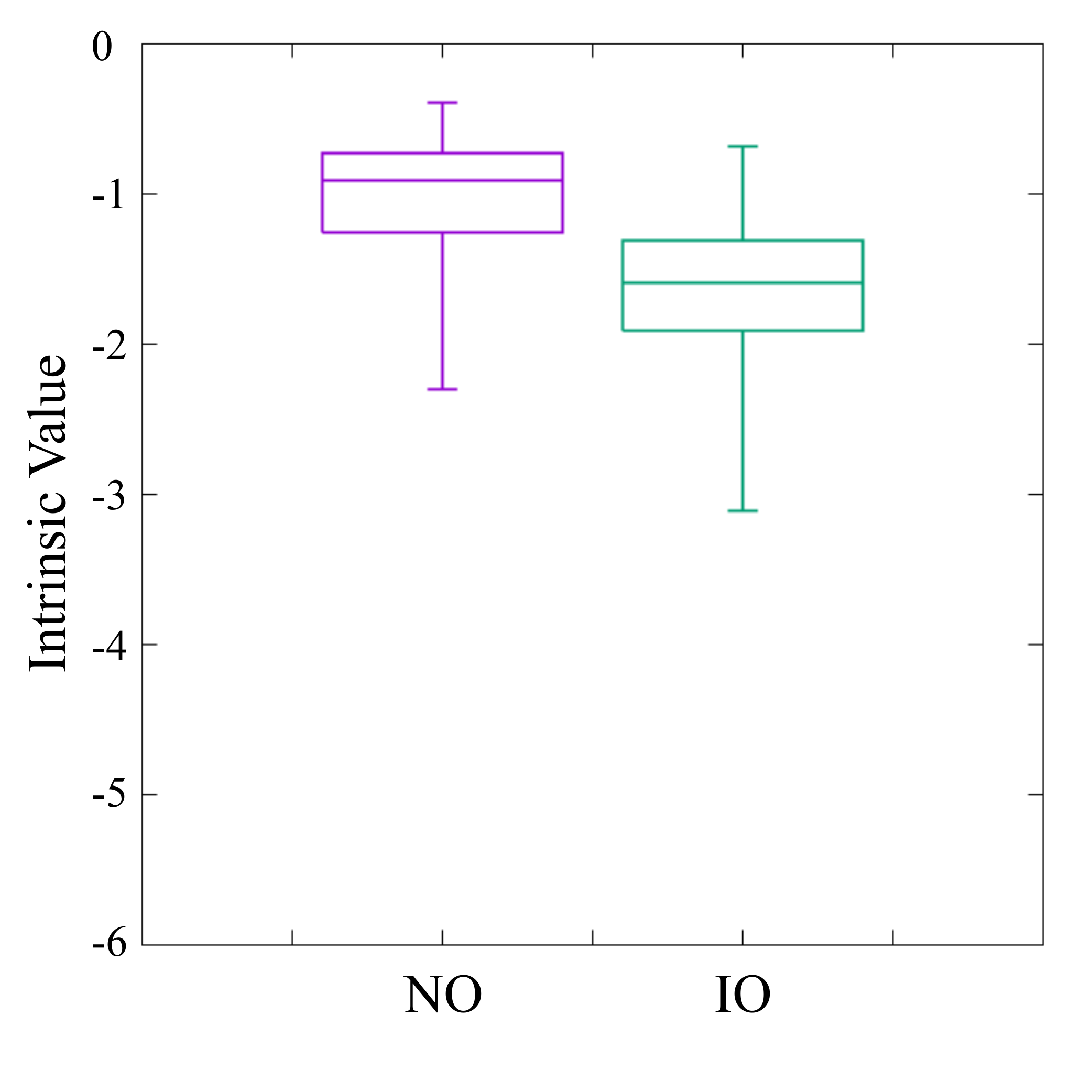}
  \subcaption{Only lepton sector.}
  \end{center}
 \end{minipage}
 \begin{minipage}{0.49\hsize}
  \begin{center}
   \includegraphics[height=60mm]{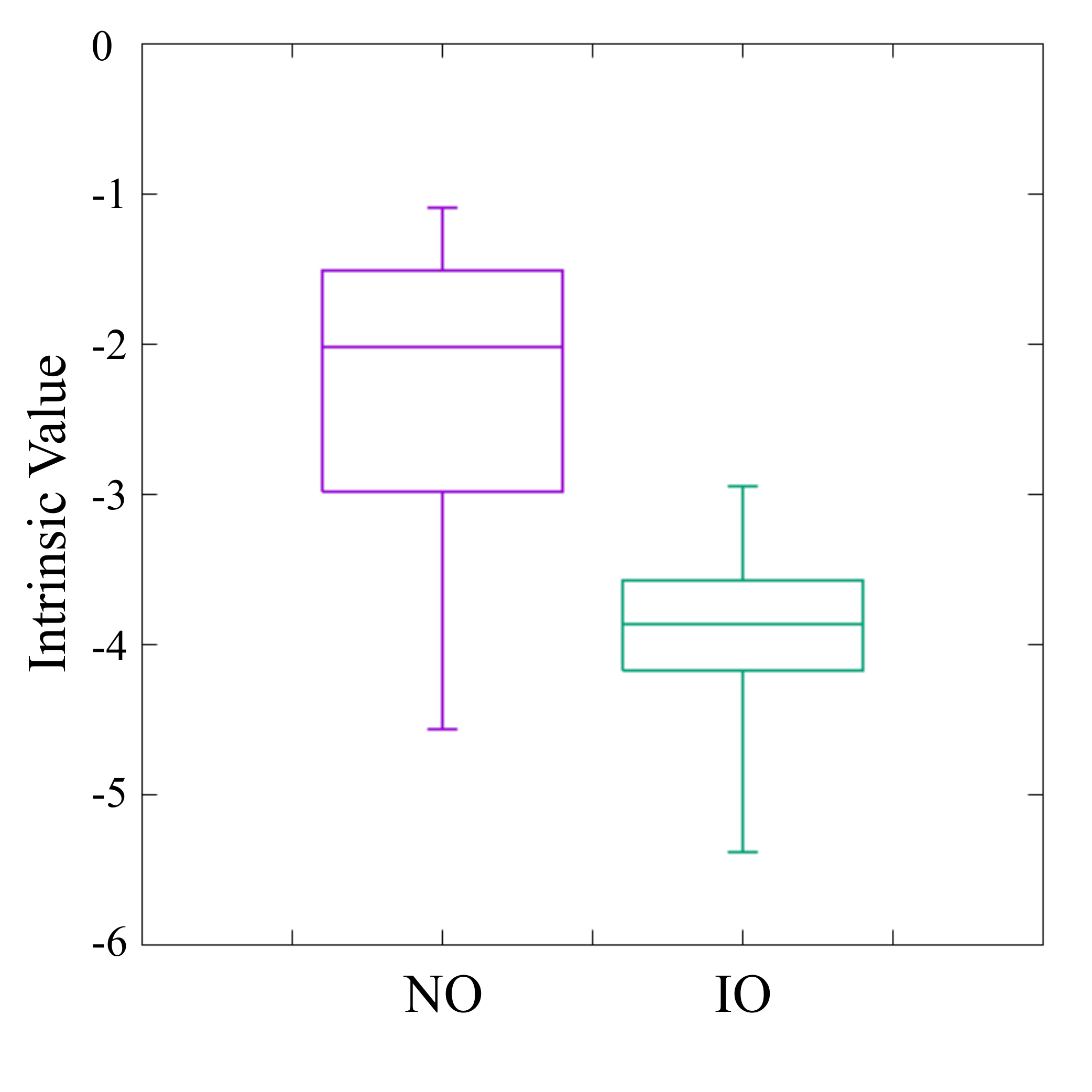}
  \subcaption{Sum with quark sector.}
  \end{center}
 \end{minipage}
  \caption{Boxplots of intrinsic values for the lepton sector, where the neutrino mass ordering is fixed in the learning of neural network. In the left panel, we show intrinsic values obtained in the RL with the lepton sector, but in the right panel, we incorporate the values of the quark sector analyzed in Sec. \ref{sec:quark}.}
\label{fig:IV_fixed}
\end{figure}

Remarkably, one cannot find the neutrino mass of inverted ordering to be consistent with the 
experimental values within $3\sigma$, although we perform the Monte-Carlo search 10 times over the ${\cal O}(1)$ coefficients $y_{ij}$ of 121 terminal states. 
It indicates that normal ordering will be favored by the autonomous behavior of the agent. 
Indeed, the intrinsic value of normal ordering after the Monte-Carlo search tends to be larger than that of inverted ordering as shown in Fig. \ref{fig:IV_fixed}. 

\newpage
\vspace*{\stretch{1}}

\begin{table}[H]
\centering
\scalebox{0.9}{
\begin{tabular}{l|c}
    \hline
    \begin{tabular}{l} Charges \end{tabular} &
    ${\cal Q}=\left(\begin{array}{ccc|ccc|ccc|cc}
        L_{1} & L_{2} & L_{3} & N_{1} & N_{2} & N_{3} & l_{1} & l_{2} & l_{3} & H & \phi \\ 
        \hline
        2 & 1 & 2 & -8 & -1 & -9 & -7 & -3 & -3 & -1 & 1 \\ 
    \end{array}\right)$ \\ 
    \hline
    \begin{tabular}{l} $\mathcal{O}\left(1\right)$ coeff. \end{tabular} & $y^{l} \simeq
    \left(\begin{array}{rrr}
        -0.889 & 2.056 & -0.299 \\
        -1.584 & -2.697 & 1.542 \\
        -0.797 & 0.918 & 1.501
    \end{array}\right) \ ,\ 
    y^{\nu} \simeq
    \left(\begin{array}{rrr}
        1.135 & -1.331 & 0.128 \\
        1.207 & -1.203 & -0.051 \\
        -0.671 & -2.639 & 0.074
    \end{array}\right)$ \\ 
    & $y^{N} \simeq
    \left(\begin{array}{rrr}
        1.125 & -0.388 & 0.950 \\
        -0.388 & 1.066 & -0.349 \\
        0.950 & -0.349 & -0.656
    \end{array}\right)$ \\ 
    \hline
    \begin{tabular}{l} VEV\end{tabular} & $v_{\phi}\simeq0.268\cdot e^{-0.166i}$\\ 
    \hline
    \begin{tabular}{l} Intrinsic value \end{tabular} & $\mathcal{V}_{\mathrm{opt}}\simeq-0.853$ \\ 
    \hline
    \begin{tabular}{l} Masses\\(output) \end{tabular} & \begin{tabular}{l} $\left(\begin{array}{lll}
    m_{e} & m_{\mu} & m_{\tau}
    \end{array}\right)
    \simeq \left(\begin{array}{lll}
    4.576\times 10^{-1}, & 1.231\times 10^{2}, & 7.571\times 10^{2}
    \end{array}\right)
    \ \mathrm{MeV}$ \\
    $\left(\begin{array}{lll}
    m_{\nu_{1}} & m_{\nu_{2}} & m_{\nu_{3}}
    \end{array}\right)
    \simeq \left(\begin{array}{lll}
    0.107, & 8.500, & 49.39
    \end{array}\right)
    \ \mathrm{meV}$ \end{tabular} \\
    \hline
    \begin{tabular}{l} Ratios\\(masses) \end{tabular}
    & $\left(\begin{array}{lll}
    E_{e} & E_{\mu} & E_{\tau}\\
    E_{\nu_{21}} & E_{\nu_{31}} &
    \end{array}\right)
    \simeq \left(\begin{array}{lll}
    0.048 & 0.066 & 0.371 \\
    0.011 & 0.012 & 
    \end{array}\right)$ \\
    \hline
    \begin{tabular}{l} PMNS matrix\\(output) \end{tabular} & $\left|V_{\mathrm{PMNS}}\right| \simeq
    \left(\begin{array}{lll}
    0.819 & 0.553 & 0.154 \\
    0.346 & 0.689 & 0.637 \\
    0.458 & 0.468 & 0.756
    \end{array}\right)$
    \\
    \hline
    \begin{tabular}{l} Ratios\\(mixings) \end{tabular}
    & $E_{\cal P} \simeq
    \left(\begin{array}{lll}
    0.003 & 0.006 & 0.013 \\
    0.030 & 0.062 & 0.041 \\
    0.064 & 0.088 & 0.038
    \end{array}\right)$ \\
    \hline
    \begin{tabular}{l} 
    Majorana phases \end{tabular} &
    $\alpha_{21} \simeq0.0,\ \alpha_{31} \simeq-0.106\pi$ \\ 
    \hline
    \begin{tabular}{l} Effective mass \end{tabular} & $m_{\beta \beta}\simeq3.793\ \mathrm{meV}$ \\ 
    \hline
\end{tabular}
}
\caption{Benchmark point for the lepton sector with NO, where the neutrino mass ordering is 
specified in the learning of the network.}
\label{tab:lepton_fixed_NO}
\end{table}

\vspace{\stretch{1}}
\newpage
\vspace*{\stretch{1}}

\vspace{\stretch{1}}
\newpage
\vspace*{\stretch{1}}

\vspace{\stretch{1}}
\newpage

\subsection{Unfixed ordering of neutrino masses}
\label{sec:neutrino2}

\begin{figure}[H]
\begin{minipage}{0.32\hsize}
  \begin{center}
  \includegraphics[height=47mm]{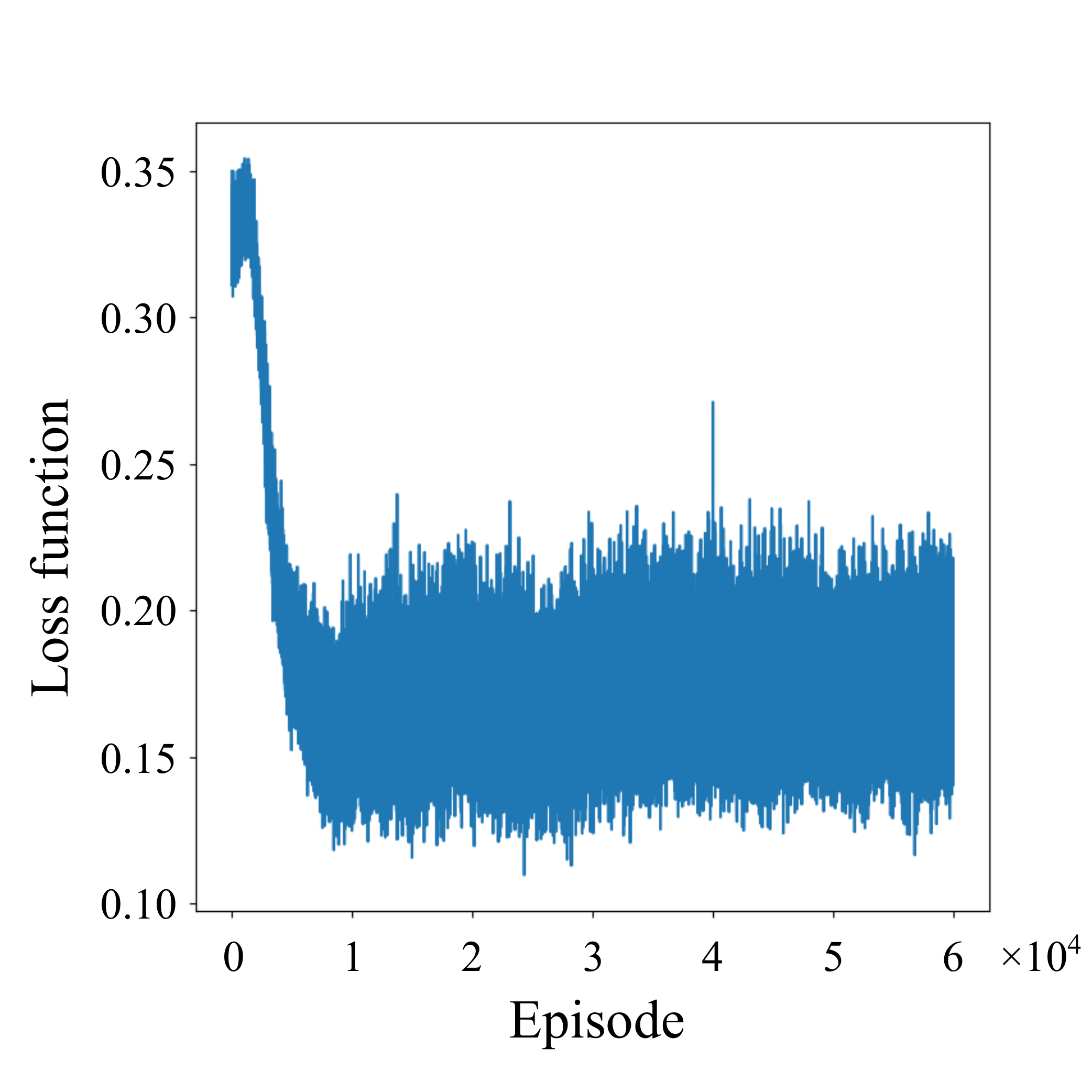}
  \end{center}
 \end{minipage}
 \begin{minipage}{0.32\hsize}
  \begin{center}
   \includegraphics[height=47mm]{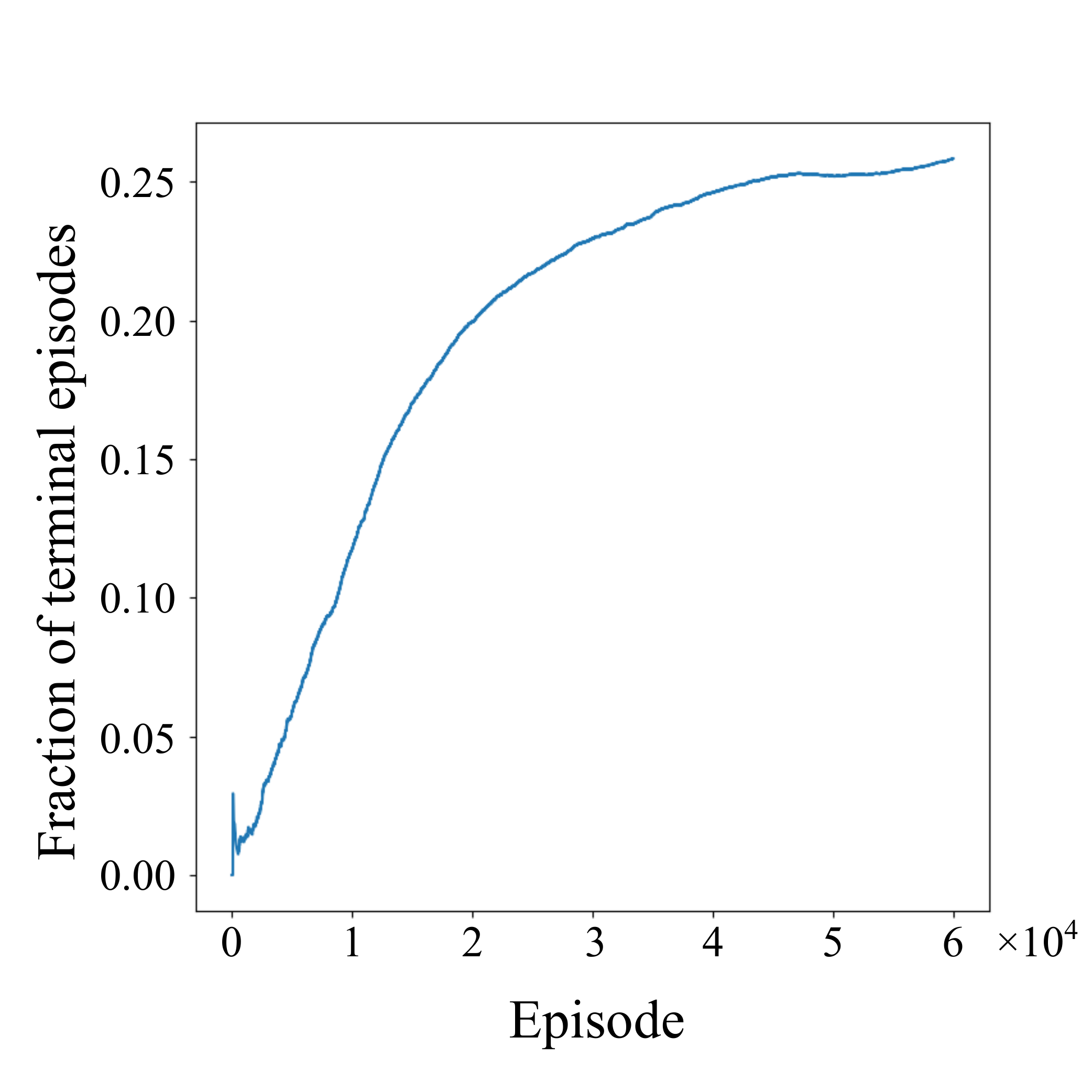}
  \end{center}
 \end{minipage}
 \begin{minipage}{0.32\hsize}
  \begin{center}
   \includegraphics[height=47mm]{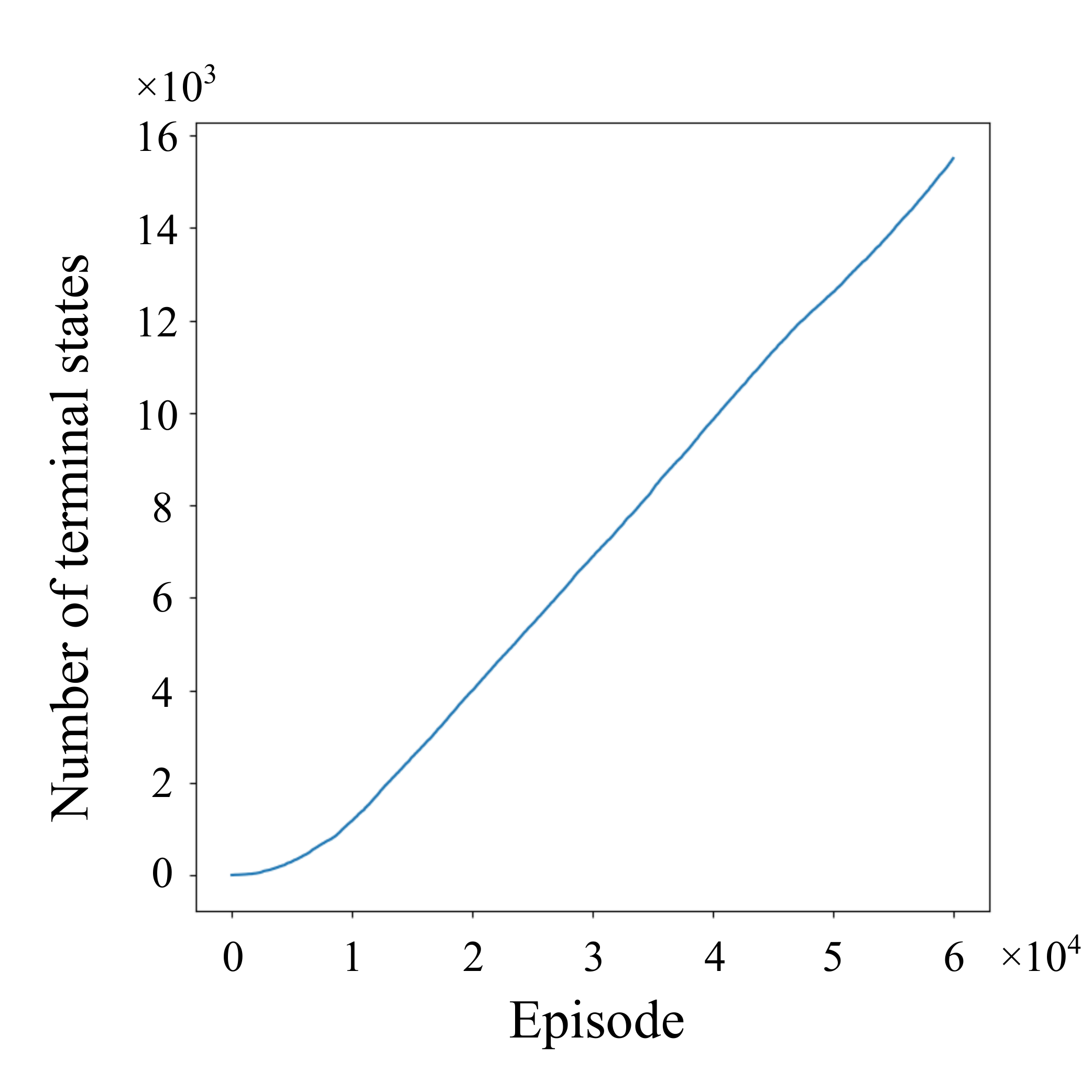}
  \end{center}
 \end{minipage}
  \caption{Learning results for the lepton sector by RL without specifying the neutrino mass ordering. The results are the output of neural network leading to the best-fit model (the diamond in Figs. \ref{fig:data_lepton_unfixed_NO1} and \ref{fig:data_lepton_unfixed_NO2}). 
  We observe a similar behavior for other outputs. 
  From left to right, three panels show (a) the loss function vs episode number (b) the fraction of terminal episodes vs episode number (c) the number of terminal states vs episode number, respectively.}
\label{fig:result_lepton_unfixed}
\end{figure}

In this subsection, we train the neural network without specifying the neutrino mass ordering. 
For each of the 21 realistic FN models in the quark sector, we performed the training twice to obtain a sufficient number of realistic models. 
Similar to the previous analyses, the neural network is trained about 12 hours on a single CPU. 
It turned out that terminal states are found after ${\cal O}(2,000)$ episodes as shown 
in Fig. \ref{fig:result_lepton_unfixed}, where the loss function tends to decrease 
until ${\cal O}(8,000)$ episodes. 
It is notable that the reward increases when the loss function decreases, and 
the lepton sector rapidly leads to the terminal states compared with the quark sector. 
The network leads to terminal states in about >60\% of all cases for total episode $N_{\rm ep}=6\times10^{4}$. 
In contrast to the previous analysis, the trained network efficiently leads to terminal states.\footnote{Note that the specifying the neutrino mass ordering in RL well reproduce the experimental values with high performance.}
After removing the negative integers of $n_{ij}$ in Eq. \eqref{eq:nij_lepton}, 
we arrive at 13,733 (13,432) and 22,430 (20,357) terminal states with normal ordering and inverted ordering in the first (second) learning, respectively
\footnote{The number of overlapping models in the first and second training was 14 for the normal ordering and 16 for the inverted ordering. 
In other words, there were 27,151 and 42,771 independent states obtained in the two training runs. 
While there are approximately $3.2\times10^{11}$ possible combinations of U(1) charges for the lepton sector, the number of episodes in this work is only $N_{\mathrm{ep}}=6\times10^{4}$. 
Therefore, the trend of little overlap in the trainings holds true even after third and fourth runs.}. 
By performing the Monte-Carlo search over the ${\cal O}(1)$ coefficients $y_{ij}$ with the Gaussian distribution shown in Fig. \ref{fig:random}, 
the lepton masses and mixings are optimized to more realistic ones, according to which the intrinsic value is also optimized. 
Specifically, we performed the Monte-Carlo search two times to search the realistic values within $3\sigma$. In the first Monte-Carlo search, we ran 10,000 trials with the Gaussian distribution shown in Fig. \ref{fig:random}. 
Then, for the ${\cal O}(1)$ coefficients with highest intrinsic value among them, we performed the second 10,000 trials with the Gaussian distribution where an average is the coefficients obtained by the first Monte-Carlo search and the standard deviation is 0.25.

\begin{figure}[H]
\begin{minipage}{0.49\hsize}
  \begin{center}
  \includegraphics[height=60mm]{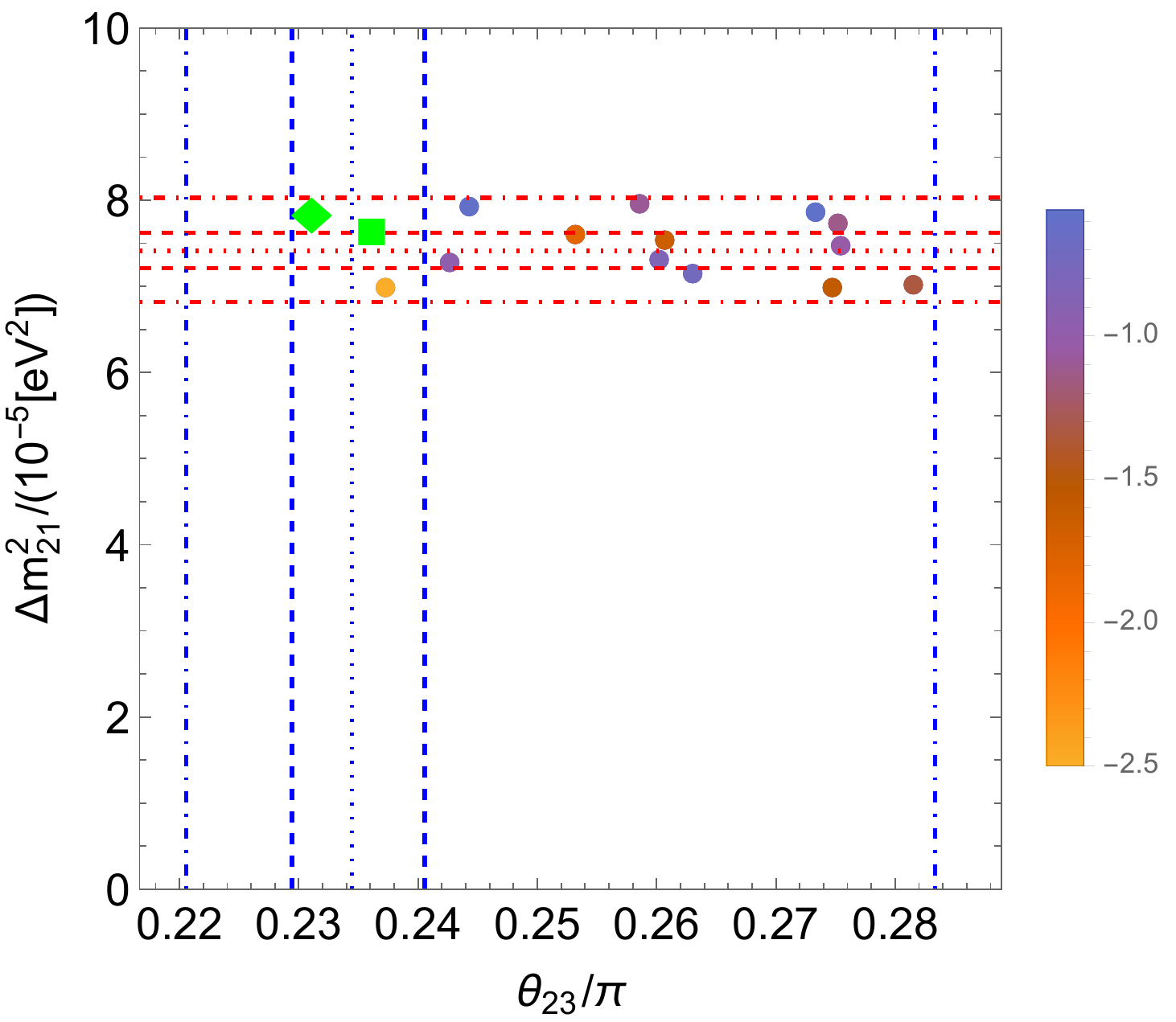}
  \end{center}
 \end{minipage}
\begin{minipage}{0.49\hsize}
  \begin{center}
  \includegraphics[height=60mm]{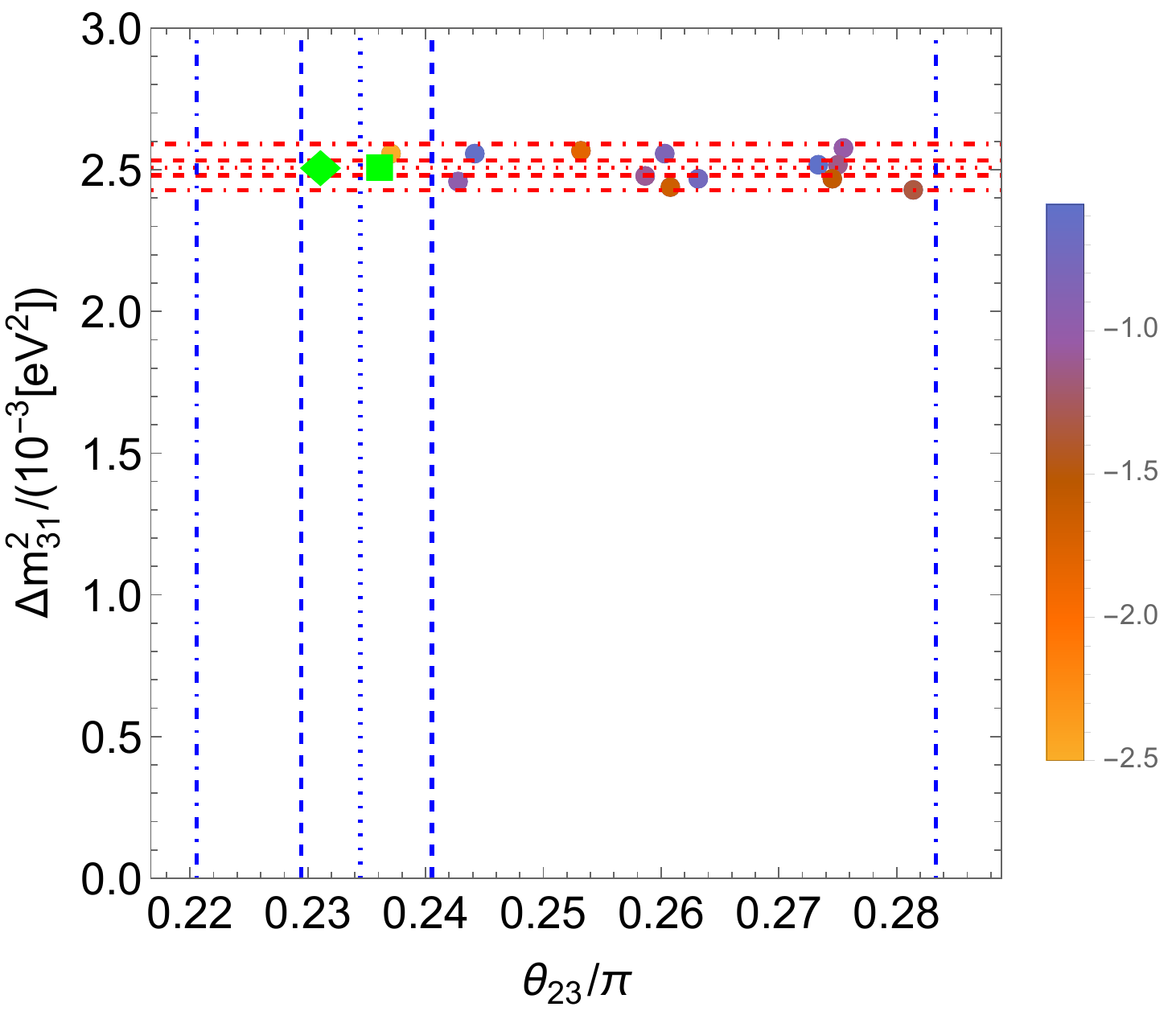}
  \end{center}
 \end{minipage}
\begin{minipage}{0.49\hsize}
  \begin{center}
  \includegraphics[height=60mm]{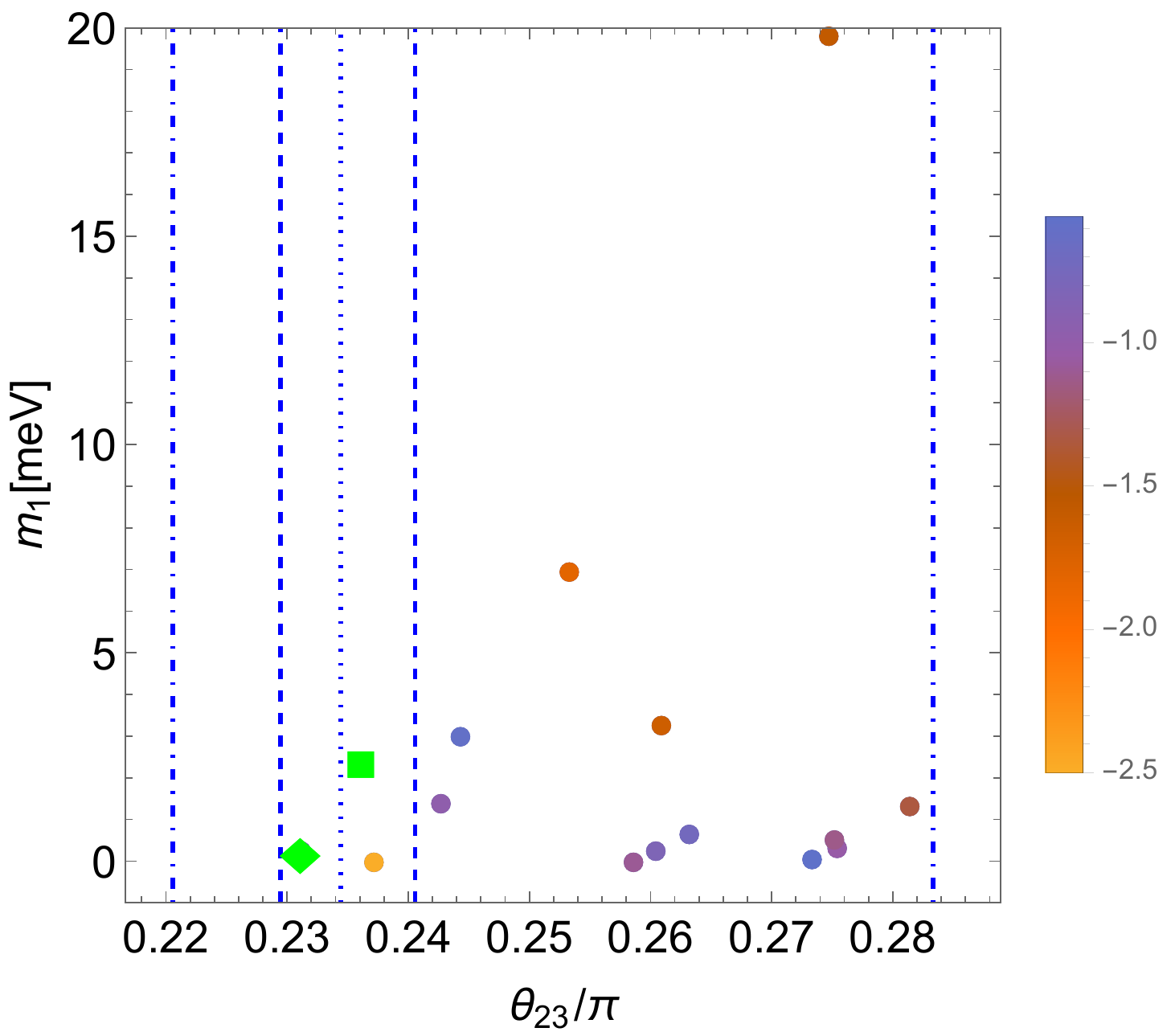}
  \end{center}
 \end{minipage}
\begin{minipage}{0.49\hsize}
  \begin{center}
  \includegraphics[height=60mm]{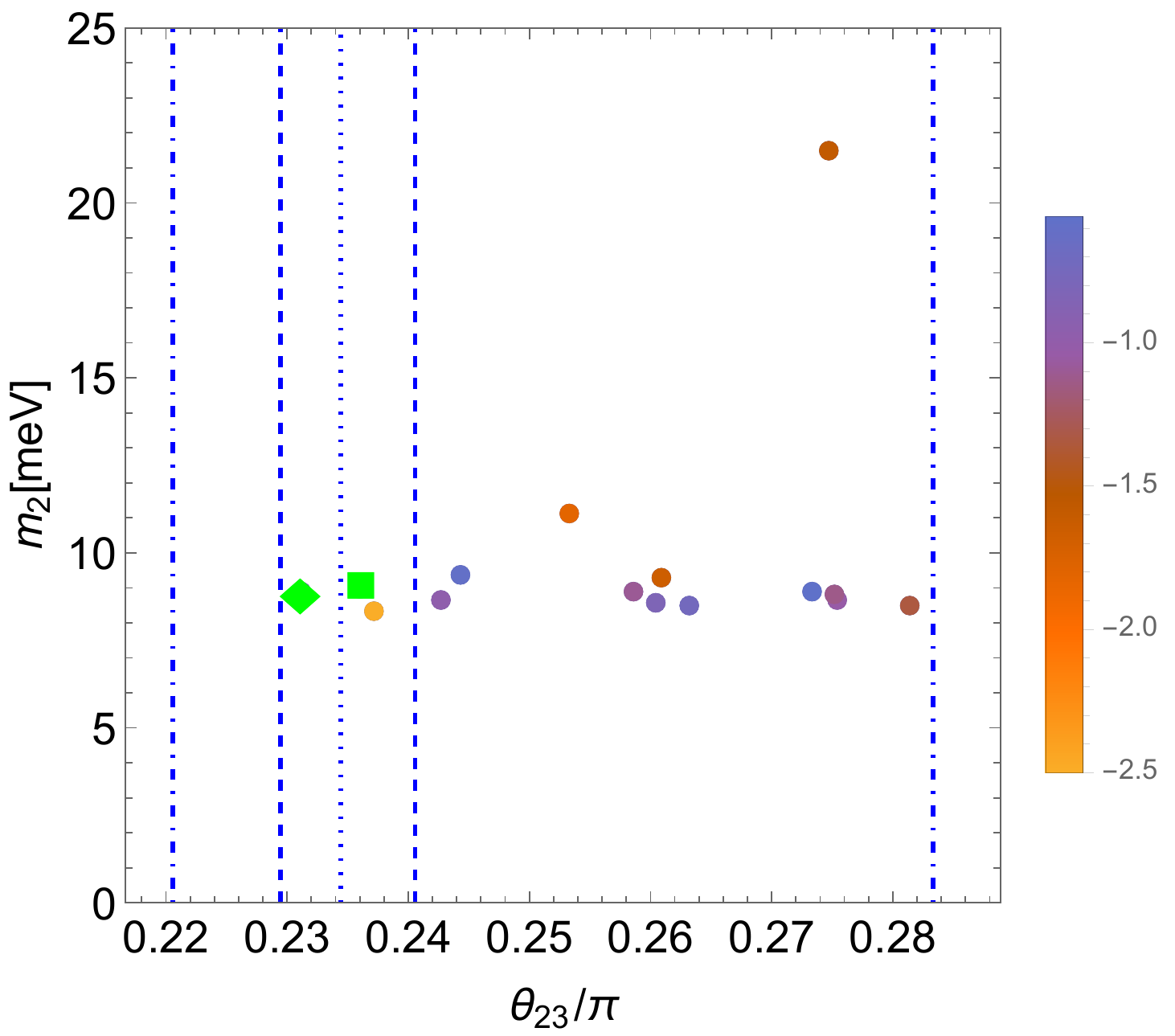}
  \end{center}
 \end{minipage}
\begin{minipage}{0.49\hsize}
  \begin{center}
  \includegraphics[height=60mm]{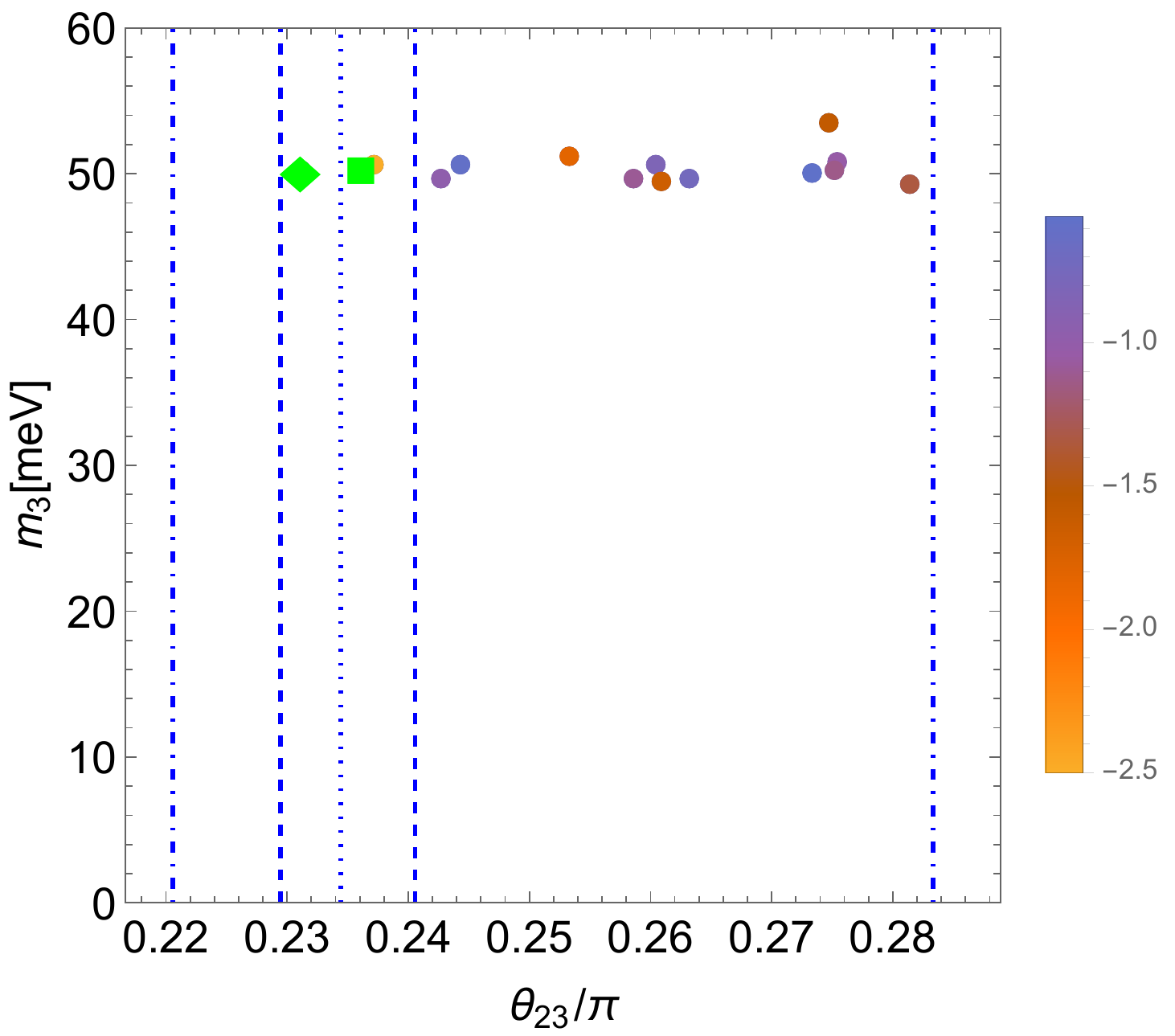}
  \end{center}
 \end{minipage}
 \begin{minipage}{0.49\hsize}
  \begin{center}
  \includegraphics[height=60mm]{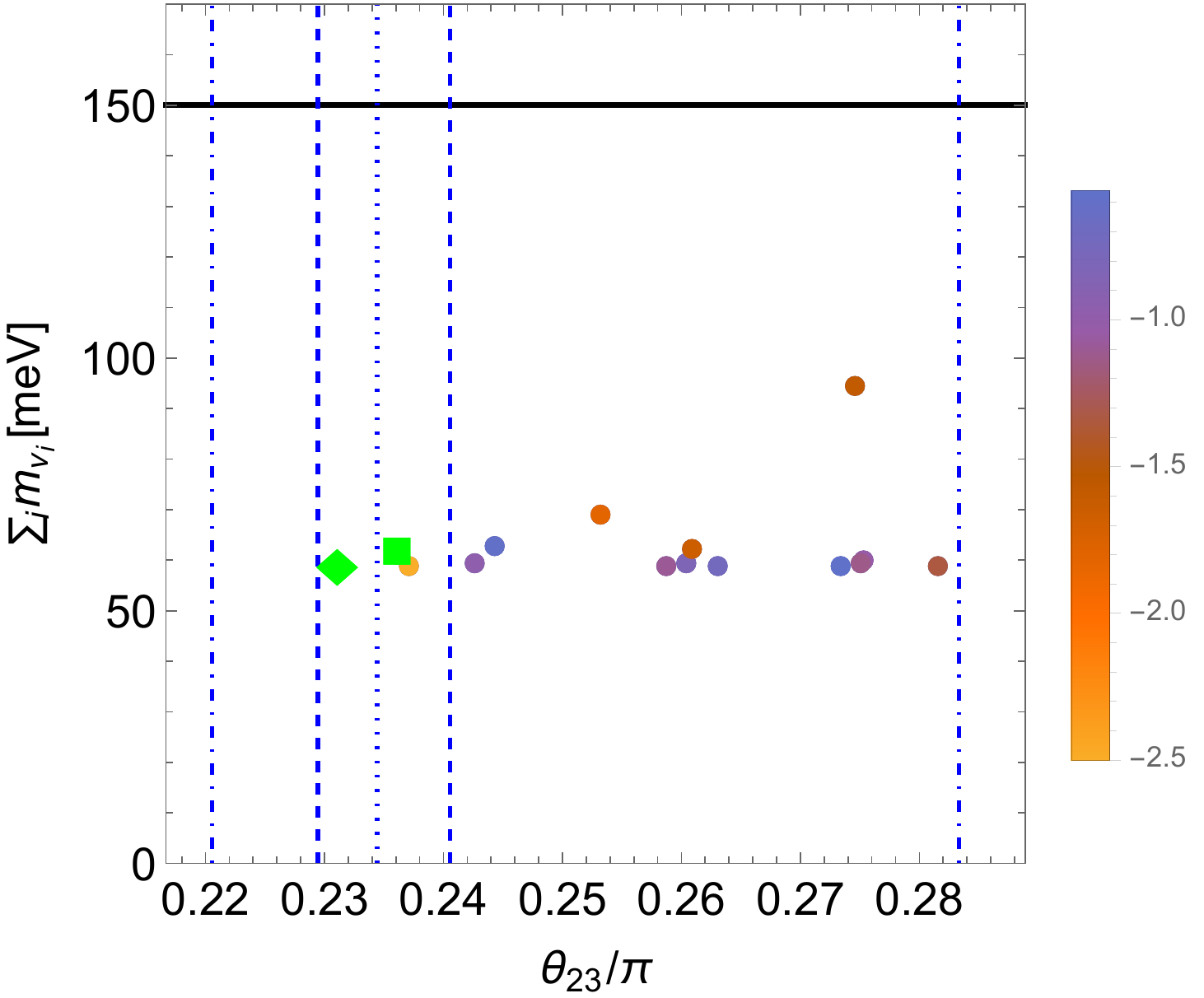}
  \end{center}
 \end{minipage}
 \caption{Neutrino masses vs mixing angle $\theta_{23}$, where the dotted line represents the global best fit value in NuFIT v5.2 results with Super-Kamiokande atmospheric data \cite{Esteban:2020cvm}, and the inside region of each line represents dashed line $\leq 1\sigma$, dotdashed line $\leq 3\sigma$ CL, respectively. 
 The sum of neutrino masses is constrained by $0.15$ eV (95\% CL) corresponding to the black solid line in the case of $\Lambda$CDM model \cite{RoyChoudhury:2019hls}. 
 We denote two best-fit points within $3\sigma$ by a diamond and a square, and the intrinsic value \eqref{eq:intrinsic_value} is written in the legend. 
 Since the neutrino mass ordering is unfixed in the training of the neural network, we show only NO results from the terminal states.}
\label{fig:data_lepton_unfixed_NO1}
\end{figure}

\begin{figure}[H]
 \begin{minipage}{0.49\hsize}
  \begin{center}
  \includegraphics[height=60mm]{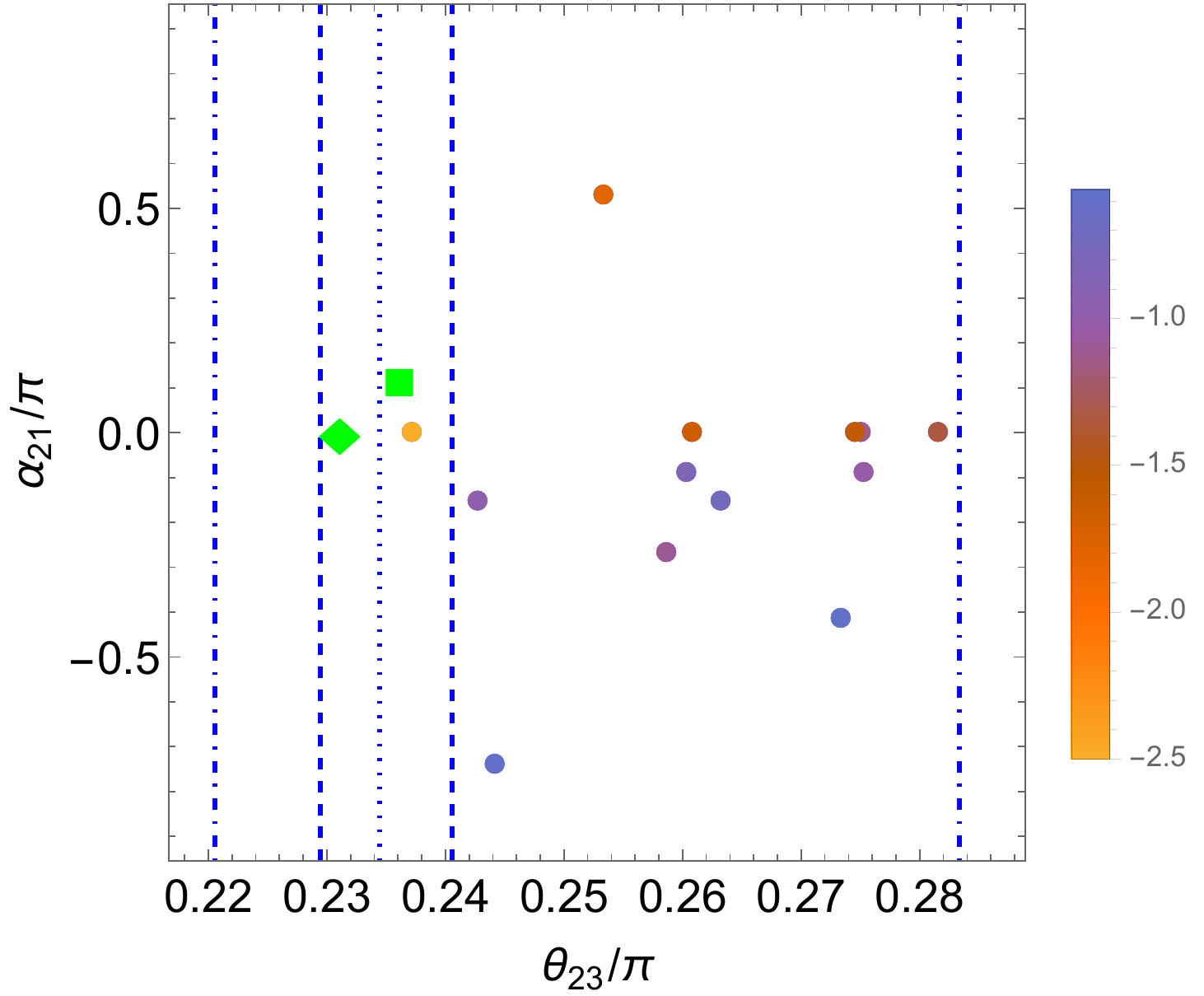}
  \end{center}
 \end{minipage} 
  \begin{minipage}{0.49\hsize}
  \begin{center}
  \includegraphics[height=60mm]{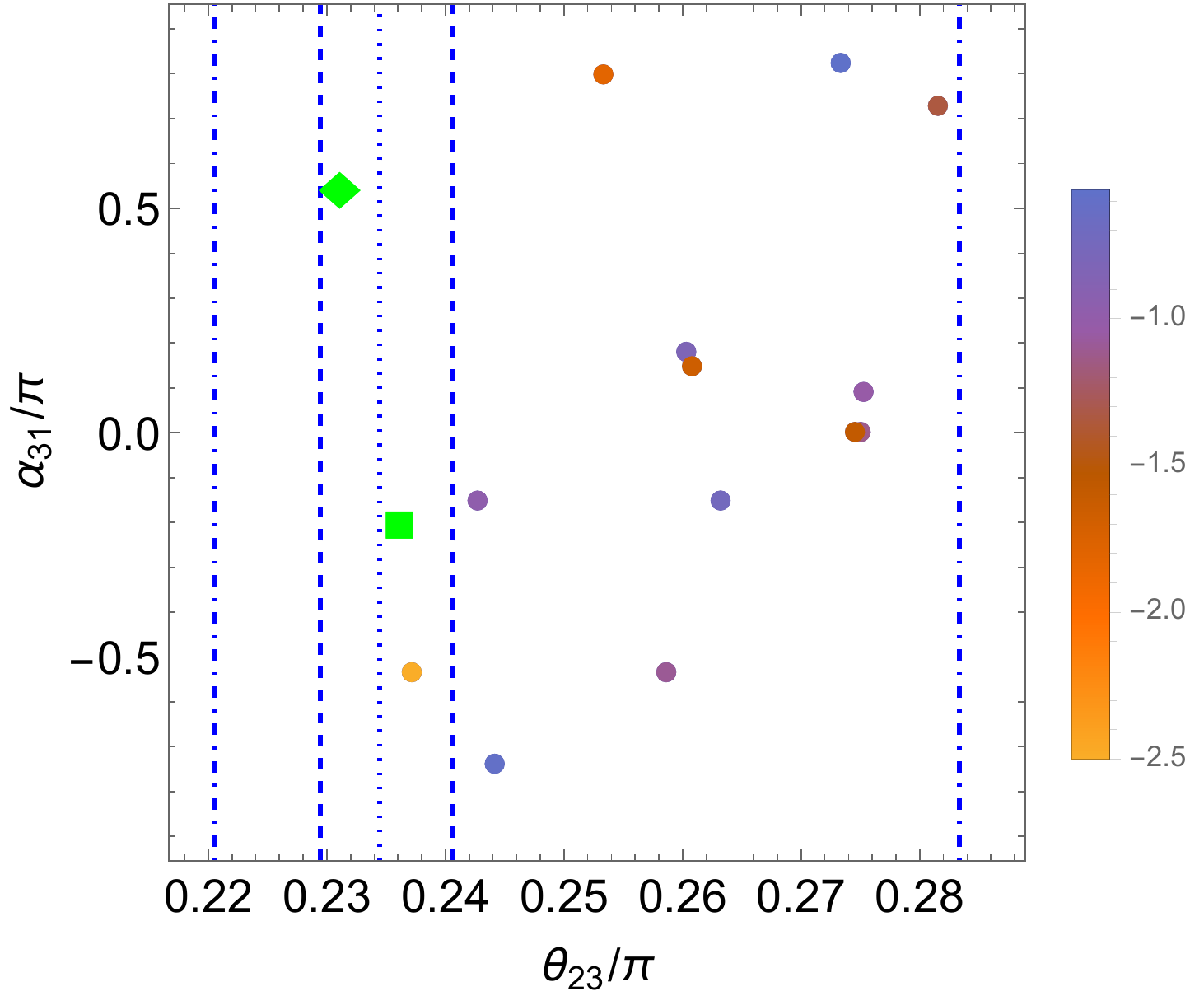}
  \end{center}
 \end{minipage}
  \begin{minipage}{0.49\hsize}
  \begin{center}
  \includegraphics[height=60mm]{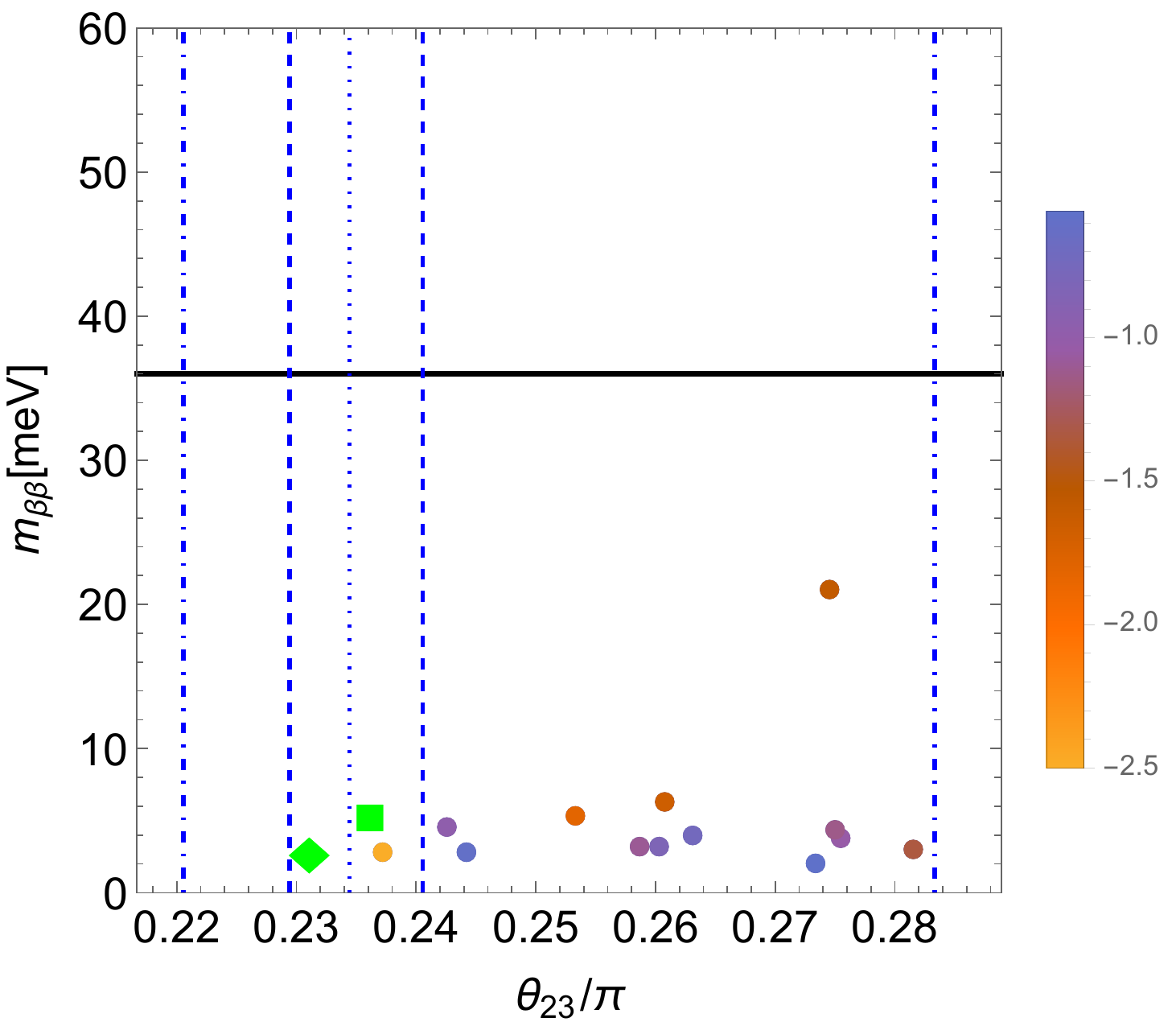}
  \end{center}
 \end{minipage}
 \caption{Majorana phases $\alpha_{21}, \alpha_{31}$ and effective Majorana neutrino mass $m_{\beta\beta}$ vs mixing angle $\theta_{23}$, where the dotted line represents the global best fit value in NuFIT v5.2 results with Super-Kamiokande atmospheric data \cite{Esteban:2020cvm}, and the inside region of each line represents dashed line $\leq 1\sigma$, dotdashed line $\leq 3\sigma$ CL, respectively. 
 The effective Majorana neutrino mass is upper bounded by $0.036$ eV (90\% CL) corresponding to the black solid line \cite{KamLAND-Zen:2022tow}. We denote two best-fit points within $3\sigma$ by a diamond and a square, and the intrinsic value \eqref{eq:intrinsic_value} is written in the legend.} 
\label{fig:data_lepton_unfixed_NO2}
\end{figure}

After carrying out the Monte-Carlo analysis, we find that the results of 15 models with normal ordering are in agreement with experimental values within $3\sigma$. 
Two best fit points with the highest intrinsic value are shown in Tables \ref{tab:lepton_unfixed_NO1} and \ref{tab:lepton_unfixed_NO2} for the normal ordering. 
As presented in the previous section, one can analyze the correlation between mixing angles and the other observed values as shown in Figs. \ref{fig:data_lepton_unfixed_NO1} and \ref{fig:data_lepton_unfixed_NO2} for the normal ordering, in which all the terminal states within $3\sigma$ are shown. 
It turned out that the Majorana CP phases are typically nonzero, and the summation of neutrino masses and the effective mass are not widely distributed but tend to be localized at $\sum_i m_{\nu_i} \sim 60\,{\rm meV} $ and $2\,{\rm meV} \leq m_{\beta\beta}\leq 6\,{\rm meV}$, respectively. 

\newpage
\vspace*{\stretch{1}}

\begin{table}[H]
\centering
\scalebox{0.9}{
\begin{tabular}{l|c}
    \hline
    \begin{tabular}{l} Charges \end{tabular} &
    ${\cal Q}=\left(\begin{array}{ccc|ccc|ccc|cc}
        L_{1} & L_{2} & L_{3} & N_{1} & N_{2} & N_{3} & l_{1} & l_{2} & l_{3} & H & \phi \\ 
        \hline
        3 & 3 & 2 & -3 & 0 & 0 & -3 & 0 & 0 & -2 & 1 \\ 
    \end{array}\right)$ \\ 
    \hline
    \begin{tabular}{l} $\mathcal{O}\left(1\right)$ coeff. \end{tabular} & $y^{l} \simeq
    \left(\begin{array}{rrr}
        1.728 & -1.717 & 1.790 \\
        1.225 & -0.456 & -1.589 \\
        -2.243 & -2.316 & -2.664
    \end{array}\right) \ ,\ 
    y^{\nu} \simeq
    \left(\begin{array}{rrr}
        -1.737 & -1.060 & 2.712 \\
        3.083 & -1.698 & -0.342 \\
        -0.396 & 0.9445 & -0.287
    \end{array}\right)$ \\ 
    & $y^{N} \simeq
    \left(\begin{array}{rrr}
        -1.031 & 2.275 & 1.453 \\
        2.275 & -0.457 & 0.333 \\
        1.453 & 0.333 & 1.559
    \end{array}\right)$ \\ 
    \hline
    \begin{tabular}{l} VEV \end{tabular} & $v_{\phi}\simeq0.181\cdot e^{-0.863i}$ \\ 
    \hline
    \begin{tabular}{l} Intrinsic value \end{tabular} & $\mathcal{V}_{\mathrm{opt}}\simeq-0.859$ \\ 
    \hline
    \begin{tabular}{l} Masses\\(output) \end{tabular} & \begin{tabular}{l} $\left(\begin{array}{lll}
    m_{e} & m_{\mu} & m_{\tau}
    \end{array}\right)
    \simeq \left(\begin{array}{lll}
    4.960\times 10^{-1}, & 8.575\times 10^{1}, & 6.553\times 10^{2}
    \end{array}\right)
    \ \mathrm{MeV}$ \\
    $\left(\begin{array}{lll}
    m_{\nu_{1}} & m_{\nu_{2}} & m_{\nu_{3}}
    \end{array}\right)
    \simeq \left(\begin{array}{lll}
    0.210, & 8.869, & 50.18
    \end{array}\right)
    \ \mathrm{meV}$ \end{tabular} \\
    \hline
    \begin{tabular}{l} Ratios\\(masses) \end{tabular}
    & $\left(\begin{array}{lll}
    E_{e} & E_{\mu} & E_{\tau}\\
    E_{\nu_{21}} & E_{\nu_{31}} &
    \end{array}\right)
    \simeq \left(\begin{array}{lll}
    0.013 & 0.091 & 0.433 \\
    0.026 & 0.002 & 
    \end{array}\right)$ \\
    \hline
    \begin{tabular}{l} PMNS matrix\\(output) \end{tabular} & $\left|V_{\mathrm{PMNS}}\right| \simeq
    \left(\begin{array}{lll}
    0.823 & 0.548 & 0.149 \\
    0.332 & 0.677 & 0.656 \\
    0.460 & 0.491 & 0.740
    \end{array}\right)$
    \\
    \hline
    \begin{tabular}{l} Ratios\\(mixings) \end{tabular}
    & $E_{\cal P} \simeq
    \left(\begin{array}{lll}
    0.000 & 0.001 & 0.001 \\
    0.048 & 0.054 & 0.028 \\
    0.066 & 0.067 & 0.029
    \end{array}\right)$ \\
    \hline
    \begin{tabular}{l} 
    Majorana phases \end{tabular} & 
    $\alpha_{21} \simeq0.0,\ \alpha_{31} \simeq0.549\pi$ \\ 
    \hline
    \begin{tabular}{l} Effective mass \end{tabular} & $m_{\beta \beta}\simeq2.850\ \mathrm{meV}$ \\ 
    \hline
\end{tabular}
}
\caption{Benchmark point for the lepton sector with NO (corresponding to the diamond in Figs. \ref{fig:data_lepton_unfixed_NO1} and \ref{fig:data_lepton_unfixed_NO2}), where the neutrino mass ordering is not specified in the learning of the network.}
\label{tab:lepton_unfixed_NO1}
\end{table}

\vspace{\stretch{1}}
\newpage

\begin{table}[H]
\centering
\scalebox{0.9}{
\begin{tabular}{l|c}
    \hline
    \begin{tabular}{l} Charges \end{tabular} &
    ${\cal Q}=\left(\begin{array}{ccc|ccc|ccc|cc}
        L_{1} & L_{2} & L_{3} & N_{1} & N_{2} & N_{3} & l_{1} & l_{2} & l_{3} & H & \phi \\ 
        \hline
        2 & 3 & 1 & -7 & -8 & -1 & -2 & -5 & -1 & -1 & 1 \\ 
    \end{array}\right)$ \\ 
    \hline
    \begin{tabular}{l} $\mathcal{O}\left(1\right)$ coeff. \end{tabular} & $y^{l} \simeq
    \left(\begin{array}{rrr}
        -0.424 & -0.567 & 0.897 \\
        -0.482 & -0.787 & 0.827 \\
        0.141 & -0.704 & 0.565
    \end{array}\right) \ ,\ 
    y^{\nu} \simeq
    \left(\begin{array}{rrr}
        -1.243 & 1.096 & 0.396 \\
        -0.898 & -1.501 & -3.224 \\
        2.361 & 2.246 & -1.668
    \end{array}\right)$ \\ 
    & $y^{N} \simeq
    \left(\begin{array}{rrr}
        2.311 & 0.877 & -1.491 \\
        0.877 & -1.746 & 0.186 \\
        -1.491 & 0.186 & -0.283
    \end{array}\right)$ \\ 
    \hline
    \begin{tabular}{l} VEV \end{tabular} & $v_{\phi}\simeq0.268\cdot e^{-0.166i}$ \\ 
    \hline
    \begin{tabular}{l} Intrinsic value \end{tabular} & $\mathcal{V}_{\mathrm{opt}}\simeq-0.720$ \\ 
    \hline
    \begin{tabular}{l} Masses\\(output) \end{tabular} & \begin{tabular}{l} $\left(\begin{array}{lll}
    m_{e} & m_{\mu} & m_{\tau}
    \end{array}\right)
    \simeq \left(\begin{array}{lll}
    4.067\times 10^{-1}, & 1.483\times 10^{2}, & 2.066\times 10^{3}
    \end{array}\right)
    \ \mathrm{MeV}$ \\
    $\left(\begin{array}{lll}
    m_{\nu_{1}} & m_{\nu_{2}} & m_{\nu_{3}}
    \end{array}\right)
    \simeq \left(\begin{array}{lll}
    2.251, & 9.006, & 50.04
    \end{array}\right)
    \ \mathrm{meV}$ \end{tabular} \\
    \hline
    \begin{tabular}{l} Ratios\\(masses) \end{tabular}
    & $\left(\begin{array}{lll}
    E_{e} & E_{\mu} & E_{\tau}\\
    E_{\nu_{21}} & E_{\nu_{31}} &
    \end{array}\right)
    \simeq \left(\begin{array}{lll}
    0.099 & 0.147 & 0.066 \\
    0.011 & 0.000 & 
    \end{array}\right)$ \\
    \hline
    \begin{tabular}{l} PMNS matrix\\(output) \end{tabular} & $\left|V_{\mathrm{PMNS}}\right| \simeq
    \left(\begin{array}{lll}
    0.817 & 0.556 & 0.151 \\
    0.499 & 0.552 & 0.668 \\
    0.288 & 0.621 & 0.729
    \end{array}\right)$
    \\
    \hline
    \begin{tabular}{l} Ratios\\(mixings) \end{tabular}
    & $E_{\cal P} \simeq
    \left(\begin{array}{lll}
    0.004 & 0.008 & 0.005 \\
    0.129 & 0.035 & 0.020 \\
    0.137 & 0.035 & 0.023
    \end{array}\right)$ \\
    \hline
    \begin{tabular}{l} 
    Majorana phases \end{tabular} &
    $\alpha_{21} \simeq0.106\pi,\ \alpha_{31} \simeq-0.211\pi$ \\ 
    \hline
    \begin{tabular}{l} Effective mass \end{tabular} & $m_{\beta \beta}\simeq5.040\ \mathrm{meV}$ \\ 
    \hline
\end{tabular}
}
\caption{Benchmark point for the lepton sector with NO (corresponding to the square in Figs. \ref{fig:data_lepton_unfixed_NO1} and \ref{fig:data_lepton_unfixed_NO2}), where the neutrino mass ordering is not specified in the learning of the network.}
\label{tab:lepton_unfixed_NO2}
\end{table}

\begin{figure}[H]
 \begin{minipage}{0.49\hsize}
  \begin{center}
   \includegraphics[height=60mm]{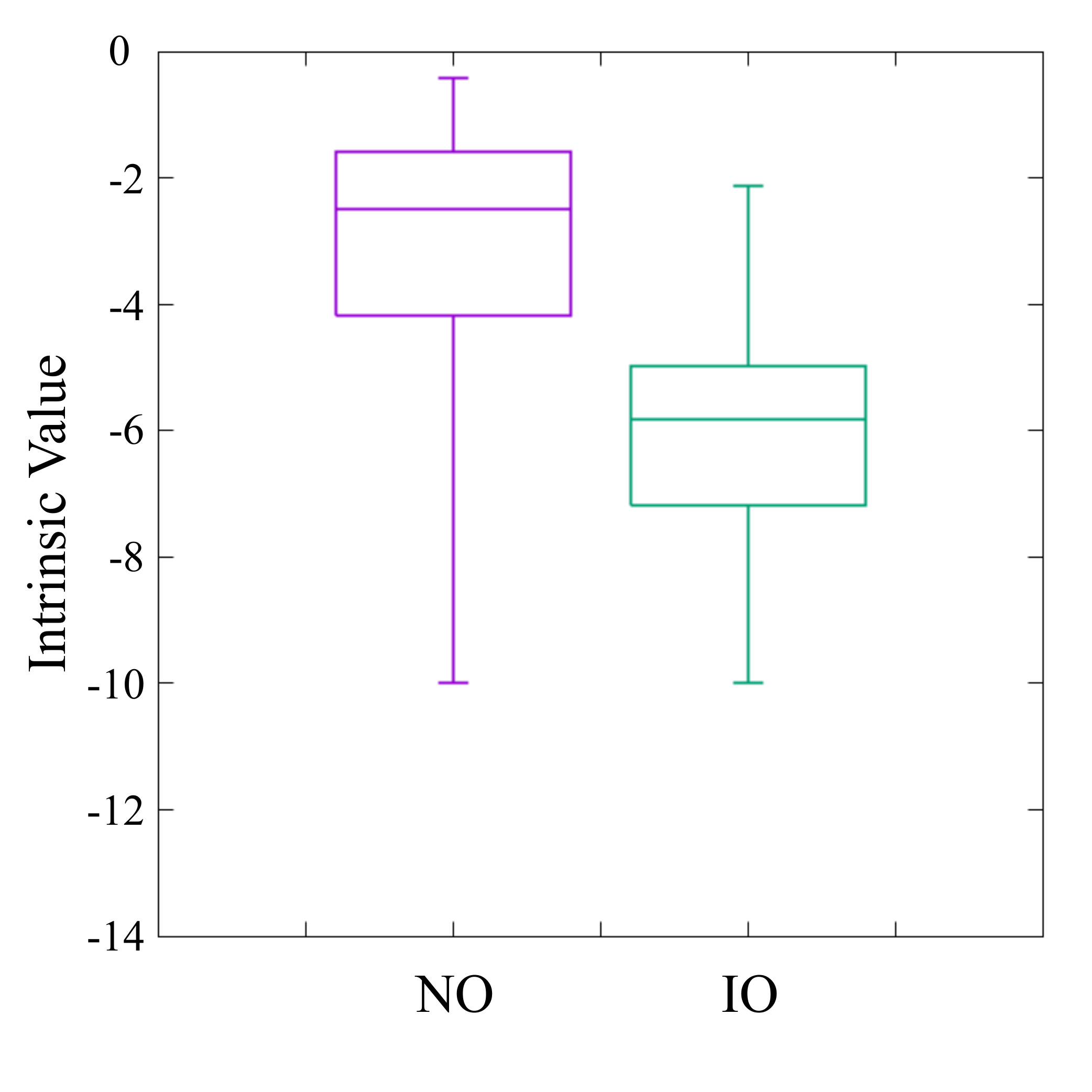}
  \subcaption{Only the lepton sector.}
  \end{center}
 \end{minipage}
 \begin{minipage}{0.49\hsize}
  \begin{center}
   \includegraphics[height=60mm]{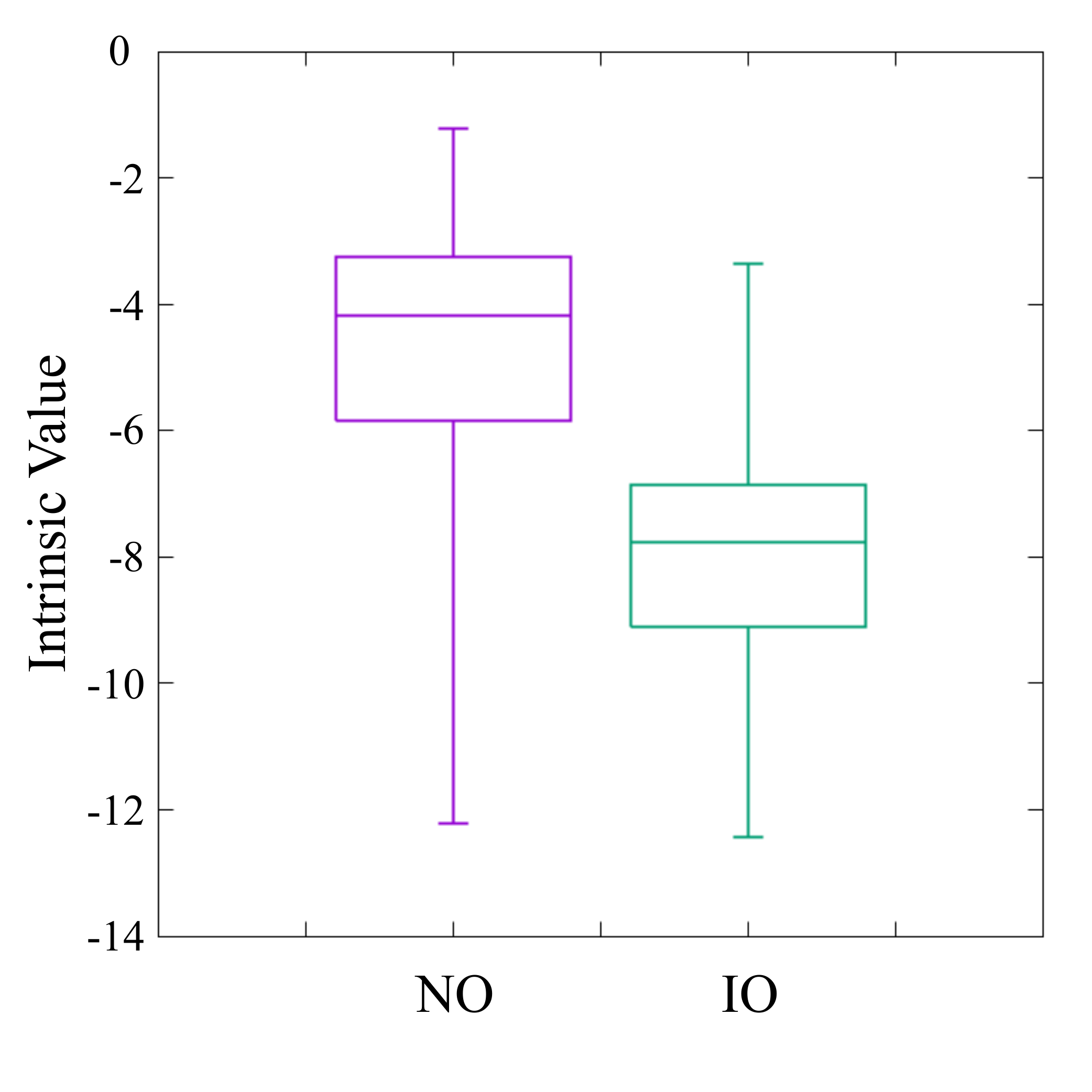}
  \subcaption{Both the lepton and quark sectors.}
  \end{center}
 \end{minipage}
  \caption{Boxplots of intrinsic values for the lepton sector from the result of second learning, where the neutrino mass ordering is unspecified in the learning of neural network. In the left panel, we show intrinsic values obtained in the RL with the lepton sector, but in the right panel, we incorporate the values of the quark sector analyzed in Sec. \ref{sec:quark}. We obtain similar results in the first learning.}
\label{fig:IV_unfixed}
\end{figure}

Remarkably, one cannot obtain the experimental values of neutrino masses and mixings within $3\sigma$ for the inverted ordering, although we perform the Monte-Carlo searches over the ${\cal O}(1)$ coefficients $y_{ij}$ of all the terminal states. 
Thus, the normal ordering of neutrino masses is also favored by the trained neural network, 
although the neural network itself was trained without any knowledge of neutrino mass ordering. 
Indeed, the intrinsic value of normal ordering after the Monte-Carlo search tends to be larger than that of inverted ordering as shown in Fig. \ref{fig:IV_unfixed}. 
This conspicuous feature can also be seen by looking at the intrinsic value 
including both the quark and lepton sectors. 

\newpage
\vspace*{\stretch{1}}

\vspace{\stretch{1}}
\newpage
\vspace*{\stretch{1}}

\vspace{\stretch{1}}
\newpage

\section{Conclusion}
\label{sec:con}

The flavor symmetries are one of attractive tools to understand the flavor structure of quarks and leptons. 
To address the flavor puzzle in the Standard Model, we have applied the reinforcement learning technique  
to flavor models with $U(1)$ horizontal symmetry. 
RL will shed new light on the phenomenological approach to scan over the parameter space of flavor models in contrast to the brute-force approach.

\medskip

In this paper, we have extended the analysis of Ref. \cite{Harvey:2021oue} to explore the flavor structure of 
quarks and leptons by employing the RL with DQN. 
Based on the neural network architectures in the framework of $U(1)$ flavor model with RL established 
in Secs. \ref{sec:DQN} and \ref{sec:model}, 
the agent is designed to exhibit autonomous behavior in the environment (parameter space of $U(1)$ 
charges). Since the parameter space of $U(1)$ charges is huge, we have performed a separate 
search for the $U(1)$ charge assignment of quarks and leptons. 
Trained neural network leads to phenomenologically promising terminal states 
in >6\% for the quark sector and >60\% for the lepton sector in the case of unfixed ordering of neutrino masses. 
In the analysis of Sec. \ref{sec:neutrino2}, we have not specified the neutrino mass ordering 
in the evaluation of intrinsic value, meaning that the agent does not have any knowledge of 
neutrino mass ordering. However, the autonomous behavior of the agent suggests us that the intrinsic value of normal ordering tends to be larger than that of inverted ordering as shown in Fig. \ref{fig:IV_unfixed}, and the normal ordering is well fitted with the current experimental data in contrast to 
the inverted ordering. 
Remarkably, the effective mass for the neutrinoless double beta decay is predicted around specific values, and the Majorana CP phases are nonzero in general.

\medskip

Before closing our paper, it is worthwhile mentioning a possible application of our analysis: 

\begin{itemize}

\item We have focused on the flavor structure of Yukawa couplings, but it is easily applicable to 
reveal the flavor structure of higher-dimensional operators (see for the Standard Model effective field theory (SMEFT) with $U(1)$ flavor symmetry \cite{Bordone:2019uzc} and discrete symmetry \cite{Kobayashi:2021pav}.). 
Since the trained neural network predicts the plausible charge assignment of quarks and leptons, 
one can also determine the flavor structure of higher-dimensional operators. 
It would be interesting to clarify whether RL technique we proposed can explore the 
flavor structure of the SMEFT.

\item On top of that, the CP-odd fluctuation of complex flavon field (flaxion) would be regarded as QCD axion as discussed in Refs. \cite{Salvio:2015cja,Ballesteros:2016euj,Ballesteros:2016xej,Ema:2016ops}, where the cosmological problems (such as the origin of dark matter, 
baryon asymmetry of the Universe, and the inflation) are simultaneously solved by the dynamics of flavon field. 
Since the flavon field has flavor changing neutral current (FCNC) interactions with quarks and leptons controlled by the $U(1)$ flavor symmetry, charge assignment of quarks and leptons plays an important role of determining 
the FCNC processes. 
It is fascinating to apply our finding charge assignment to such an axion physics, which 
left for future work. 

\item We have focused on the $U(1)$ horizontal symmetry, but it is easily applicable to other flavor symmetries 
such as discrete flavor symmetries. We hope to elucidate a comprehensive study about the global structure of flavor models in an upcoming paper. 

\item 
By analyzing the underlying factors that characterize reasonable flavor models from neural networks, we will figure out general patterns behind the flavor models. 
The search for the flavor structure by RL is expected not only to explore new physics beyond the Standard Model by validating flavor models, but also to unravel the black box of machine learning itself.

\end{itemize}

\acknowledgments

This work was supported in part by Kyushu University’s Innovator Fellowship Program (S. N., C. M.), JSPS KAKENHI Grant Numbers JP20K14477 (H. O.) and JP23H04512 (H.O).

\appendix

\section{Formulation of reinforcement learning}
\label{app_RL}

This Appendix provides a brief review about the formulation of RL.
Note that the notation used here is independent of that in the main text. 
In the following, we denote the state space $\mathcal{S}$ and the action space $\mathcal{A}$, respectively. 

The most fundamental assumption in RL is to utilize the Markov Decision Processes (MDPs) for solving the problem.
MDPs are state transitions in which the state $s_{t}$ and reward $r_{t}$ at time $t$ are completely determined by the state $s_{t-1}$ and action $a_{t-1}$ at the previous time $t-1$. 

Considering a specific policy $\pi \left(\mathfrak{a}|s\right)$ under the MDPs, the total reward $R_{t}$, the state-value function $V\left(s\right)$ and the action-value function $Q\left(s,\mathfrak{a}\right)$ are defined as follows: 
\begin{align}
    R_{t} &= \sum_{t'=0}^{\infty} \gamma^{t'} r_{t+t'}, \\
    V^{\pi}\left(s\right) &= \mathbb{E} \left[ R_{t} ; s_{t}=s \right], \label{eq:def_Vpi} \\
    Q^{\pi}\left(s,\mathfrak{a}\right) &= \mathbb{E} \left[ R_{t} ; s_{t}=s, \mathfrak{a}_{t}=\mathfrak{a} \right], \label{eq:def_Qpi}
\end{align}
where $\mathbb{E}$ means an expected value. 
Note that the discount rate $\gamma$ is a real number satisfying $0<\gamma<1$, and it weights the rewards at each time to prevent divergence of the values. 
These equations define the value functions as the expected value of the total reward that could be obtained at times after $t$. 
The relation between $V\left(s\right)$ and $Q\left(s,\mathfrak{a}\right)$ is as follows:
\begin{align}
    V^{\pi}\left(s\right) = \sum_{\mathfrak{a}\in \mathcal{A}} \pi \left(\mathfrak{a}|s\right) Q^{\pi}\left(s,\mathfrak{a}\right).
\end{align}

The goal of RL is to determine an optimal policy $\pi^{*}$ that maximizes the expected value of $R_{t}$. 
This is equivalent to deriving $\pi^{*}$ such that $V\left(s\right)$ and $Q\left(s,\mathfrak{a}\right)$ are maximized. 
To solve this optimization problem, we first rewrite Eq.\eqref{eq:def_Vpi} (the definition of $V\left(s\right)$) into a recursive expression as follows: 
\begin{align}
    V^{\pi}\left(s\right) &= \mathbb{E} \left[ r_{t} + \gamma \sum_{t'=1}^{\infty} \gamma^{t'-1} r_{t+t'} ; s_{t}=s \right] \\
    &= \sum_{\mathfrak{a}\in\mathcal{A}} \pi \left(\mathfrak{a}|s\right) \sum_{s'\in\mathcal{S}} p\left(s'|s,\mathfrak{a}\right) \left[ r_{t} + \gamma  V^{\pi}\left(s'\right) \right].
\end{align}
Here, $p\left(s'|s,\mathfrak{a}\right)$ is the probability distribution for ``choosing the action $\mathfrak{a}_{t}$ in state $s_{t}$, and transition to state $s_{t+1}$'', which completely characterizes the time evolution of the environment under MDPs. 
The similar deformation for Eq.\eqref{eq:def_Qpi} (the definition of $Q\left(s,\mathfrak{a}\right)$) leads to the following equation: 
\begin{align}
    Q^{\pi}\left(s,\mathfrak{a}\right) = \sum_{s'\in\mathcal{S}} p\left(s'|s,\mathfrak{a}\right) \left[ r_{t} + \gamma \sum_{\mathfrak{a'}\in\mathcal{A}} \pi\left(\mathfrak{a'}|s'\right) Q^{\pi}\left(s',\mathfrak{a'}\right) \right].
\end{align}
These equations represent the expected reward obtained under a specific policy $\pi$. 
On the other hand, the probability of choosing an action is deterministic if the agent has acquired the optimal policy $\pi^{*}$. 
That is, the probability of taking a particular action is 1, and 0 otherwise. 
This makes the optimal solutions $\{V^{*},Q^{*}\}$ satisfy the following equations: 
\begin{align}
    V^{*}\left(s\right) &= \max_{\mathfrak{a}\in\mathcal{A}} \sum_{s'\in\mathcal{S}} p\left(s'|s,\mathfrak{a}\right) \left[ r_{t} + \gamma  V^{*}\left(s'\right) \right], \\
    Q^{*}\left(s,\mathfrak{a}\right) &= \sum_{s'\in\mathcal{S}} p\left(s'|s,\mathfrak{a}\right) \left[ r_{t} + \gamma \max_{\mathfrak{a'}\in\mathcal{A}} Q^{*}\left(s',\mathfrak{a'}\right) \right]. \label{eq:Bellman_Q}
\end{align}
These recursive equations are called as Bellman equations. 
When the state transition probability $p\left(s'|s,\mathfrak{a}\right)$ is known, the equations can be solved by the sequential assignment method. 
In reality, however, complete information on $p\left(s'|s,\mathfrak{a}\right)$ is rarely known. 
Furthermore, even if the state-value $V$ is calculated without such information, the expected value of rewards from a particular action cannot be calculated, so an appropriate action cannot be determined. 
Therefore, in many realistic problems, the action-value $Q$ is approximated by some mathematical models such as a neural network, and parameters are determined so that $Q$ is maximized. 
This method is called as the value-based approach\footnote{The method of approximating the policy $\pi$ itself with a mathematical model is called as the policy-based approach, which directly seeks parameters that maximize the expected value of the total reward $R_{t}$.}.

Rewriting Eq.\eqref{eq:Bellman_Q} based on the sequential assignment method, we obtain the following expression where $\alpha$ is the learning rate:
\begin{align}
    Q\left(s,\mathfrak{a}\right) \leftarrow Q\left(s,\mathfrak{a}\right) + \alpha \left[ r\left(s,\mathfrak{a}\right) + \gamma \sum_{s'\in\mathcal{S}} p\left(s'|s,\mathfrak{a}\right) \max_{\mathfrak{a'}\in\mathcal{A}} Q\left(s',\mathfrak{a'}\right) - Q\left(s,\mathfrak{a}\right) \right].
\end{align}
The expected reward $r\left(s,\mathfrak{a}\right)$ is defined as: 
\begin{align}
    r\left(s,\mathfrak{a}\right) = \sum_{s'\in\mathcal{S}} p\left(s'|s,\mathfrak{a}\right)r_{t}.
\end{align}
When the concrete expression of $p\left(s'|s,\mathfrak{a}\right)$ is not known, first, the action $\mathfrak{a}$ is selected based on the currently expected action-value. 
Next, the maximized action-value is achieved by minimizing the difference from the actually obtained action-value. 
In this process, $Q\left(s_{t},\mathfrak{a}_{t}\right)$ is updated as follows, and this method of learning is called as Q learning:
\begin{align}
    Q\left(s_{t},\mathfrak{a}_{t}\right) \leftarrow Q\left(s_{t},\mathfrak{a}_{t}\right) + \alpha \left[ r_{t} + \gamma \max_{\mathfrak{a}\in\mathcal{A}} Q\left(s_{t+1},\mathfrak{a}\right) - Q\left(s_{t},\mathfrak{a}_{t}\right) \right].
\end{align}
In DQN, the term $\max_{\mathfrak{a}\in\mathcal{A}} Q\left(s_{t+1},\mathfrak{a}\right)$ is approximated by the neural network. 
The action-value is normalized by using a softmax function in the final layer of the neural network. 
Then, the output value can also be interpreted as probabilities. 
This allows the output of the neural network to be interpreted as a probability indicating which action is most plausible. 
In addition, DQN uses Q network (with parameters $\Theta$) and target network (with parameters $\Theta'$) to update the action-value as follows to improve the stability of learning:
\begin{align}
    Q_{\Theta}\left(s_{t},\mathfrak{a}_{t}\right) \leftarrow Q_{\Theta}\left(s_{t},\mathfrak{a}_{t}\right) + \alpha \left[ r_{t} + \gamma \max_{\mathfrak{a}\in\mathcal{A}} Q_{\Theta'}\left(s_{t+1},\mathfrak{a}\right) - Q_{\Theta}\left(s_{t},\mathfrak{a}_{t}\right) \right].
\end{align}

\section{FN charges}
\label{app}

We list our finding charge assignment of quarks in Appendix \ref{app:quarks}. 
For the lepton sector, we present the results of RL by picking the models up only when theoretical values of neutrinos with the normal ordering are within $3\sigma$ considering $(\Delta m_{21}^2, \Delta m_{31}^2, \sin^2 \theta_{12}, \sin^2 \theta_{13}, \sin^2 \theta_{23})$. 
These are summarized in Appendices \ref{app:lepton_NO_fixed} and \ref{app:lepton_unfixed}, where the neutrino mass ordering is specified and unspecified in the learning of neural networks, respectively.

\subsection{Quark sector}
\label{app:quarks}

\begin{table}[H]
\label{table_quark_list_1}
\centering
\begin{tabular}{l|c}
    \hline
    Charges &
    ${\cal Q}=\left(\begin{array}{ccc|ccc|ccc|cc}
        Q_{1} & Q_{2} & Q_{3} & u_{1} & u_{2} & u_{3} & d_{1} & d_{2} & d_{3} & H & \phi \\ 
        \hline
        0 & -2 & 3 & 1 & 7 & 1 & 8 & 9 & 7 & -2 & -1 \\ 
    \end{array}\right)$ \\ 
    \hline
    $\mathcal{O}\left(1\right)$ coeff. & $y^{u} \simeq
    \left(\begin{array}{rrr}
        0.856 & 1.537 & 0.895 \\ 
        -1.071 & 0.833 & -1.377 \\ 
        1.181 & -1.507 & 0.805
    \end{array}\right) \ ,\ 
    y^{d} \simeq
    \left(\begin{array}{rrr}
        -0.915 & 1.347 & -0.746 \\ 
        -0.559 & 1.108 & 0.884 \\ 
        0.898 & 1.056 & 1.111
    \end{array}\right)$ \\ 
    \hline
    VEV, Value & $v_{\phi}\simeq0.294\cdot e^{-0.930i}\ ,\ \mathcal{V}_{\mathrm{opt}}\simeq-1.834$ \\ 
    \hline
    \hline
    Charges &
    ${\cal Q}=\left(\begin{array}{ccc|ccc|ccc|cc}
        Q_{1} & Q_{2} & Q_{3} & u_{1} & u_{2} & u_{3} & d_{1} & d_{2} & d_{3} & H & \phi \\ 
        \hline
        -7 & -9 & -4 & -6 & -5 & -7 & 3 & 4 & 2 & -3 & -1 \\ 
    \end{array}\right)$ \\ 
    \hline
    $\mathcal{O}\left(1\right)$ coeff. & $y^{u} \simeq
    \left(\begin{array}{rrr}
        -1.001 & -0.857 & 0.989 \\ 
        0.791 & 0.868 & 0.889 \\ 
        -0.889 & -0.885 & -0.940
    \end{array}\right) \ ,\ 
    y^{d} \simeq
    \left(\begin{array}{rrr}
        -0.995 & 0.870 & -0.985 \\ 
        1.199 & -1.127 & -1.412 \\ 
        -1.325 & -0.882 & -1.020
    \end{array}\right)$ \\ 
    \hline
    VEV, Value & $v_{\phi}\simeq0.283\cdot e^{-0.860i}\ ,\ \mathcal{V}_{\mathrm{opt}}\simeq-1.416$ \\ 
    \hline
    \hline
    Charges &
    ${\cal Q}=\left(\begin{array}{ccc|ccc|ccc|cc}
        Q_{1} & Q_{2} & Q_{3} & u_{1} & u_{2} & u_{3} & d_{1} & d_{2} & d_{3} & H & \phi \\ 
        \hline
        -6 & -5 & -3 & 5 & 4 & 2 & -5 & -7 & -5 & 5 & -1 \\ 
    \end{array}\right)$ \\ 
    \hline
    $\mathcal{O}\left(1\right)$ coeff. & $y^{u} \simeq
    \left(\begin{array}{rrr}
        -1.451 & 1.161 & -0.722 \\ 
        -1.475 & 0.953 & -1.262 \\ 
        0.798 & 0.865 & 1.182
    \end{array}\right) \ ,\ 
    y^{d} \simeq
    \left(\begin{array}{rrr}
        1.229 & -0.976 & 0.756 \\ 
        -1.280 & -0.985 & -1.159 \\ 
        1.235 & 0.355 & 0.825
    \end{array}\right)$ \\ 
    \hline
    VEV, Value & $v_{\phi}\simeq0.171\cdot e^{0.649i}\ ,\ \mathcal{V}_{\mathrm{opt}}\simeq-2.066$ \\ 
    \hline
    \hline
    Charges &
    ${\cal Q}=\left(\begin{array}{ccc|ccc|ccc|cc}
        Q_{1} & Q_{2} & Q_{3} & u_{1} & u_{2} & u_{3} & d_{1} & d_{2} & d_{3} & H & \phi \\ 
        \hline
        -6 & -8 & -3 & 3 & 7 & 1 & -5 & -3 & -5 & 4 & -1 \\ 
    \end{array}\right)$ \\ 
    \hline
    $\mathcal{O}\left(1\right)$ coeff. & $y^{u} \simeq
    \left(\begin{array}{rrr}
        1.498 & 1.767 & -1.223 \\ 
        1.090 & -1.531 & -1.080 \\ 
        -1.035 & -1.383 & -0.864
    \end{array}\right) \ ,\ 
    y^{d} \simeq
    \left(\begin{array}{rrr}
        0.778 & 1.466 & 0.921 \\ 
        -0.945 & -0.858 & -1.249 \\ 
        -1.077 & 0.555 & -1.006
    \end{array}\right)$ \\ 
    \hline
    VEV, Value & $v_{\phi}\simeq0.300\cdot e^{0.912i}\ ,\ \mathcal{V}_{\mathrm{opt}}\simeq-1.881$ \\ 
    \hline
    \hline
    Charges &
    ${\cal Q}=\left(\begin{array}{ccc|ccc|ccc|cc}
        Q_{1} & Q_{2} & Q_{3} & u_{1} & u_{2} & u_{3} & d_{1} & d_{2} & d_{3} & H & \phi \\ 
        \hline
        -3 & -2 & 0 & 6 & 3 & 2 & 1 & -1 & 0 & 2 & -1 \\ 
    \end{array}\right)$ \\ 
    \hline
    $\mathcal{O}\left(1\right)$ coeff. & $y^{u} \simeq
    \left(\begin{array}{rrr}
        -0.791 & 1.769 & 1.053 \\ 
        1.177 & 1.045 & 0.807 \\ 
        1.280 & 0.924 & -0.984
    \end{array}\right) \ ,\ 
    y^{d} \simeq
    \left(\begin{array}{rrr}
        1.112 & -0.976 & 0.964 \\ 
        -1.332 & -0.817 & 1.121 \\ 
        1.327 & -0.774 & 0.910
    \end{array}\right)$ \\ 
    \hline
    VEV, Value & $v_{\phi}\simeq0.166\cdot e^{-2.199i}\ ,\ \mathcal{V}_{\mathrm{opt}}\simeq-1.584$ \\ 
    \hline
\end{tabular}
\end{table}

\newpage
\vspace*{\stretch{1}}

\begin{table}[H]
\label{table_quark_list_2}
\centering
\begin{tabular}{l|c}
    \hline
    Charges &
    ${\cal Q}=\left(\begin{array}{ccc|ccc|ccc|cc}
        Q_{1} & Q_{2} & Q_{3} & u_{1} & u_{2} & u_{3} & d_{1} & d_{2} & d_{3} & H & \phi \\ 
        \hline
        -5 & -4 & -2 & 3 & 2 & 0 & -1 & -2 & -1 & 2 & -1 \\ 
    \end{array}\right)$ \\ 
    \hline
    $\mathcal{O}\left(1\right)$ coeff. & $y^{u} \simeq
    \left(\begin{array}{rrr}
        -0.812 & -1.382 & 1.015 \\ 
        0.949 & 0.924 & -1.598 \\ 
        -1.319 & -1.511 & -0.935
    \end{array}\right) \ ,\ 
    y^{d} \simeq
    \left(\begin{array}{rrr}
        -1.129 & -0.680 & -0.789 \\ 
        0.505 & -0.857 & -0.994 \\ 
        -1.033 & -1.134 & 1.287
    \end{array}\right)$ \\ 
    \hline
    VEV, Value & $v_{\phi}\simeq0.172\cdot e^{1.985i}\ ,\ \mathcal{V}_{\mathrm{opt}}\simeq-1.240$ \\ 
    \hline
    \hline
    Charges &
    ${\cal Q}=\left(\begin{array}{ccc|ccc|ccc|cc}
        Q_{1} & Q_{2} & Q_{3} & u_{1} & u_{2} & u_{3} & d_{1} & d_{2} & d_{3} & H & \phi \\ 
        \hline
        3 & 5 & 0 & 1 & -3 & 1 & -4 & -5 & -3 & 1 & 1 \\ 
    \end{array}\right)$ \\ 
    \hline
    $\mathcal{O}\left(1\right)$ coeff. & $y^{u} \simeq
    \left(\begin{array}{rrr}
        1.030 & -0.155 & 1.152 \\ 
        -0.934 & 1.382 & -0.996 \\ 
        -1.349 & 1.185 & -1.036
    \end{array}\right) \ ,\ 
    y^{d} \simeq
    \left(\begin{array}{rrr}
        0.323 & -1.270 & 1.333 \\ 
        -0.945 & 1.084 & 0.987 \\ 
        0.925 & -1.042 & 0.852
    \end{array}\right)$ \\ 
    \hline
    VEV, Value & $v_{\phi}\simeq0.289\cdot e^{-1.223i}\ ,\ \mathcal{V}_{\mathrm{opt}}\simeq-1.405$ \\ 
    \hline
    \hline
    Charges &
    ${\cal Q}=\left(\begin{array}{ccc|ccc|ccc|cc}
        Q_{1} & Q_{2} & Q_{3} & u_{1} & u_{2} & u_{3} & d_{1} & d_{2} & d_{3} & H & \phi \\ 
        \hline
        1 & 0 & -2 & 1 & -2 & 2 & -8 & -8 & -8 & 4 & 1 \\ 
    \end{array}\right)$ \\ 
    \hline
    $\mathcal{O}\left(1\right)$ coeff. & $y^{u} \simeq
    \left(\begin{array}{rrr}
        0.999 & 0.909 & -1.021 \\ 
        -0.385 & 1.263 & 0.820 \\ 
        -1.117 & 0.825 & -0.940
    \end{array}\right) \ ,\ 
    y^{d} \simeq
    \left(\begin{array}{rrr}
        -1.019 & -0.966 & -1.312 \\ 
        1.064 & 0.979 & 0.800 \\ 
        1.052 & 0.824 & 1.242
    \end{array}\right)$ \\ 
    \hline
    VEV, Value & $v_{\phi}\simeq0.173\cdot e^{1.916i}\ ,\ \mathcal{V}_{\mathrm{opt}}\simeq-1.158$ \\ 
    \hline
    \hline
    Charges &
    ${\cal Q}=\left(\begin{array}{ccc|ccc|ccc|cc}
        Q_{1} & Q_{2} & Q_{3} & u_{1} & u_{2} & u_{3} & d_{1} & d_{2} & d_{3} & H & \phi \\ 
        \hline
        -4 & -6 & -1 & 5 & 9 & 3 & -2 & -1 & -3 & 4 & -1 \\ 
    \end{array}\right)$ \\ 
    \hline
    $\mathcal{O}\left(1\right)$ coeff. & $y^{u} \simeq
    \left(\begin{array}{rrr}
        1.222 & -1.247 & 1.028 \\ 
        -1.169 & -1.066 & -1.216 \\ 
        1.172 & 0.916 & -1.204
    \end{array}\right) \ ,\ 
    y^{d} \simeq
    \left(\begin{array}{rrr}
        -1.021 & 0.900 & 0.793 \\ 
        0.804 & 1.260 & 0.375 \\ 
        -1.029 & 0.973 & 0.401
    \end{array}\right)$ \\ 
    \hline
    VEV, Value & $v_{\phi}\simeq0.294\cdot e^{-2.999i}\ ,\ \mathcal{V}_{\mathrm{opt}}\simeq-1.676$ \\ 
    \hline
\end{tabular}
\end{table}

\vspace{\stretch{1}}
\newpage
\vspace*{\stretch{1}}

\begin{table}[H]
\label{table_quark_list_3}
\centering
\begin{tabular}{l|c}
    \hline
    Charges &
    ${\cal Q}=\left(\begin{array}{ccc|ccc|ccc|cc}
        Q_{1} & Q_{2} & Q_{3} & u_{1} & u_{2} & u_{3} & d_{1} & d_{2} & d_{3} & H & \phi \\ 
        \hline
        -3 & -2 & 0 & 1 & 0 & -2 & 3 & 6 & 6 & -2 & -1 \\ 
    \end{array}\right)$ \\ 
    \hline
    $\mathcal{O}\left(1\right)$ coeff. & $y^{u} \simeq
    \left(\begin{array}{rrr}
        -1.127 & -0.701 & 0.878 \\ 
        -1.028 & -1.237 & -1.274 \\ 
        0.841 & 0.807 & -0.576
    \end{array}\right) \ ,\ 
    y^{d} \simeq
    \left(\begin{array}{rrr}
        -1.081 & -0.886 & -1.166 \\ 
        0.536 & 1.347 & -1.446 \\ 
        0.610 & -1.644 & -0.976
    \end{array}\right)$ \\ 
    \hline
    VEV, Value & $v_{\phi}\simeq0.184\cdot e^{1.097i}\ ,\ \mathcal{V}_{\mathrm{opt}}\simeq-2.452$ \\ 
    \hline
    \hline
    Charges &
    ${\cal Q}=\left(\begin{array}{ccc|ccc|ccc|cc}
        Q_{1} & Q_{2} & Q_{3} & u_{1} & u_{2} & u_{3} & d_{1} & d_{2} & d_{3} & H & \phi \\ 
        \hline
        -7 & -7 & -4 & 6 & 9 & 4 & -9 & -9 & -8 & 8 & -1 \\ 
    \end{array}\right)$ \\ 
    \hline
    $\mathcal{O}\left(1\right)$ coeff. & $y^{u} \simeq
    \left(\begin{array}{rrr}
        1.117 & 1.057 & -0.942 \\ 
        -1.244 & -0.815 & 1.181 \\ 
        -1.097 & 0.892 & -0.963
    \end{array}\right) \ ,\ 
    y^{d} \simeq
    \left(\begin{array}{rrr}
        -1.027 & -0.966 & -1.200 \\ 
        1.054 & 0.364 & 0.947 \\ 
        0.949 & -1.305 & 1.355
    \end{array}\right)$ \\ 
    \hline
    VEV, Value & $v_{\phi}\simeq0.268\cdot e^{1.800i}\ ,\ \mathcal{V}_{\mathrm{opt}}\simeq-1.056$ \\ 
    \hline
    \hline
    Charges &
    ${\cal Q}=\left(\begin{array}{ccc|ccc|ccc|cc}
        Q_{1} & Q_{2} & Q_{3} & u_{1} & u_{2} & u_{3} & d_{1} & d_{2} & d_{3} & H & \phi \\ 
        \hline
        9 & 8 & 6 & 1 & 3 & 4 & 6 & 5 & 5 & -2 & 1 \\ 
    \end{array}\right)$ \\ 
    \hline
    $\mathcal{O}\left(1\right)$ coeff. & $y^{u} \simeq
    \left(\begin{array}{rrr}
        0.430 & 0.806 & -1.220 \\ 
        -0.962 & -0.598 & -0.747 \\ 
        -1.328 & -1.172 & 1.018
    \end{array}\right) \ ,\ 
    y^{d} \simeq
    \left(\begin{array}{rrr}
        -0.996 & -0.747 & -1.068 \\ 
        0.958 & 1.441 & 1.033 \\ 
        0.765 & -0.500 & -1.029
    \end{array}\right)$ \\ 
    \hline
    VEV, Value & $v_{\phi}\simeq0.181\cdot e^{-0.863i}\ ,\ \mathcal{V}_{\mathrm{opt}}\simeq-0.701$ \\ 
    \hline
    \hline
    Charges &
    ${\cal Q}=\left(\begin{array}{ccc|ccc|ccc|cc}
        Q_{1} & Q_{2} & Q_{3} & u_{1} & u_{2} & u_{3} & d_{1} & d_{2} & d_{3} & H & \phi \\ 
        \hline
        -8 & -7 & -5 & 0 & -2 & -3 & -5 & -3 & -4 & 2 & -1 \\ 
    \end{array}\right)$ \\ 
    \hline
    $\mathcal{O}\left(1\right)$ coeff. & $y^{u} \simeq
    \left(\begin{array}{rrr}
        1.040 & 1.098 & -1.452 \\ 
        -1.006 & -0.864 & -0.603 \\ 
        -1.176 & -0.977 & 1.008
    \end{array}\right) \ ,\ 
    y^{d} \simeq
    \left(\begin{array}{rrr}
        -1.403 & 1.057 & -0.865 \\ 
        1.033 & 1.000 & 0.913 \\ 
        0.824 & 0.971 & -1.104
    \end{array}\right)$ \\ 
    \hline
    VEV, Value & $v_{\phi}\simeq0.185\cdot e^{-0.236i}\ ,\ \mathcal{V}_{\mathrm{opt}}\simeq-1.060$ \\ 
    \hline
\end{tabular}
\end{table}

\vspace{\stretch{1}}
\newpage
\vspace*{\stretch{1}}

\begin{table}[H]
\label{table_quark_list_5}
\centering
\begin{tabular}{l|c}
    \hline
    Charges &
    ${\cal Q}=\left(\begin{array}{ccc|ccc|ccc|cc}
        Q_{1} & Q_{2} & Q_{3} & u_{1} & u_{2} & u_{3} & d_{1} & d_{2} & d_{3} & H & \phi \\ 
        \hline
        -9 & -7 & -5 & 5 & 1 & 0 & -7 & -5 & -7 & 5 & -1 \\ 
    \end{array}\right)$ \\ 
    \hline
    $\mathcal{O}\left(1\right)$ coeff. & $y^{u} \simeq
    \left(\begin{array}{rrr}
        -1.186 & -1.082 & 1.178 \\ 
        -1.059 & 0.830 & -0.657 \\ 
        -0.826 & -1.086 & 1.490
    \end{array}\right) \ ,\ 
    y^{d} \simeq
    \left(\begin{array}{rrr}
        -1.125 & 1.069 & 0.827 \\ 
        0.881 & -1.207 & 0.748 \\ 
        -1.128 & 0.701 & -0.512
    \end{array}\right)$ \\ 
    \hline
    VEV, Value & $v_{\phi}\simeq0.288\cdot e^{-1.881i}\ ,\ \mathcal{V}_{\mathrm{opt}}\simeq-1.294$ \\ 
    \hline
    \hline
    Charges &
    ${\cal Q}=\left(\begin{array}{ccc|ccc|ccc|cc}
        Q_{1} & Q_{2} & Q_{3} & u_{1} & u_{2} & u_{3} & d_{1} & d_{2} & d_{3} & H & \phi \\ 
        \hline
        -1 & -3 & 2 & 9 & 9 & 7 & 1 & 3 & 0 & 5 & -1 \\ 
    \end{array}\right)$ \\ 
    \hline
    $\mathcal{O}\left(1\right)$ coeff. & $y^{u} \simeq
    \left(\begin{array}{rrr}
        -0.908 & 0.457 & -0.657 \\ 
        1.220 & -0.777 & -0.895 \\ 
        1.553 & -1.077 & -1.022
    \end{array}\right) \ ,\ 
    y^{d} \simeq
    \left(\begin{array}{rrr}
        0.588 & 0.583 & 1.024 \\ 
        -1.275 & -0.942 & 1.162 \\ 
        -1.376 & 1.200 & -0.808
    \end{array}\right)$ \\ 
    \hline
    VEV, Value & $v_{\phi}\simeq0.298\cdot e^{-1.217i}\ ,\ \mathcal{V}_{\mathrm{opt}}\simeq-1.400$ \\ 
    \hline
    \hline
    Charges &
    ${\cal Q}=\left(\begin{array}{ccc|ccc|ccc|cc}
        Q_{1} & Q_{2} & Q_{3} & u_{1} & u_{2} & u_{3} & d_{1} & d_{2} & d_{3} & H & \phi \\ 
        \hline
        -4 & -6 & -1 & 5 & 9 & 3 & -3 & -1 & -3 & 4 & -1 \\ 
    \end{array}\right)$ \\ 
    \hline
    $\mathcal{O}\left(1\right)$ coeff. & $y^{u} \simeq
    \left(\begin{array}{rrr}
        -1.315 & -1.071 & -0.778 \\ 
        -0.857 & 0.827 & -0.989 \\ 
        1.047 & 0.655 & -1.147
    \end{array}\right) \ ,\ 
    y^{d} \simeq
    \left(\begin{array}{rrr}
        -1.095 & 0.651 & -1.004 \\ 
        0.974 & -0.805 & 1.206 \\ 
        1.404 & 0.828 & 1.007
    \end{array}\right)$ \\ 
    \hline
    VEV, Value & $v_{\phi}\simeq0.300\cdot e^{-1.998i}\ ,\ \mathcal{V}_{\mathrm{opt}}\simeq-1.938$ \\ 
    \hline
    \hline
    Charges &
    ${\cal Q}=\left(\begin{array}{ccc|ccc|ccc|cc}
        Q_{1} & Q_{2} & Q_{3} & u_{1} & u_{2} & u_{3} & d_{1} & d_{2} & d_{3} & H & \phi \\ 
        \hline
        4 & 4 & 1 & 0 & -4 & 0 & -2 & -2 & -3 & -1 & 1 \\ 
    \end{array}\right)$ \\ 
    \hline
    $\mathcal{O}\left(1\right)$ coeff. & $y^{u} \simeq
    \left(\begin{array}{rrr}
        0.656 & -0.920 & 1.291 \\ 
        1.032 & -1.123 & 1.159 \\ 
        0.917 & -1.091 & 0.737
    \end{array}\right) \ ,\ 
    y^{d} \simeq
    \left(\begin{array}{rrr}
        -0.998 & -0.948 & -1.265 \\ 
        0.546 & -0.845 & 0.978 \\ 
        1.421 & -1.241 & 0.857
    \end{array}\right)$ \\ 
    \hline
    VEV, Value & $v_{\phi}\simeq0.268\cdot e^{-0.166i}\ ,\ \mathcal{V}_{\mathrm{opt}}\simeq-2.092$ \\ 
    \hline
\end{tabular}
\end{table}

\vspace{\stretch{1}}
\newpage
\vspace*{\stretch{1}}

\begin{table}[H]
\label{table_quark_list_4}
\centering
\begin{tabular}{l|c}
    \hline
    Charges &
    ${\cal Q}=\left(\begin{array}{ccc|ccc|ccc|cc}
        Q_{1} & Q_{2} & Q_{3} & u_{1} & u_{2} & u_{3} & d_{1} & d_{2} & d_{3} & H & \phi \\ 
        \hline
        -6 & -5 & -3 & -5 & -6 & -8 & 5 & 3 & 5 & -5 & -1 \\ 
    \end{array}\right)$ \\ 
    \hline
    $\mathcal{O}\left(1\right)$ coeff. & $y^{u} \simeq
    \left(\begin{array}{rrr}
        0.977 & -0.755 & 0.909 \\ 
        -1.460 & -1.720 & -1.092 \\ 
        0.731 & -0.836 & 1.109
    \end{array}\right) \ ,\ 
    y^{d} \simeq
    \left(\begin{array}{rrr}
        1.290 & -0.798 & -0.694 \\ 
        -0.461 & 1.375 & -1.045 \\ 
        -0.560 & -0.628 & 0.843
    \end{array}\right)$ \\ 
    \hline
    VEV, Value & $v_{\phi}\simeq0.171\cdot e^{-2.525i}\ ,\ \mathcal{V}_{\mathrm{opt}}\simeq-2.236$ \\ 
    \hline
    \hline
    Charges &
    ${\cal Q}=\left(\begin{array}{ccc|ccc|ccc|cc}
        Q_{1} & Q_{2} & Q_{3} & u_{1} & u_{2} & u_{3} & d_{1} & d_{2} & d_{3} & H & \phi \\ 
        \hline
        -5 & -7 & -2 & 6 & 4 & 3 & -3 & -3 & -4 & 5 & -1 \\ 
    \end{array}\right)$ \\ 
    \hline
    $\mathcal{O}\left(1\right)$ coeff. & $y^{u} \simeq
    \left(\begin{array}{rrr}
        -1.137 & -1.003 & -1.111 \\ 
        1.483 & 1.324 & 1.763 \\ 
        -1.162 & 1.346 & -1.035
    \end{array}\right) \ ,\ 
    y^{d} \simeq
    \left(\begin{array}{rrr}
        -0.758 & -0.830 & -1.025 \\ 
        1.349 & -1.234 & -0.979 \\ 
        1.101 & -1.417 & 1.060
    \end{array}\right)$ \\ 
    \hline
    VEV, Value & $v_{\phi}\simeq0.290\cdot e^{-2.680i}\ ,\ \mathcal{V}_{\mathrm{opt}}\simeq-1.077$ \\ 
    \hline
    \hline
    Charges &
    ${\cal Q}=\left(\begin{array}{ccc|ccc|ccc|cc}
        Q_{1} & Q_{2} & Q_{3} & u_{1} & u_{2} & u_{3} & d_{1} & d_{2} & d_{3} & H & \phi \\ 
        \hline
        -4 & -6 & -1 & 2 & 4 & -1 & 3 & 5 & 3 & 0 & -1 \\ 
    \end{array}\right)$ \\ 
    \hline
    $\mathcal{O}\left(1\right)$ coeff. & $y^{u} \simeq
    \left(\begin{array}{rrr}
        -0.672 & -0.788 & 1.315 \\ 
        1.186 & -0.642 & 0.970 \\ 
        1.060 & 0.924 & 0.604
    \end{array}\right) \ ,\ 
    y^{d} \simeq
    \left(\begin{array}{rrr}
        -0.796 & -1.026 & -1.109 \\ 
        1.127 & 1.424 & -1.072 \\ 
        -1.364 & 1.139 & -1.078
    \end{array}\right)$ \\ 
    \hline
    VEV, Value & $v_{\phi}\simeq0.300\cdot e^{2.723i}\ ,\ \mathcal{V}_{\mathrm{opt}}\simeq-2.409$ \\ 
    \hline
    \hline
    Charges &
    ${\cal Q}=\left(\begin{array}{ccc|ccc|ccc|cc}
        Q_{1} & Q_{2} & Q_{3} & u_{1} & u_{2} & u_{3} & d_{1} & d_{2} & d_{3} & H & \phi \\ 
        \hline
        -4 & -6 & -1 & 2 & 5 & -1 & 2 & 5 & 2 & 0 & -1 \\ 
    \end{array}\right)$ \\ 
    \hline
    $\mathcal{O}\left(1\right)$ coeff. & $y^{u} \simeq
    \left(\begin{array}{rrr}
        -1.334 & 1.343 & -0.781 \\ 
        -1.592 & -1.149 & -1.334 \\ 
        -1.565 & -0.969 & 1.063
    \end{array}\right) \ ,\ 
    y^{d} \simeq
    \left(\begin{array}{rrr}
        1.598 & -1.740 & 1.046 \\ 
        -1.041 & -1.034 & 0.984 \\ 
        0.727 & 0.933 & 1.393
    \end{array}\right)$ \\ 
    \hline
    VEV, Value & $v_{\phi}\simeq0.300\cdot e^{2.498i}\ ,\ \mathcal{V}_{\mathrm{opt}}\simeq-2.267$ \\ 
    \hline
\end{tabular}
\end{table}

\vspace{\stretch{1}}
\newpage

\subsection{Lepton sector (RL with NO designated)}
\label{app:lepton_NO_fixed}

\vspace{\stretch{1}}

\begin{table}[H]
\label{table_fixedNO_1}
\centering
\begin{tabular}{l|c}
    \hline
    Charges &
    ${\cal Q}=\left(\begin{array}{ccc|ccc|ccc|cc}
        L_{1} & L_{2} & L_{3} & N_{1} & N_{2} & N_{3} & l_{1} & l_{2} & l_{3} & H & \phi \\ 
        \hline
        -1 & -1 & 0 & 2 & 4 & 1 & 5 & 6 & 4 & 0 & -1 \\ 
    \end{array}\right)$ \\ 
    \hline
    $\mathcal{O}\left(1\right)$ coeff. & $y^{l} \simeq
    \left(\begin{array}{rrr}
        -0.705 & -0.713 & 0.799 \\ 
        -1.439 & -1.472 & 1.516 \\ 
        -0.157 & 0.186 & -2.133
    \end{array}\right) \ ,\ 
    y^{\nu} \simeq
    \left(\begin{array}{rrr}
        0.659 & -0.836 & -1.219 \\ 
        -1.264 & -2.891 & 0.651 \\ 
        -0.465 & -1.062 & 1.138
    \end{array}\right)$ \\ 
    & $y^{N} \simeq
    \left(\begin{array}{rrr}
        2.553 & 1.290 & 1.402 \\ 
        1.290 & 1.224 & -0.902 \\ 
        1.402 & -0.902 & 0.105
    \end{array}\right)$ \\ 
    \hline
    VEV, Value & $v_{\phi}\simeq0.300\cdot e^{2.723i}\ ,\ \mathcal{V}_{\mathrm{opt}}\simeq-0.915$ \\ 
    \hline
    \hline
    Charges &
    ${\cal Q}=\left(\begin{array}{ccc|ccc|ccc|cc}
        L_{1} & L_{2} & L_{3} & N_{1} & N_{2} & N_{3} & l_{1} & l_{2} & l_{3} & H & \phi \\ 
        \hline
        2 & 1 & 2 & -2 & -7 & -5 & -7 & -3 & -3 & -1 & 1 \\ 
    \end{array}\right)$ \\ 
    \hline
    $\mathcal{O}\left(1\right)$ coeff. & $y^{l} \simeq
    \left(\begin{array}{rrr}
        3.386 & -0.205 & -2.696 \\ 
        -0.523 & -1.396 & 3.760 \\ 
        1.375 & -0.561 & -2.044
    \end{array}\right) \ ,\ 
    y^{\nu} \simeq
    \left(\begin{array}{rrr}
        -0.774 & -2.694 & 1.101 \\ 
        -0.949 & -0.905 & -0.432 \\ 
        -2.286 & -0.314 & 1.325
    \end{array}\right)$ \\ 
    & $y^{N} \simeq
    \left(\begin{array}{rrr}
        1.246 & 0.199 & 1.121 \\ 
        0.199 & 1.280 & -0.879 \\ 
        1.121 & -0.879 & -0.214
    \end{array}\right)$ \\ 
    \hline
    VEV, Value & $v_{\phi}\simeq0.268\cdot e^{-0.166i}\ ,\ \mathcal{V}_{\mathrm{opt}}\simeq-0.611$ \\ 
    \hline
    \hline
    Charges &
    ${\cal Q}=\left(\begin{array}{ccc|ccc|ccc|cc}
        L_{1} & L_{2} & L_{3} & N_{1} & N_{2} & N_{3} & l_{1} & l_{2} & l_{3} & H & \phi \\ 
        \hline
        2 & 1 & 2 & -8 & -1 & -9 & -7 & -3 & -3 & -1 & 1 \\ 
    \end{array}\right)$ \\ 
    \hline
    $\mathcal{O}\left(1\right)$ coeff. & $y^{l} \simeq
    \left(\begin{array}{rrr}
        -0.889 & 2.056 & -0.299 \\ 
        -1.584 & -2.697 & 1.542 \\ 
        -0.797 & 0.918 & 1.501
    \end{array}\right) \ ,\ 
    y^{\nu} \simeq
    \left(\begin{array}{rrr}
        1.135 & -1.331 & 0.128 \\ 
        1.207 & -1.203 & -0.051 \\ 
        -0.671 & -2.639 & 0.074
    \end{array}\right)$ \\ 
    & $y^{N} \simeq
    \left(\begin{array}{rrr}
        1.125 & -0.388 & 0.950 \\ 
        -0.388 & 1.066 & -0.349 \\ 
        0.950 & -0.349 & -0.656
    \end{array}\right)$ \\ 
    \hline
    VEV, Value & $v_{\phi}\simeq0.268\cdot e^{-0.166i}\ ,\ \mathcal{V}_{\mathrm{opt}}\simeq-0.853$ \\ 
    \hline
\end{tabular}
\end{table}

\vspace{\stretch{1}}
\newpage
\vspace*{\stretch{1}}

\begin{table}[H]
\label{table_fixedNO_2}
\centering
\begin{tabular}{l|c}
    \hline
    Charges &
    ${\cal Q}=\left(\begin{array}{ccc|ccc|ccc|cc}
        L_{1} & L_{2} & L_{3} & N_{1} & N_{2} & N_{3} & l_{1} & l_{2} & l_{3} & H & \phi \\ 
        \hline
        3 & 3 & 2 & -3 & -4 & -7 & 1 & -2 & 0 & -2 & 1 \\ 
    \end{array}\right)$ \\ 
    \hline
    $\mathcal{O}\left(1\right)$ coeff. & $y^{l} \simeq
    \left(\begin{array}{rrr}
        -1.003 & 0.149 & -1.704 \\ 
        0.287 & -0.165 & 2.051 \\ 
        2.022 & 1.639 & -1.034
    \end{array}\right) \ ,\ 
    y^{\nu} \simeq
    \left(\begin{array}{rrr}
        -0.905 & -1.445 & 1.882 \\ 
        -1.829 & 1.376 & -1.951 \\ 
        0.481 & -0.595 & -1.632
    \end{array}\right)$ \\ 
    & $y^{N} \simeq
    \left(\begin{array}{rrr}
        1.122 & -1.332 & -1.843 \\ 
        -1.332 & -2.152 & -1.901 \\ 
        -1.843 & -1.901 & -2.376
    \end{array}\right)$ \\ 
    \hline
    VEV, Value & $v_{\phi}\simeq0.181\cdot e^{-0.863i}\ ,\ \mathcal{V}_{\mathrm{opt}}\simeq-0.565$ \\ 
    \hline
    \hline
    Charges &
    ${\cal Q}=\left(\begin{array}{ccc|ccc|ccc|cc}
        L_{1} & L_{2} & L_{3} & N_{1} & N_{2} & N_{3} & l_{1} & l_{2} & l_{3} & H & \phi \\ 
        \hline
        2 & 1 & 2 & -1 & -2 & -8 & -6 & -2 & -2 & -1 & 1 \\ 
    \end{array}\right)$ \\ 
    \hline
    $\mathcal{O}\left(1\right)$ coeff. & $y^{l} \simeq
    \left(\begin{array}{rrr}
        -0.509 & 0.762 & 0.382 \\ 
        1.310 & -1.751 & -0.428 \\ 
        -0.301 & 1.804 & 0.823
    \end{array}\right) \ ,\ 
    y^{\nu} \simeq
    \left(\begin{array}{rrr}
        -0.693 & 1.295 & -1.990 \\ 
        1.573 & 1.228 & 1.282 \\ 
        -1.986 & 1.580 & 0.391
    \end{array}\right)$ \\ 
    & $y^{N} \simeq
    \left(\begin{array}{rrr}
        1.377 & 1.972 & -1.113 \\ 
        1.972 & 1.472 & 0.982 \\ 
        -1.113 & 0.982 & -1.331
    \end{array}\right)$ \\ 
    \hline
    VEV, Value & $v_{\phi}\simeq0.268\cdot e^{-0.166i}\ ,\ \mathcal{V}_{\mathrm{opt}}\simeq-0.457$ \\ 
    \hline
    \hline
    Charges &
    ${\cal Q}=\left(\begin{array}{ccc|ccc|ccc|cc}
        L_{1} & L_{2} & L_{3} & N_{1} & N_{2} & N_{3} & l_{1} & l_{2} & l_{3} & H & \phi \\ 
        \hline
        3 & 3 & 2 & -9 & -6 & 0 & 1 & -3 & 0 & -2 & 1 \\ 
    \end{array}\right)$ \\ 
    \hline
    $\mathcal{O}\left(1\right)$ coeff. & $y^{l} \simeq
    \left(\begin{array}{rrr}
        -3.207 & 0.771 & -1.370 \\ 
        -1.542 & -0.925 & 1.347 \\ 
        1.859 & -1.534 & -1.198
    \end{array}\right) \ ,\ 
    y^{\nu} \simeq
    \left(\begin{array}{rrr}
        2.273 & 0.779 & -0.955 \\ 
        0.936 & 1.809 & 2.821 \\ 
        0.714 & -0.719 & -1.239
    \end{array}\right)$ \\ 
    & $y^{N} \simeq
    \left(\begin{array}{rrr}
        -1.312 & 0.729 & 0.441 \\ 
        0.729 & 0.913 & 1.028 \\ 
        0.441 & 1.028 & 0.524
    \end{array}\right)$ \\ 
    \hline
    VEV, Value & $v_{\phi}\simeq0.181\cdot e^{-0.863i}\ ,\ \mathcal{V}_{\mathrm{opt}}\simeq-0.529$ \\ 
    \hline
\end{tabular}
\end{table}

\vspace{\stretch{1}}
\newpage

\subsection{Lepton sector (RL without specifying the neutrino mass ordering)}
\label{app:lepton_unfixed}

\vspace{\stretch{1}}

\begin{table}[H]
\label{table_unfixedNO_1}
\centering
\begin{tabular}{l|c}
    \hline
    Charges &
    ${\cal Q}=\left(\begin{array}{ccc|ccc|ccc|cc}
        L_{1} & L_{2} & L_{3} & N_{1} & N_{2} & N_{3} & l_{1} & l_{2} & l_{3} & H & \phi \\ 
        \hline
        -6 & -5 & -7 & 5 & 4 & 2 & -4 & -6 & -8 & 5 & -1 \\ 
    \end{array}\right)$ \\ 
    \hline
    $\mathcal{O}\left(1\right)$ coeff. & $y^{l} \simeq
    \left(\begin{array}{rrr}
        -0.688 & -1.090 & -1.149 \\ 
        0.459 & -1.353 & -0.229 \\ 
        1.044 & -0.597 & -3.286
    \end{array}\right) \ ,\ 
    y^{\nu} \simeq
    \left(\begin{array}{rrr}
        -1.270 & -1.387 & 3.625 \\ 
        -0.230 & 1.512 & 0.826 \\ 
        -1.327 & 0.590 & -1.473
    \end{array}\right)$ \\ 
    & $y^{N} \simeq
    \left(\begin{array}{rrr}
        -1.328 & -1.209 & -0.765 \\ 
        -1.209 & 0.714 & 0.571 \\ 
        -0.765 & 0.571 & 2.154
    \end{array}\right)$ \\ 
    \hline
    VEV, Value & $v_{\phi}\simeq0.171\cdot e^{0.649i}\ ,\ \mathcal{V}_{\mathrm{opt}}\simeq-0.559$ \\ 
    \hline
    \hline
    Charges &
    ${\cal Q}=\left(\begin{array}{ccc|ccc|ccc|cc}
        L_{1} & L_{2} & L_{3} & N_{1} & N_{2} & N_{3} & l_{1} & l_{2} & l_{3} & H & \phi \\ 
        \hline
        -5 & -5 & -4 & 0 & 9 & 7 & 2 & -5 & -1 & 4 & -1 \\ 
    \end{array}\right)$ \\ 
    \hline
    $\mathcal{O}\left(1\right)$ coeff. & $y^{l} \simeq
    \left(\begin{array}{rrr}
        -1.146 & 0.729 & -0.022 \\ 
        0.954 & 1.968 & -1.317 \\ 
        -1.070 & 0.476 & -1.263
    \end{array}\right) \ ,\ 
    y^{\nu} \simeq
    \left(\begin{array}{rrr}
        -1.236 & 0.755 & 0.613 \\ 
        1.308 & 0.545 & -1.086 \\ 
        0.567 & -1.310 & -1.059
    \end{array}\right)$ \\ 
    & $y^{N} \simeq
    \left(\begin{array}{rrr}
        0.741 & 1.092 & 0.703 \\ 
        1.092 & -0.498 & -0.999 \\ 
        0.703 & -0.999 & -0.438
    \end{array}\right)$ \\ 
    \hline
    VEV, Value & $v_{\phi}\simeq0.300\cdot e^{-1.998i}\ ,\ \mathcal{V}_{\mathrm{opt}}\simeq-1.349$ \\ 
    \hline
    \hline
    Charges &
    ${\cal Q}=\left(\begin{array}{ccc|ccc|ccc|cc}
        L_{1} & L_{2} & L_{3} & N_{1} & N_{2} & N_{3} & l_{1} & l_{2} & l_{3} & H & \phi \\ 
        \hline
        -5 & -6 & -4 & 9 & 6 & 0 & -3 & 0 & -2 & 4 & -1 \\ 
    \end{array}\right)$ \\ 
    \hline
    $\mathcal{O}\left(1\right)$ coeff. & $y^{l} \simeq
    \left(\begin{array}{rrr}
        -2.733 & -0.172 & 2.087 \\ 
        -0.443 & -0.578 & 0.215 \\ 
        1.717 & -0.553 & 0.961
    \end{array}\right) \ ,\ 
    y^{\nu} \simeq
    \left(\begin{array}{rrr}
        0.027 & -0.994 & 1.790 \\ 
        1.267 & 1.215 & 1.901 \\ 
        -1.875 & -1.609 & 1.474
    \end{array}\right)$ \\ 
    & $y^{N} \simeq
    \left(\begin{array}{rrr}
        1.318 & -1.208 & 1.567 \\ 
        -1.208 & 0.724 & 0.904 \\ 
        1.567 & 0.904 & -1.328
    \end{array}\right)$ \\ 
    \hline
    VEV, Value & $v_{\phi}\simeq0.294\cdot e^{-2.999i}\ ,\ \mathcal{V}_{\mathrm{opt}}\simeq-0.829$ \\ 
    \hline
\end{tabular}
\end{table}

\vspace{\stretch{1}}
\newpage
\vspace*{\stretch{1}}

\begin{table}[H]
\label{table_unfixedNO_2}
\centering
\begin{tabular}{l|c}
    \hline
    Charges &
    ${\cal Q}=\left(\begin{array}{ccc|ccc|ccc|cc}
        L_{1} & L_{2} & L_{3} & N_{1} & N_{2} & N_{3} & l_{1} & l_{2} & l_{3} & H & \phi \\ 
        \hline
        -2 & 1 & 0 & 1 & 4 & 5 & 8 & 5 & 4 & 0 & -1 \\ 
    \end{array}\right)$ \\ 
    \hline
    $\mathcal{O}\left(1\right)$ coeff. & $y^{l} \simeq
    \left(\begin{array}{rrr}
        -0.355 & 1.038 & 1.347 \\ 
        -0.897 & -0.474 & -1.544 \\ 
        -1.169 & 1.683 & 1.084
    \end{array}\right) \ ,\ 
    y^{\nu} \simeq
    \left(\begin{array}{rrr}
        -0.745 & -1.062 & -1.450 \\ 
        -0.600 & 0.622 & 0.681 \\ 
        0.783 & 0.617 & -1.079
    \end{array}\right)$ \\ 
    & $y^{N} \simeq
    \left(\begin{array}{rrr}
        1.647 & 0.446 & -0.805 \\ 
        0.446 & 0.671 & -1.362 \\ 
        -0.805 & -1.362 & -0.538
    \end{array}\right)$ \\ 
    \hline
    VEV, Value & $v_{\phi}\simeq0.300\cdot e^{2.723i}\ ,\ \mathcal{V}_{\mathrm{opt}}\simeq-1.833$ \\ 
    \hline
    \hline
    Charges &
    ${\cal Q}=\left(\begin{array}{ccc|ccc|ccc|cc}
        L_{1} & L_{2} & L_{3} & N_{1} & N_{2} & N_{3} & l_{1} & l_{2} & l_{3} & H & \phi \\ 
        \hline
        -1 & -2 & 0 & 2 & 6 & 5 & 5 & 2 & 8 & 0 & -1 \\ 
    \end{array}\right)$ \\ 
    \hline
    $\mathcal{O}\left(1\right)$ coeff. & $y^{l} \simeq
    \left(\begin{array}{rrr}
        1.624 & -0.465 & -0.270 \\ 
        0.966 & -0.539 & 1.162 \\ 
        1.110 & -0.184 & -1.372
    \end{array}\right) \ ,\ 
    y^{\nu} \simeq
    \left(\begin{array}{rrr}
        2.013 & -0.359 & 1.176 \\ 
        -0.662 & 0.541 & 0.742 \\ 
        0.343 & -0.624 & -0.954
    \end{array}\right)$ \\ 
    & $y^{N} \simeq
    \left(\begin{array}{rrr}
        -1.164 & 0.857 & -1.060 \\ 
        0.857 & 1.496 & -0.562 \\ 
        -1.060 & -0.562 & 1.259
    \end{array}\right)$ \\ 
    \hline
    VEV, Value & $v_{\phi}\simeq0.300\cdot e^{2.723i}\ ,\ \mathcal{V}_{\mathrm{opt}}\simeq-1.100$ \\ 
    \hline
    \hline
    Charges &
    ${\cal Q}=\left(\begin{array}{ccc|ccc|ccc|cc}
        L_{1} & L_{2} & L_{3} & N_{1} & N_{2} & N_{3} & l_{1} & l_{2} & l_{3} & H & \phi \\ 
        \hline
        -3 & -3 & -2 & 7 & 8 & 4 & -2 & -1 & 2 & 2 & -1 \\ 
    \end{array}\right)$ \\ 
    \hline
    $\mathcal{O}\left(1\right)$ coeff. & $y^{l} \simeq
    \left(\begin{array}{rrr}
        1.383 & -1.292 & 0.824 \\ 
        0.843 & 1.180 & 1.222 \\ 
        1.289 & -1.438 & 1.018
    \end{array}\right) \ ,\ 
    y^{\nu} \simeq
    \left(\begin{array}{rrr}
        -0.566 & 1.811 & -1.107 \\ 
        1.388 & -2.973 & -0.638 \\ 
        -1.065 & -0.216 & -0.756
    \end{array}\right)$ \\ 
    & $y^{N} \simeq
    \left(\begin{array}{rrr}
        1.406 & -1.156 & 1.351 \\ 
        -1.156 & -1.314 & -1.341 \\ 
        1.351 & -1.341 & 1.503
    \end{array}\right)$ \\ 
    \hline
    VEV, Value & $v_{\phi}\simeq0.185\cdot e^{-0.236i}\ ,\ \mathcal{V}_{\mathrm{opt}}\simeq-1.675$ \\ 
    \hline
\end{tabular}
\end{table}

\vspace{\stretch{1}}
\newpage
\vspace*{\stretch{1}}

\begin{table}[H]
\label{table_unfixedNO_3}
\centering
\begin{tabular}{l|c}
    \hline
    Charges &
    ${\cal Q}=\left(\begin{array}{ccc|ccc|ccc|cc}
        L_{1} & L_{2} & L_{3} & N_{1} & N_{2} & N_{3} & l_{1} & l_{2} & l_{3} & H & \phi \\ 
        \hline
        3 & 3 & 2 & -3 & 0 & 0 & -3 & 0 & 0 & -2 & 1 \\ 
    \end{array}\right)$ \\ 
    \hline
    $\mathcal{O}\left(1\right)$ coeff. & $y^{l} \simeq
    \left(\begin{array}{rrr}
        1.728 & -1.717 & 1.790 \\ 
        1.225 & -0.456 & -1.589 \\ 
        -2.243 & -2.316 & -2.664
    \end{array}\right) \ ,\ 
    y^{\nu} \simeq
    \left(\begin{array}{rrr}
        -1.737 & -1.060 & 2.712 \\ 
        3.083 & -1.698 & -0.342 \\ 
        -0.396 & 0.945 & -0.287
    \end{array}\right)$ \\ 
    & $y^{N} \simeq
    \left(\begin{array}{rrr}
        -1.031 & 2.275 & 1.453 \\ 
        2.275 & -0.457 & 0.333 \\ 
        1.453 & 0.333 & 1.559
    \end{array}\right)$ \\ 
    \hline
    VEV, Value & $v_{\phi}\simeq0.181\cdot e^{-0.863i}\ ,\ \mathcal{V}_{\mathrm{opt}}\simeq-0.859$ \\ 
    \hline
    \hline
    Charges &
    ${\cal Q}=\left(\begin{array}{ccc|ccc|ccc|cc}
        L_{1} & L_{2} & L_{3} & N_{1} & N_{2} & N_{3} & l_{1} & l_{2} & l_{3} & H & \phi \\ 
        \hline
        -2 & -2 & 0 & 0 & 7 & 5 & 2 & 3 & 6 & 0 & -1 \\ 
    \end{array}\right)$ \\ 
    \hline
    $\mathcal{O}\left(1\right)$ coeff. & $y^{l} \simeq
    \left(\begin{array}{rrr}
        0.600 & 0.635 & 1.101 \\ 
        0.662 & 1.039 & 1.355 \\ 
        0.607 & -0.658 & 0.463
    \end{array}\right) \ ,\ 
    y^{\nu} \simeq
    \left(\begin{array}{rrr}
        2.163 & 0.502 & 1.764 \\ 
        1.899 & 0.519 & -1.095 \\ 
        -0.358 & -0.097 & 0.560
    \end{array}\right)$ \\ 
    & $y^{N} \simeq
    \left(\begin{array}{rrr}
        0.811 & -1.049 & 0.843 \\ 
        -1.049 & 1.112 & -0.813 \\ 
        0.843 & -0.813 & 1.163
    \end{array}\right)$ \\ 
    \hline
    VEV, Value & $v_{\phi}\simeq0.300\cdot e^{2.723i}\ ,\ \mathcal{V}_{\mathrm{opt}}\simeq-2.523$ \\ 
    \hline
    \hline
    Charges &
    ${\cal Q}=\left(\begin{array}{ccc|ccc|ccc|cc}
        L_{1} & L_{2} & L_{3} & N_{1} & N_{2} & N_{3} & l_{1} & l_{2} & l_{3} & H & \phi \\ 
        \hline
        -5 & -4 & -4 & 3 & 7 & 0 & 0 & -3 & 2 & 4 & -1 \\ 
    \end{array}\right)$ \\ 
    \hline
    $\mathcal{O}\left(1\right)$ coeff. & $y^{l} \simeq
    \left(\begin{array}{rrr}
        1.894 & 0.508 & 1.913 \\ 
        -1.790 & -1.682 & 0.160 \\ 
        2.956 & -4.174 & -0.467
    \end{array}\right) \ ,\ 
    y^{\nu} \simeq
    \left(\begin{array}{rrr}
        1.475 & 0.386 & 0.832 \\ 
        -0.092 & 0.107 & 1.297 \\ 
        1.944 & -0.483 & -1.370
    \end{array}\right)$ \\ 
    & $y^{N} \simeq
    \left(\begin{array}{rrr}
        -0.678 & 0.961 & -3.538 \\ 
        0.961 & 2.518 & 1.031 \\ 
        -3.538 & 1.031 & 0.980
    \end{array}\right)$ \\ 
    \hline
    VEV, Value & $v_{\phi}\simeq0.294\cdot e^{-2.999i}\ ,\ \mathcal{V}_{\mathrm{opt}}\simeq-1.045$ \\ 
    \hline
\end{tabular}
\end{table}

\vspace{\stretch{1}}
\newpage
\vspace*{\stretch{1}}

\begin{table}[H]
\label{table_unfixedNO_4}
\centering
\begin{tabular}{l|c}
    \hline
    Charges &
    ${\cal Q}=\left(\begin{array}{ccc|ccc|ccc|cc}
        L_{1} & L_{2} & L_{3} & N_{1} & N_{2} & N_{3} & l_{1} & l_{2} & l_{3} & H & \phi \\ 
        \hline
        -5 & -5 & -5 & 4 & 9 & 1 & -7 & -3 & -1 & 5 & -1 \\ 
    \end{array}\right)$ \\ 
    \hline
    $\mathcal{O}\left(1\right)$ coeff. & $y^{l} \simeq
    \left(\begin{array}{rrr}
        1.023 & -1.307 & -1.767 \\ 
        0.477 & 1.943 & 1.116 \\ 
        0.684 & -0.425 & -0.585
    \end{array}\right) \ ,\ 
    y^{\nu} \simeq
    \left(\begin{array}{rrr}
        -0.821 & -0.611 & -1.179 \\ 
        -0.908 & 0.623 & 0.373 \\ 
        0.548 & -0.481 & -0.604
    \end{array}\right)$ \\ 
    & $y^{N} \simeq
    \left(\begin{array}{rrr}
        -0.494 & -1.719 & 1.384 \\ 
        -1.719 & -0.665 & -1.049 \\ 
        1.384 & -1.049 & -0.629
    \end{array}\right)$ \\ 
    \hline
    VEV, Value & $v_{\phi}\simeq0.288\cdot e^{-1.881i}\ ,\ \mathcal{V}_{\mathrm{opt}}\simeq-1.138$ \\ 
    \hline
    \hline
    Charges &
    ${\cal Q}=\left(\begin{array}{ccc|ccc|ccc|cc}
        L_{1} & L_{2} & L_{3} & N_{1} & N_{2} & N_{3} & l_{1} & l_{2} & l_{3} & H & \phi \\ 
        \hline
        -2 & -3 & -2 & 3 & 1 & 5 & 1 & 3 & 0 & 2 & -1 \\ 
    \end{array}\right)$ \\ 
    \hline
    $\mathcal{O}\left(1\right)$ coeff. & $y^{l} \simeq
    \left(\begin{array}{rrr}
        -2.724 & 1.474 & 0.620 \\ 
        1.587 & 0.908 & -2.094 \\ 
        -0.151 & 0.751 & 3.821
    \end{array}\right) \ ,\ 
    y^{\nu} \simeq
    \left(\begin{array}{rrr}
        -0.646 & -0.613 & 1.543 \\ 
        2.919 & 0.630 & 0.799 \\ 
        1.475 & -0.413 & 1.272
    \end{array}\right)$ \\ 
    & $y^{N} \simeq
    \left(\begin{array}{rrr}
        2.432 & 2.685 & 0.485 \\ 
        2.685 & -1.013 & 1.135 \\ 
        0.485 & 1.135 & -2.252
    \end{array}\right)$ \\ 
    \hline
    VEV, Value & $v_{\phi}\simeq0.185\cdot e^{-0.236i}\ ,\ \mathcal{V}_{\mathrm{opt}}\simeq-0.734$ \\ 
    \hline
    \hline
    Charges &
    ${\cal Q}=\left(\begin{array}{ccc|ccc|ccc|cc}
        L_{1} & L_{2} & L_{3} & N_{1} & N_{2} & N_{3} & l_{1} & l_{2} & l_{3} & H & \phi \\ 
        \hline
        -3 & -2 & -2 & 2 & 4 & 6 & 2 & 1 & -1 & 2 & -1 \\ 
    \end{array}\right)$ \\ 
    \hline
    $\mathcal{O}\left(1\right)$ coeff. & $y^{l} \simeq
    \left(\begin{array}{rrr}
        -0.389 & 1.140 & 1.134 \\ 
        -1.416 & 2.166 & 0.899 \\ 
        -0.599 & -1.828 & 1.735
    \end{array}\right) \ ,\ 
    y^{\nu} \simeq
    \left(\begin{array}{rrr}
        1.377 & 0.500 & 0.271 \\ 
        -0.940 & 0.977 & -1.874 \\ 
        -0.621 & -0.057 & 1.075
    \end{array}\right)$ \\ 
    & $y^{N} \simeq
    \left(\begin{array}{rrr}
        -1.131 & 0.135 & 0.917 \\ 
        0.135 & -0.788 & 1.361 \\ 
        0.917 & 1.361 & 0.830
    \end{array}\right)$ \\ 
    \hline
    VEV, Value & $v_{\phi}\simeq0.172\cdot e^{1.985i}\ ,\ \mathcal{V}_{\mathrm{opt}}\simeq-0.592$ \\ 
    \hline
\end{tabular}
\end{table}

\vspace{\stretch{1}}
\newpage
\vspace*{\stretch{1}}

\begin{table}[H]
\label{table_unfixedNO_5}
\centering
\begin{tabular}{l|c}
    \hline
    Charges &
    ${\cal Q}=\left(\begin{array}{ccc|ccc|ccc|cc}
        L_{1} & L_{2} & L_{3} & N_{1} & N_{2} & N_{3} & l_{1} & l_{2} & l_{3} & H & \phi \\ 
        \hline
        2 & 3 & 1 & -7 & -8 & -1 & -2 & -5 & -1 & -1 & 1 \\ 
    \end{array}\right)$ \\ 
    \hline
    $\mathcal{O}\left(1\right)$ coeff. & $y^{l} \simeq
    \left(\begin{array}{rrr}
        -0.424 & -0.567 & 0.897 \\ 
        -0.482 & -0.787 & 0.827 \\ 
        0.141 & -0.704 & 0.565
    \end{array}\right) \ ,\ 
    y^{\nu} \simeq
    \left(\begin{array}{rrr}
        -1.243 & 1.096 & 0.396 \\ 
        -0.898 & -1.501 & -3.224 \\ 
        2.361 & 2.246 & -1.668
    \end{array}\right)$ \\ 
    & $y^{N} \simeq
    \left(\begin{array}{rrr}
        2.311 & 0.877 & -1.491 \\ 
        0.877 & -1.746 & 0.186 \\ 
        -1.491 & 0.186 & -0.283
    \end{array}\right)$ \\ 
    \hline
    VEV, Value & $v_{\phi}\simeq0.268\cdot e^{-0.166i}\ ,\ \mathcal{V}_{\mathrm{opt}}\simeq-0.720$ \\ 
    \hline
    \hline
    Charges &
    ${\cal Q}=\left(\begin{array}{ccc|ccc|ccc|cc}
        L_{1} & L_{2} & L_{3} & N_{1} & N_{2} & N_{3} & l_{1} & l_{2} & l_{3} & H & \phi \\ 
        \hline
        -3 & -2 & -2 & 4 & 5 & 2 & 1 & -1 & 0 & 2 & -1 \\ 
    \end{array}\right)$ \\ 
    \hline
    $\mathcal{O}\left(1\right)$ coeff. & $y^{l} \simeq
    \left(\begin{array}{rrr}
        -1.042 & 1.484 & 1.100 \\ 
        -0.867 & -0.756 & 1.175 \\ 
        1.026 & -0.978 & -1.210
    \end{array}\right) \ ,\ 
    y^{\nu} \simeq
    \left(\begin{array}{rrr}
        -0.563 & -1.584 & -0.739 \\ 
        -1.078 & -0.343 & -0.655 \\ 
        0.798 & 0.888 & -0.997
    \end{array}\right)$ \\ 
    & $y^{N} \simeq
    \left(\begin{array}{rrr}
        1.764 & -1.266 & -1.134 \\ 
        -1.266 & 0.375 & -1.738 \\ 
        -1.134 & -1.738 & -0.823
    \end{array}\right)$ \\ 
    \hline
    VEV, Value & $v_{\phi}\simeq0.185\cdot e^{-0.236i}\ ,\ \mathcal{V}_{\mathrm{opt}}\simeq-0.972$ \\ 
    \hline
    \hline
    Charges &
    ${\cal Q}=\left(\begin{array}{ccc|ccc|ccc|cc}
        L_{1} & L_{2} & L_{3} & N_{1} & N_{2} & N_{3} & l_{1} & l_{2} & l_{3} & H & \phi \\ 
        \hline
        -2 & -2 & -2 & 8 & 1 & 7 & 4 & 2 & -1 & 2 & -1 \\ 
    \end{array}\right)$ \\ 
    \hline
    $\mathcal{O}\left(1\right)$ coeff. & $y^{l} \simeq
    \left(\begin{array}{rrr}
        -1.475 & 0.798 & 2.082 \\ 
        -0.475 & 0.975 & -1.885 \\ 
        -1.395 & -1.548 & -0.236
    \end{array}\right) \ ,\ 
    y^{\nu} \simeq
    \left(\begin{array}{rrr}
        1.104 & 1.783 & 1.208 \\ 
        0.906 & 0.454 & -0.456 \\ 
        1.244 & -0.511 & 0.218
    \end{array}\right)$ \\ 
    & $y^{N} \simeq
    \left(\begin{array}{rrr}
        1.836 & 1.151 & -1.546 \\ 
        1.151 & 2.194 & 0.400 \\ 
        -1.546 & 0.400 & -0.853
    \end{array}\right)$ \\ 
    \hline
    VEV, Value & $v_{\phi}\simeq0.185\cdot e^{-0.236i}\ ,\ \mathcal{V}_{\mathrm{opt}}\simeq-1.596$ \\ 
    \hline
\end{tabular}
\end{table}

\vspace{\stretch{1}}
\newpage

\bibliography{ref}{}

\providecommand{\href}[2]{#2}\begingroup\raggedright\begin{thebibliography}{10}

\bibitem{Altarelli:2010gt}
G.~Altarelli and F.~Feruglio, \emph{{Discrete Flavor Symmetries and Models of
  Neutrino Mixing}},
  \href{https://doi.org/10.1103/RevModPhys.82.2701}{\emph{Rev. Mod. Phys.}
  {\bfseries 82} (2010) 2701}
  [\href{https://arxiv.org/abs/1002.0211}{{\ttfamily 1002.0211}}].

\bibitem{Ishimori:2010au}
H.~Ishimori, T.~Kobayashi, H.~Ohki, Y.~Shimizu, H.~Okada and M.~Tanimoto,
  \emph{{Non-Abelian Discrete Symmetries in Particle Physics}},
  \href{https://doi.org/10.1143/PTPS.183.1}{\emph{Prog. Theor. Phys. Suppl.}
  {\bfseries 183} (2010) 1} [\href{https://arxiv.org/abs/1003.3552}{{\ttfamily
  1003.3552}}].

\bibitem{Hernandez:2012ra}
D.~Hernandez and A.Y.~Smirnov, \emph{{Lepton mixing and discrete symmetries}},
  \href{https://doi.org/10.1103/PhysRevD.86.053014}{\emph{Phys. Rev. D}
  {\bfseries 86} (2012) 053014}
  [\href{https://arxiv.org/abs/1204.0445}{{\ttfamily 1204.0445}}].

\bibitem{King:2013eh}
S.F.~King and C.~Luhn, \emph{{Neutrino Mass and Mixing with Discrete
  Symmetry}}, \href{https://doi.org/10.1088/0034-4885/76/5/056201}{\emph{Rept.
  Prog. Phys.} {\bfseries 76} (2013) 056201}
  [\href{https://arxiv.org/abs/1301.1340}{{\ttfamily 1301.1340}}].

\bibitem{King:2014nza}
S.F.~King, A.~Merle, S.~Morisi, Y.~Shimizu and M.~Tanimoto, \emph{{Neutrino
  Mass and Mixing: from Theory to Experiment}},
  \href{https://doi.org/10.1088/1367-2630/16/4/045018}{\emph{New J. Phys.}
  {\bfseries 16} (2014) 045018}
  [\href{https://arxiv.org/abs/1402.4271}{{\ttfamily 1402.4271}}].

\bibitem{Tanimoto:2015nfa}
M.~Tanimoto, \emph{{Neutrinos and flavor symmetries}},
  \href{https://doi.org/10.1063/1.4915578}{\emph{AIP Conf. Proc.} {\bfseries
  1666} (2015) 120002}.

\bibitem{King:2017guk}
S.F.~King, \emph{{Unified Models of Neutrinos, Flavour and CP Violation}},
  \href{https://doi.org/10.1016/j.ppnp.2017.01.003}{\emph{Prog. Part. Nucl.
  Phys.} {\bfseries 94} (2017) 217}
  [\href{https://arxiv.org/abs/1701.04413}{{\ttfamily 1701.04413}}].

\bibitem{Petcov:2017ggy}
S.T.~Petcov, \emph{{Discrete Flavour Symmetries, Neutrino Mixing and Leptonic
  CP Violation}},
  \href{https://doi.org/10.1140/epjc/s10052-018-6158-5}{\emph{Eur. Phys. J. C}
  {\bfseries 78} (2018) 709}
  [\href{https://arxiv.org/abs/1711.10806}{{\ttfamily 1711.10806}}].

\bibitem{Feruglio:2019ybq}
F.~Feruglio and A.~Romanino, \emph{{Lepton flavor symmetries}},
  \href{https://doi.org/10.1103/RevModPhys.93.015007}{\emph{Rev. Mod. Phys.}
  {\bfseries 93} (2021) 015007}
  [\href{https://arxiv.org/abs/1912.06028}{{\ttfamily 1912.06028}}].

\bibitem{Kobayashi:2022moq}
T.~Kobayashi, H.~Ohki, H.~Okada, Y.~Shimizu and M.~Tanimoto, \emph{{An
  Introduction to Non-Abelian Discrete Symmetries for Particle Physicists}} (1,
  2022),
  \href{https://doi.org/10.1007/978-3-662-64679-3}{10.1007/978-3-662-64679-3}.

\bibitem{Froggatt:1978nt}
C.D.~Froggatt and H.B.~Nielsen, \emph{{Hierarchy of Quark Masses, Cabibbo
  Angles and CP Violation}},
  \href{https://doi.org/10.1016/0550-3213(79)90316-X}{\emph{Nucl. Phys. B}
  {\bfseries 147} (1979) 277}.

\bibitem{Harvey:2021oue}
T.R.~Harvey and A.~Lukas, \emph{{Quark Mass Models and Reinforcement
  Learning}}, \href{https://doi.org/10.1007/JHEP08(2021)161}{\emph{JHEP}
  {\bfseries 08} (2021) 161}
  [\href{https://arxiv.org/abs/2103.04759}{{\ttfamily 2103.04759}}].

\bibitem{RL}
R.S.~Sutton and A.G.~Barto, \emph{Reinforcement learning: An introduction}, MIT
  press (2018).

\bibitem{Alonso:2018bcg}
R.~Alonso, A.~Carmona, B.M.~Dillon, J.F.~Kamenik, J.~Martin~Camalich and
  J.~Zupan, \emph{{A clockwork solution to the flavor puzzle}},
  \href{https://doi.org/10.1007/JHEP10(2018)099}{\emph{JHEP} {\bfseries 10}
  (2018) 099} [\href{https://arxiv.org/abs/1807.09792}{{\ttfamily
  1807.09792}}].

\bibitem{DBLP:journals/corr/AbadiABBCCCDDDG16}
M.~Abadi, A.~Agarwal, P.~Barham, E.~Brevdo, Z.~Chen, C.~Citro et~al.,
  \emph{Tensorflow: Large-scale machine learning on heterogeneous distributed
  systems}, {\emph{CoRR} {\bfseries abs/1603.04467} (2016) }
  [\href{https://arxiv.org/abs/1603.04467}{{\ttfamily 1603.04467}}].

\bibitem{ParticleDataGroup:2022pth}
{\scshape Particle Data Group} collaboration, \emph{{Review of Particle
  Physics}}, \href{https://doi.org/10.1093/ptep/ptac097}{\emph{PTEP} {\bfseries
  2022} (2022) 083C01}.

\bibitem{Esteban:2020cvm}
I.~Esteban, M.C.~Gonzalez-Garcia, M.~Maltoni, T.~Schwetz and A.~Zhou,
  \emph{{The fate of hints: updated global analysis of three-flavor neutrino
  oscillations}}, \href{https://doi.org/10.1007/JHEP09(2020)178}{\emph{JHEP}
  {\bfseries 09} (2020) 178}
  [\href{https://arxiv.org/abs/2007.14792}{{\ttfamily 2007.14792}}].

\bibitem{Leurer:1993gy}
M.~Leurer, Y.~Nir and N.~Seiberg, \emph{{Mass matrix models: The Sequel}},
  \href{https://doi.org/10.1016/0550-3213(94)90074-4}{\emph{Nucl. Phys. B}
  {\bfseries 420} (1994) 468}
  [\href{https://arxiv.org/abs/hep-ph/9310320}{{\ttfamily hep-ph/9310320}}].

\bibitem{KamLAND-Zen:2022tow}
{\scshape KamLAND-Zen} collaboration, \emph{{Search for the Majorana Nature of
  Neutrinos in the Inverted Mass Ordering Region with KamLAND-Zen}},
  \href{https://doi.org/10.1103/PhysRevLett.130.051801}{\emph{Phys. Rev. Lett.}
  {\bfseries 130} (2023) 051801}
  [\href{https://arxiv.org/abs/2203.02139}{{\ttfamily 2203.02139}}].

\bibitem{RoyChoudhury:2019hls}
S.~Roy~Choudhury and S.~Hannestad, \emph{{Updated results on neutrino mass and
  mass hierarchy from cosmology with Planck 2018 likelihoods}},
  \href{https://doi.org/10.1088/1475-7516/2020/07/037}{\emph{JCAP} {\bfseries
  07} (2020) 037} [\href{https://arxiv.org/abs/1907.12598}{{\ttfamily
  1907.12598}}].

\bibitem{Bordone:2019uzc}
M.~Bordone, O.~Cat\`a and T.~Feldmann, \emph{{Effective Theory Approach to New
  Physics with Flavour: General Framework and a Leptoquark Example}},
  \href{https://doi.org/10.1007/JHEP01(2020)067}{\emph{JHEP} {\bfseries 01}
  (2020) 067} [\href{https://arxiv.org/abs/1910.02641}{{\ttfamily
  1910.02641}}].

\bibitem{Kobayashi:2021pav}
T.~Kobayashi, H.~Otsuka, M.~Tanimoto and K.~Yamamoto, \emph{{Modular symmetry
  in the SMEFT}},
  \href{https://doi.org/10.1103/PhysRevD.105.055022}{\emph{Phys. Rev. D}
  {\bfseries 105} (2022) 055022}
  [\href{https://arxiv.org/abs/2112.00493}{{\ttfamily 2112.00493}}].

\bibitem{Salvio:2015cja}
A.~Salvio, \emph{{A Simple Motivated Completion of the Standard Model below the
  Planck Scale: Axions and Right-Handed Neutrinos}},
  \href{https://doi.org/10.1016/j.physletb.2015.03.015}{\emph{Phys. Lett. B}
  {\bfseries 743} (2015) 428}
  [\href{https://arxiv.org/abs/1501.03781}{{\ttfamily 1501.03781}}].

\bibitem{Ballesteros:2016euj}
G.~Ballesteros, J.~Redondo, A.~Ringwald and C.~Tamarit, \emph{{Unifying
  inflation with the axion, dark matter, baryogenesis and the seesaw
  mechanism}},
  \href{https://doi.org/10.1103/PhysRevLett.118.071802}{\emph{Phys. Rev. Lett.}
  {\bfseries 118} (2017) 071802}
  [\href{https://arxiv.org/abs/1608.05414}{{\ttfamily 1608.05414}}].

\bibitem{Ballesteros:2016xej}
G.~Ballesteros, J.~Redondo, A.~Ringwald and C.~Tamarit, \emph{{Standard
  Model\textemdash{}axion\textemdash{}seesaw\textemdash{}Higgs portal
  inflation. Five problems of particle physics and cosmology solved in one
  stroke}}, \href{https://doi.org/10.1088/1475-7516/2017/08/001}{\emph{JCAP}
  {\bfseries 08} (2017) 001}
  [\href{https://arxiv.org/abs/1610.01639}{{\ttfamily 1610.01639}}].

\bibitem{Ema:2016ops}
Y.~Ema, K.~Hamaguchi, T.~Moroi and K.~Nakayama, \emph{{Flaxion: a minimal
  extension to solve puzzles in the standard model}},
  \href{https://doi.org/10.1007/JHEP01(2017)096}{\emph{JHEP} {\bfseries 01}
  (2017) 096} [\href{https://arxiv.org/abs/1612.05492}{{\ttfamily
  1612.05492}}].

\end{thebibliography}\endgroup
\bibliographystyle{JHEP} 

\end{document}